\pdfoutput=1
\documentclass[%
 reprint,
 superscriptaddress,
 amsmath,amssymb,
 aps
]{revtex4-2}

\usepackage{array}
\usepackage{braket}
\usepackage{amsmath}
\usepackage{mathtools}
\usepackage{here}
\mathtoolsset{showonlyrefs} 
\usepackage{latexsym}
\usepackage{amsfonts}
\usepackage{amsthm}
\usepackage{mathrsfs}
\usepackage{graphicx}
\usepackage{xcolor}
\usepackage{comment}
\usepackage{bm}
\usepackage{physics}
\usepackage{hyperref}
\hypersetup{
    colorlinks=true,
    linkcolor=blue,
    filecolor=magenta,   
    citecolor=magenta,
    urlcolor=cyan
    }
\usepackage{url}
\usepackage{amssymb}
\usepackage{qcircuit}
\usepackage{afterpage}
\usepackage{qcircuit}
\usepackage[ruled,linesnumbered]{algorithm2e}

\newcommand{\beb}{\begin{itembox}}
\newcommand{\enb}{\end{itembox}}

\newcommand{\rhohat}{\hat{\rho}}
\newcommand{\zhat}{\hat{Z}}

\begin{document}

\title{Practical quantum advantage on partially fault-tolerant quantum computer}

\author{Riki Toshio}
 \email{toshio.riki@fujitsu.com}

\affiliation{
Quantum Laboratory, Fujitsu Research, Fujitsu Limited,
4-1-1 Kawasaki, Kanagawa 211-8588, Japan
}
\affiliation{
Fujitsu Quantum Computing Joint Research Division,
Center for Quantum Information and Quantum Biology, Osaka University, 1-2 Machikaneyama, Toyonaka, Osaka, 565-8531, Japan
}

\author{Yutaro Akahoshi}
\affiliation{
Quantum Laboratory, Fujitsu Research, Fujitsu Limited,
4-1-1 Kawasaki, Kanagawa 211-8588, Japan
}
\affiliation{
Fujitsu Quantum Computing Joint Research Division,
Center for Quantum Information and Quantum Biology, Osaka University, 1-2 Machikaneyama, Toyonaka, Osaka, 565-8531, Japan
}

\author{Jun Fujisaki}
\affiliation{
Quantum Laboratory, Fujitsu Research, Fujitsu Limited,
4-1-1 Kawasaki, Kanagawa 211-8588, Japan
}
\affiliation{
Fujitsu Quantum Computing Joint Research Division,
Center for Quantum Information and Quantum Biology, Osaka University, 1-2 Machikaneyama, Toyonaka, Osaka, 565-8531, Japan
}

\author{Hirotaka Oshima}
\affiliation{
Quantum Laboratory, Fujitsu Research, Fujitsu Limited,
4-1-1 Kawasaki, Kanagawa 211-8588, Japan
}
\affiliation{
Fujitsu Quantum Computing Joint Research Division,
Center for Quantum Information and Quantum Biology, Osaka University, 1-2 Machikaneyama, Toyonaka, Osaka, 565-8531, Japan
}

\author{Shintaro Sato}
\affiliation{
Quantum Laboratory, Fujitsu Research, Fujitsu Limited,
4-1-1 Kawasaki, Kanagawa 211-8588, Japan
}
\affiliation{
Fujitsu Quantum Computing Joint Research Division,
Center for Quantum Information and Quantum Biology, Osaka University, 1-2 Machikaneyama, Toyonaka, Osaka, 565-8531, Japan
}

\author{Keisuke Fujii}
\affiliation{
Fujitsu Quantum Computing Joint Research Division,
Center for Quantum Information and Quantum Biology, Osaka University, 1-2 Machikaneyama, Toyonaka, Osaka, 565-8531, Japan
}

\affiliation{
Graduate School of Engineering Science, Osaka University,
1-3 Machikaneyama, Toyonaka, Osaka, 560-8531, Japan
}
\affiliation{
Center for Quantum Information and Quantum Biology, Osaka University, 560-0043, Japan
}
\affiliation{
RIKEN Center for Quantum Computing (RQC), Wako Saitama 351-0198, Japan
}

\date{\today}

\begin{abstract}
Achieving quantum speedups in practical tasks remains challenging for current noisy intermediate-scale quantum (NISQ) devices. These devices always encounter significant obstacles such as inevitable physical errors and the limited scalability of current near-term algorithms.
Meanwhile, assuming a typical architecture for fault-tolerant quantum computing (FTQC), realistic applications
inevitably require a vast number of qubits, typically exceeding $10^6$, which seems far beyond near-term realization.
In this work, to bridge the gap between the NISQ and FTQC eras, we propose an alternative approach to achieve practical quantum advantages on early-FTQC devices. Our framework is based on partially fault-tolerant logical operations to minimize spatial overhead and avoids the costly distillation techniques typically required for executing non-Clifford gates. 
To this end, we develop a space-time efficient state preparation protocol to generate an ancillary non-Clifford state consumed for implementing an analog rotation gate with an arbitrary small angle $\theta$ and a remarkably low worst-case error rate below $\order{|\theta| p_{\text{ph}}}$, where $p_{\text{ph}}$ is the physical error rate. 
Furthermore, we propose several error suppression schemes tailored to our preparation protocol, which are essential to minimize the overhead for mitigating errors.
Based on this framework, we present several promising applications that leverage the potential of our framework, including the Trotter simulation and quantum phase estimation (QPE). 
Notably, we demonstrate that our framework allows us to perform the QPE for $(8\times 8)$-site Hubbard model with fewer than $4.9\times 10^4$ qubits and an execution time of 9 days (or 12 minutes with full parallelization) under $p_{\text{ph}}=10^{-4}$, which is significantly faster than recent classical estimation with tensor network techniques (DMRG and PEPS). 
\end{abstract}

\maketitle


\section{Introduction}

Today, full-fledged quantum computers are widely expected to enable exponential speedups in several applications including prime factoring~\cite{Shor1994,Shor1999}, simulation of materials~\cite{Lloyd1996,Abrams1999,Aspuru-Guzik2005}, and linear algebraic operations~\cite{Harrow2009}.
However, in realistic quantum devices, interactions with the environment always perturb the state of qubits, preventing us from benefiting from these quantum advantages.
To overcome such difficulties, fault-tolerant quantum computing (FTQC) architectures are designed to employ sophisticated techniques of quantum error correction and to achieve fault-tolerant implementations of logical unitary gates by utilizing, for example, magic state distillation~\cite{Bravyi2005} and lattice surgery~\cite{Horsman2012,Litinski2019}.
Unfortunately, these architectures are known to require a huge number of qubits to ensure a long lifetime of quantum coherence~\cite{Dennis2002,Fowler2012} and to provide non-Clifford operations with high gate fidelity and supply rate~\cite{Fowler2012,Gidney2019,Litinski2019magic}.
For example, in the context of materials simulations, current studies~\cite{Babbush2018qubitization, Kivlichan2020improved,Campbell2021early,Lee2021,Goings2022,Yoshioka2022hunting} have suggested that cutting-edge quantum algorithms require more than $10^5$ physical qubits to estimate the ground state energy of simple theoretical models like the Hubbard model~\cite{Hubbard1964,Arovas2022hubbard}, and more than $10^6$ physical qubits for the same task for exotic chemical systems such as the FeMo cofactor of nitrogenase~\cite{Spatzal2011evidence,Lancaster2011,Reiher2017} and the active site of cytochrome P450 enzymes~\cite{Nelson2018CYP}.
From a practical viewpoint, quantum devices satisfying these requirements seem to be far beyond near-term realization.

\begin{figure*}[tb]
    \centering
    \includegraphics[width=\linewidth]{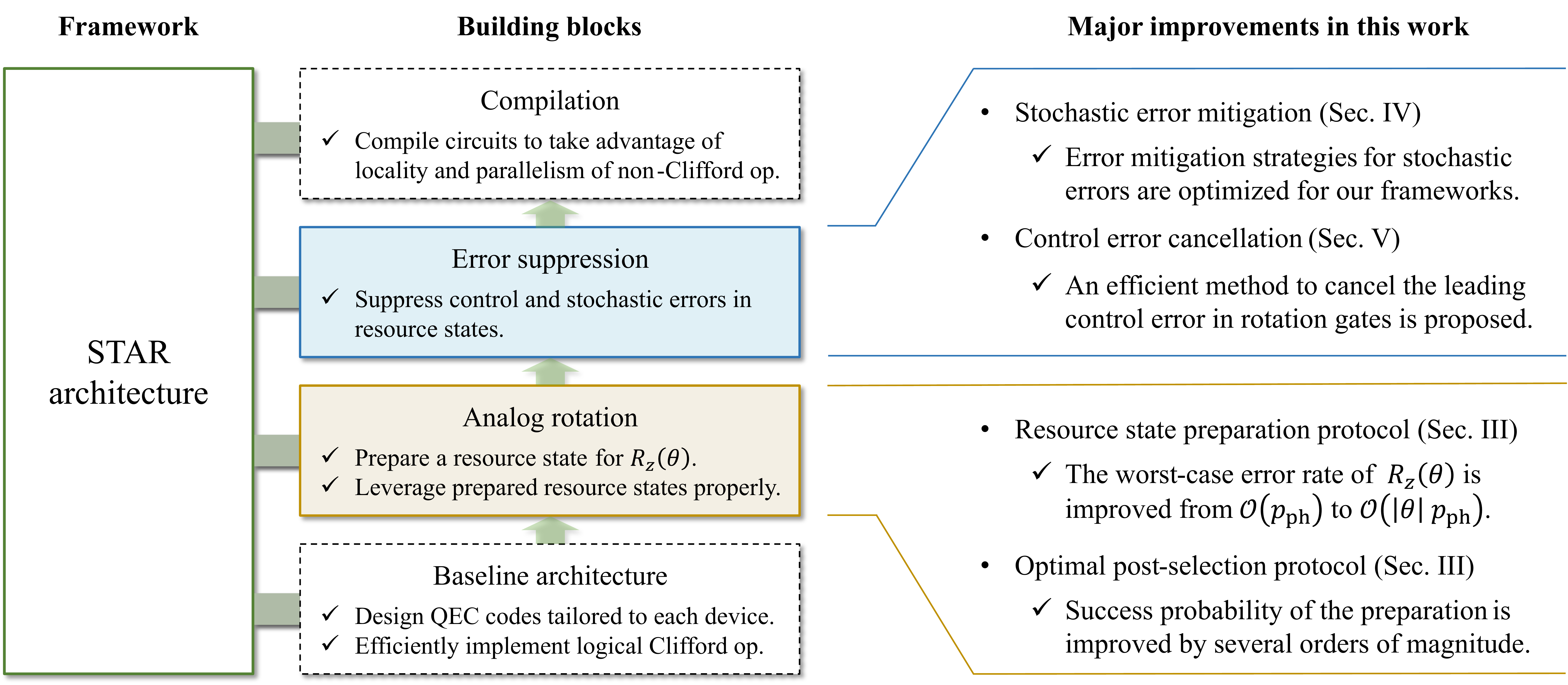}
    \caption{Overall picture of the STAR architecture and major improvements realized in this work. ({\bf Left}) The essence of the STAR architecture (or its generalization) can be summarized into four building blocks: (1) baseline architecture, (2) direct implementation of logical analog rotation gates, (3) error mitigation schemes to suppress control/stochastic errors occurring in prepared resource states, and (4) optimal compilation to take advantage of locally parallelizable rotation gates. The figure outlines the key issues that must be addressed in each building block. ({\bf Right}) In this work, we realize some remarkable improvements, especially in the second and third building blocks, updating the STAR architecture to the point where quantum speedups in practical tasks can be achieved on early-FTQC devices with only tens of thousands of qubits. Further, We present some promising scenarios for its application and a detailed resource analysis for estimating the ground state energy of the 2D Hubbard model in Sec.~\ref{sec:application} and Sec.~\ref{sec:resource estimation}, respectively.} 
    \label{fig:overview}
\end{figure*}

In the past decade, many researchers have devoted significant efforts to developing an alternative framework that works well even on noisy intermediate-scale quantum (NISQ)~\cite{Preskill2018,Bhatri2022} devices.
In such devices, we can no longer protect qubits from physical errors.
Consequently, most of the existing NISQ algorithms employ the so-called variational quantum algorithms (VQAs)~\cite{Cerezo2021review}, where we repeatedly perform quantum measurements after applying shallow parameterized quantum circuits and then perform the post-processing of measurement outcomes on classical computers.
If the circuit is sufficiently shallow, we can suppress the effect of noises in measurement outcomes utilizing quantum error mitigation techniques~\cite{Endo2021,Cai2022}, such as probabilistic error cancellation (PEC)~\cite{Temme2017,Endo2018} and zero noise extrapolation~\cite{Li2017,Temme2017,Kurita2023}. These techniques typically require additional measurement costs to suppress the amplified variance of the modified estimator.

However, these variational approaches usually face the problems of scalability in various aspects. These include the enormous measurement cost required to evaluate energy  ~\cite{Wecker2015,Elfving2020,Gonthier2022,Tilly2022}, the exponentially vanishing gradient of cost functions~\cite{McClean2018,Wang2021,Cerezo2022,Ragone2023,Larocca2024review}, the NP-hardness of variational optimization~\cite{Bittel2021NP-Hard}, and the universal cost bound of error mitigation~\cite{Takagi2022,Takagi2022_2,Tsubouchi2022}.
These challenges strongly motivate us to explore a novel framework that (partially) corrects quantum states perturbed by errors while keeping the spatial overhead as low as possible, as addressed in this study.
Such a framework would alleviate some aforementioned issues, such as exponentially growing error-mitigation costs~\cite{Takagi2022,Takagi2022_2,Tsubouchi2022} and noise-induced barren plateaus~\cite{Wang2021}, and allow us to explore more optimal approaches for estimating the expectation values of various physical quantities~\cite{Knill2007,Lin2022,Huggins2022optimal}.



To bridge the gap between the NISQ and FTQC era, researchers
have explored alternative frameworks to fully utilize quantum devices with around $10^{3}$--$10^{5}$ qubits~\cite{Piveteau2021,Suzuki2022, Akahoshi2023,Bultrini2023,Koukoulekidis2023,Katabarwa2023}. Here, we refer to such mid-sized quantum devices as ``early-FTQC" devices.
For example, Refs.~\cite{Piveteau2021,Suzuki2022} discuss a framework for such devices, where each encoded qubit has an imperfect ability to correct physical errors due to the limitation of space resources. In their framework, inevitable logical errors were suppressed via typical error-mitigation techniques such as the PEC.
More recently, another quantum computing architecture was proposed for early-FTQC devices, called the “space-time efficient analog rotation quantum computing (STAR) architecture”~\cite{Akahoshi2023}.
In the left-side of Fig.~\ref{fig:overview}, we show the overall picture of this architecture.
Within this architecture, Clifford operations are implemented fault-tolerantly using lattice surgery techniques on surface codes~\cite{Horsman2012,Litinski2019}, while noisy arbitrary rotation gates are implemented by the gate teleportation~\cite{Zhou2000} with carefully prepared non-Clifford states, referred to as resource states in this paper.
In particular, the authors carefully designed a space-time efficient state preparation protocol for the resource states to minimize logical errors in these states.
These ideas enable the execution of universal quantum computation with remarkably low space-time overhead, thereby avoiding lengthy Solovay-Kitaev decomposition~\cite{Kitaev1997_Review,Dawson2005,Ross2016} and costly magic state distillation~\cite{Fowler2012,Gidney2019,Litinski2019magic}.

However, within the STAR architecture, we still suffer from the time overhead for mitigating residual errors arising in analog rotation gates in the order of $\order{p_{\text{ph}}}$. Here, $p_{\text{ph}}$ is the physical error rate.
This imposes a clear limitation on the size of executable quantum circuits. 
While the resulting bound allows for classically intractable circuit simulation, it falls short of enabling typical practical quantum algorithms, such as quantum phase estimation (QPE).
Therefore, it is worthwhile to verify if early-FTQC devices truly possess a quantum advantage  in practical tasks such as materials simulation.

In this study, we aim to present a promising avenue for achieving a practical quantum advantage on early-FTQC devices. 
To this end, we first develop a novel quantum architecture specifically for analog rotation gate-based quantum algorithms, such as Trotter simulation, in the spirit of the STAR architecture. 
We then present promising scenarios for fully utilizing our framework and provide a detailed evaluation of the spatial (and temporal) cost for those tasks.
Notably, our architecture is well-suited not only for long-term algorithms such as the Trotter simulation or QPE but also for near-term algorithms including VQAs and other modern approaches~\cite{Huang2020measurement,Huggins2022,Xu2023QCQMC,Layden2023,Kanno2023qsci,Robledo2024QSCI}.

In what follows, we briefly illustrate our achievements in the remodeling of the STAR architecture (right side of Fig.~\ref{fig:overview}).
First, we propose a novel state preparation protocol to improve the quality of analog rotation gates, inspired by the idea presented in Ref.~\cite{Choi2023}.
Remarkably, this protocol can generate resource states for implementing analog rotation gates with an arbitrary small angle $\theta$ and a significantly low worst-case error rate below $\order{|\theta| p_{\text{ph}}}$. 
This contrasts with the original preparation protocol proposed in Ref.~\cite{Akahoshi2023}, which leads to a worse 
error rate of the order of $\order{p_{\text{ph}}}$.
This improvement in the error rate offers significant benefits for various applications, including the Trotter simulation of materials and variational quantum eigensolvers with unitary-coupled cluster ansatz. This is because these algorithms utilize a quantum circuit comprising a large number of Clifford gates and analog rotation gates with fairly small angles ($\theta \ll 1$). 
For example, in the QPE for the Hubbard model, we set each rotation angle in the Trotter circuit to be roughly $|\theta|\simeq10^{-3}$ [rad] to ensure the energy accuracy of $\epsilon=0.01$~\cite{Kivlichan2020improved}.
In such a situation, our protocol 
improves the worst-case error rate by a factor of $|\theta|\simeq 10^{-3}$ compared with the original one reported in Ref.~\cite{Akahoshi2023}.
Furthermore, the proposed framework can be applicable to generic error-correcting architectures beyond surface codes as our preparation protocol works on any stabilizer codes.
We call our preparation protocol the {\it transversal multi-rotation protocol}, because it utilizes a type of transversal rotation gates over multiple qubits and then projects the state into a desired resource state via stabilizer measurement.

Furthermore, to avoid the accumulation of errors that occur in prepared resource states, we must develop an appropriate strategy to cancel or mitigate logical errors.
For stochastic errors, we propose a randomized method that properly post-processes data qubit after gate teleportation by applying an inverse rotation probabilistically.
This method efficiently cancels the coherent (off-diagonal) part of stochastic errors without incurring exponentially growing mitigation costs.
Subsequently, we employ a standard PEC method to mitigate residual incoherent errors.
We also demonstrate that switching between different state preparation protocols depending on the target rotation angle can minimizes error accumulation during the repeat-until-success process for the gate teleportation.
In conclusion, we present a clear formula that relates the error mitigation cost to the total analog angles rotated throughout the entire circuit.
Furthermore, for systematic control errors like over-rotation, we propose another randomized method called {\it randomized transversal rotation}.
This method allows us to eliminate the leading contribution of systematic control errors and suppress the relative error in the analog rotation angle by several orders of magnitude.

Finally, we discuss several promising applications of our framework and the associated resource requirements.  
In particular, we provide a detailed estimation of the space-time resources---namely, the number of qubits and execution time---required to estimate the ground state energy of the Hubbard model~\cite{Hubbard1964,Arovas2022hubbard}.
This task has been widely studied as a benchmark of the practical quantum advantage~\cite{Babbush2018qubitization, Yoshioka2022hunting, Kivlichan2020improved}.
Remarkably, we show that our framework enables us to perform the QPE for the $(8\times 8)$-site Hubbard model with fewer than $4.9\times 10^4$ qubits and an execution time of 9 days under $p_{\text{ph}}=10^{-4}$. This is significantly faster than recent runtime estimations performed on classical computers using tensor network methods (DMRG and PEPS)~\cite{Yoshioka2022hunting}, and requires only a fraction of the number of physical qubits compared with previous FTQC studies~\cite{Babbush2018qubitization, Yoshioka2022hunting, Kivlichan2020improved}.
In principle, assuming fully parallel computation with a large number of quantum processing units, the execution time can be further reduced to $7.1\times 10^2$ seconds $\simeq 12$ minutes.
These analyses are readily applicable not only to the Hubbard model, but also to more generic systems such as extended Hubbard models derived via the ab-initio down-folding method~\cite{Kanno2022,Ivanov2023,Clinton2024,Yoshida2024} and electronic structure problems for quantum chemistry~\cite{McArdle2020}.
Furthermore, we illustrate the utility of our framework in the quantum simulation of disordered spin systems, which could be applied to understand the nature of self-thermalization in closed quantum systems~\cite{Childs2018speedup}, and discuss possible applications of near-term algorithms such as the VQAs and the quantum-selected configuration interaction~\cite{Kanno2023qsci}.
These results strongly suggest that our framework has great potential to achieve quantum speedups in several practical tasks before the arrival of full-fledged FTQC devices.

This paper is organized as follows. In Sec.~\ref{sec:Preliminary}, we review the framework of the STAR architecture proposed in Ref.~\cite{Akahoshi2023}, and slightly generalize their idea by introducing a more formal definition to include the framework that we propose. In Sec.~\ref{sec:preparation protocol}, we discuss the resource state preparation protocol for implementing logical rotation gates with an arbitrary rotation angle.
We review the recent proposal in Ref.~\cite{Choi2023} and then, develop a novel preparation protocol that generalizes the two independent ideas in Ref.~\cite{Choi2023} and Ref.~\cite{Akahoshi2023}.
We also discuss methods to optimize the post-selection process after stabilizer measurements to maximize the success rate of the protocol without worsening the error rate of the selected states. 
In Sec.~\ref{sec:Stochastic error mitigation}, we formulate the error model of the noisy logical rotation gate produced using the above preparation protocol and gate-teleportation scheme. We then propose several error mitigation strategies optimized for the error model.
In Sec.~\ref{sec:Control error suppression}, we present a randomized method to suppress the systematic control errors that arise in our preparation protocol.
In Sec.~\ref{sec:application}, we showcase several promising applications of our framework, which includes the Trotter simulation, the QPE, and some near-term quantum algorithms.
In Sec.~\ref{sec:resource estimation}, we present resource estimation for QPE for many-body Hamiltonian like the Hubbard model. 
Finally, we conclude our study in Sec.~\ref{sec:conclusion}, outlining some remaining open issues and providing an outlook on future directions. In Appendix.~\ref{Appendix:notations}, we list notations that are frequently used in this paper. Appendix.~\ref{Appendix:Akahoshi injection} and subsequent sections offer detailed technical discussions of the concepts presented in the main text.


\section{Preliminary: STAR architecture}
\label{sec:Preliminary}

The original concept of the STAR architecture was initially proposed as a promising framework for early-FTQC devices in Ref.~\cite{Akahoshi2023}.
In our paper, one of the goals is to significantly improve the performance of this architecture, by replacing their original gadgets for resource state preparation and error mitigation schemes with more sophisticated ones.
In addition, we offer a solution to a severe issue concerning systematic control errors, which was not addressed in the original work~\cite{Akahoshi2023}.
To this end, in this section, we provide an overview of the preliminary details of the STAR architecture, outlining challenges it encounters and our accomplishments realized in this study.

\subsection{Original construction}

In the original STAR architecture~\cite{Akahoshi2023}, each logical qubit is constructed based on the rotated planar surface code~\cite{Horsman2012}, and the quantum computation running on it is implemented with two types of operations: (i) fault-tolerant Clifford operations with lattice surgery and (ii) analog rotation gates with reasonably clean ancilla state preparation.
The first one is familiar also in conventional FTQC architectures~\cite{Horsman2012,Litinski2019,Fowler2018}. 
The lattice surgery techniques comprise two patch merging, splitting, and patch deformation operations, and these techniques enable us to implement any Clifford operations on planar surface codes even under the constraints of nearest-neighbor connectivity.
Readers unfamiliar with these concepts can refer to the original works~\cite{Horsman2012,Litinski2019,Fowler2018} or concise introduction provided in Ref.~\cite{Akahoshi2023}, which will be enough to understand this work.

On the other hand, the second operation is notably in contrast to the one used in the usual FTQC architecture, where usually the basic gate set of Clifford$+T$ or Clifford$+$Toffoli is usually used. In the case of FTQC, we need to implement several magic state factories, which require large amounts of physical qubits and a long latency time to prepare magic states successfully, to implement a high-fidelity non-Clifford gate such as $T$ or Toffoli gate~\cite{Bravyi2005,Fowler2012,Gidney2019,Litinski2019magic}.
Moreover, the parallel implementation of non-Clifford gates necessitates further scaling up of magic state factories and adequate routing areas.
In addition, implementing analog rotation gates with $T$ gates requires a gate synthesis via lengthy Solovay-Kitaev decomposition~\cite{Kitaev1997_Review,Dawson2005,Ross2016}.
The state-of-the-art optimal Clifford+$T$ decomposition~\cite{Ross2016} still requires several tens or hundreds
of $T$ gates to achieve a highly accurate arbitrary single-qubit rotational gate.

The STAR architecture avoids these costly processes by preparing a special ancilla state for the direct implementation of an analog rotation gate in a non-fault-tolerant manner.
This successfully reduces the computational cost of executing intermediate-scale quantum circuits with a limited number of physical qubits.
Ref.~\cite{Akahoshi2023} suggested that the STAR architecture allows for the reliable implementation of $3.75\times 10^4$ arbitrary rotation gates and $1.72\times
10^7$ Clifford gates on 64 logical qubits by assuming devices with only $10^4$ physical qubits and the physical error rate of $p_{\text{ph}}=10^{-4}$. Such computations cannot be simulated on classical computers, and the existing NISQ
and FTQC architectures on the same device still cannot realize this amount of computational power.

In the following subsections, we will illustrate how to implement analog rotation gates in the original STAR architecture more specifically.

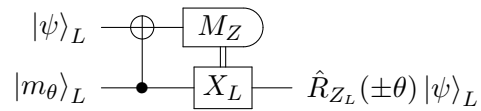
\begin{figure}[tbp]
\hspace{-10mm}
\centering
\fontsize{11pt}{11pt}\selectfont
  \mbox{
  \Qcircuit @C=1em @R=.7em {
    \lstick{\ket{\psi}_L}                      & \targ     & \measureD{M_{Z}}  &  \\
    \lstick{\ket{m_\theta}_L} & \ctrl{-1} & \gate{X_L} \cwx  & \rstick{\hat{R}_{Z_L}(\pm\theta) \ket{\psi}_L } \qw 
    }
  }
  \caption{Quantum circuit for implementing an analog $Z$ rotation gate $\hat{R}_{z,L}(\theta)$. $M_Z$ denotes destructive $Z_L$ measurement on a encoded qubit.}
  \label{fig:GT_circ}
\end{figure}

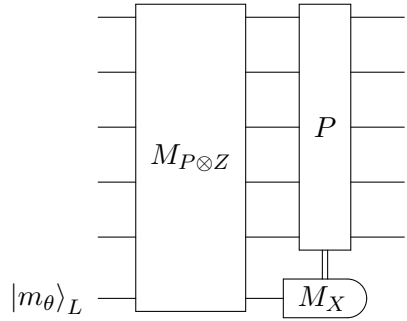
\begin{figure}[tbp]
  \centering
  \fontsize{11pt}{11pt}\selectfont
  \mbox
  {
  \Qcircuit @C=1.3em @R=1.0em {
    & \multigate{5}{M_{P \otimes Z}} & \multigate{4}{P} & \qw\\
    & \ghost{M_{P \otimes Z}} & \ghost{P} & \qw\\
    & \ghost{M_{P \otimes Z}} & \ghost{P} & \qw\\
    & \ghost{M_{P \otimes Z}} & \ghost{P} & \qw\\
    & \ghost{M_{P \otimes Z}} & \ghost{P} & \qw\\
    \lstick{\ket{m_\theta}_L} & \ghost{M_{P \otimes Z}} & \measureD{M_X} \cwx & 
  }
  }
  \caption{Quantum circuit for implementing an analog multi-Pauli rotation gate $\hat{R}_{P,L}(\theta)$. In this setup, we can choose any Pauli string operator $\hat{P}$.}
  \label{fig:mprot_circ}
\end{figure}

\subsubsection{Repeat-until-success implementation of analog rotation gate}
\label{sec:RUS}

In the STAR architecture, we exploit the following type of ancillary non-Clifford states instead of magic states~\cite{Bravyi2005}:
\begin{equation}
\label{eq:resource_state}
    \ket{m_\theta}_L \equiv \hat{R}_{z,L}(\theta)\ket{+}_L = \cos\theta \ket{+}_L +i\sin\theta \ket{-}_L,
\end{equation}
where any encoded quantum state is denoted as $\ket{\cdots}_L$ and a logical Pauli-$Z$ rotation gate as $\hat{R}_{z,L}(\theta)=e^{i\theta \hat{Z}_L}$~\cite{comment}. The rotation angle $\theta$ is arbitrarily chosen. In what follows, we will refer to this type of non-Clifford state as a {\it resource state}.
As illustrated in Fig.~\ref{fig:GT_circ}, we execute an analog Pauli-$Z$ rotation gate on any target state $\ket{\psi}_L$ non-deterministically by entangling it with a resource state $\ket{m_\theta}_L$ through the gate-teleportation circuit~\cite{Zhou2000}.
The output state becomes a correctly rotated state $\hat{R}_{z,L}(\theta)\ket{\psi}_L$ if
the measurement outcome is $+1$ with a probability of $1/2$; otherwise, the
output becomes an inversely rotated state $\hat{R}_{z,L}(-\theta)\ket{\psi}_L$.
If the inversely rotated state is obtained, we can repeat the teleportation process to correct its rotation direction by doubling the rotation angle of the input resource state. 
This procedure is repeated until we obtain a measurement outcome of $+1$, thereby yielding the desired state $\hat{R}_{z,L}(\theta)\ket{\psi}_L$~\cite{Jones2012}. We will refer to this procedure as {\it repeat-until-success} (RUS) procedure.
As is easily checked, this procedure succeeds in two trials on average.

More generally, we can implement any multi-Pauli rotation gates via the quantum circuit in Fig.~\ref{fig:mprot_circ}, consuming a single resource state $\ket{m_\theta}_L$. 
This circuit is based on multi-Pauli measurement rather than multiple $CNOT$ gates. This is preferable for an efficient implementation of multi-Pauli rotations via the lattice surgery techniques~\cite{Litinski2019}. Therefore, we usually assume the circuit in Fig.~\ref{fig:mprot_circ} for implementing these rotation gates.

{\renewcommand{\arraystretch}{1.5}
\begin{table*}
    \centering
    \caption{Comparison between a typical FTQC architecture and an alternative architecture proposed for early-FTQC devices in this paper. The most notable feature of the latter architecture is to utilize arbitrary analog rotation gates as a non-Clifford part of the logical gate set, instead of the $T$ or Toffoli gate. This is achieved by applying space-time efficient resource state preparation protocols for small analog rotations, called as transversal multi-rotation. Because the rotation gates obtained by the protocol have quite high but limited fidelity, our architecture requires error mitigation to suppress the effect of stochastic errors and coherent errors.}
    \begin{tabular}{p{4.8cm}p{0.1cm}p{6cm}p{0.1cm}p{6cm}p{0.1cm}}
    \hline\hline
         && Typical FTQC arcihtecture && Our arcihtecture for early-FTQC&\\
        \hline
        Basic logical gate set && $\{H,S,T,\text{CNOT}\}$ or $\{H,S,\text{Toffoli},\text{CNOT}\}$ && $\{H,S,R_z(\theta),\text{CNOT}\}$ &\\ 
        Error correcting code && (Rotated) Planar surface codes {\it et al.} && (Rotated) Planar surface codes {\it et al.} &\\ 
        Implementation of Clifford gates && Lattice surgery {\it et al.} & &Lattice surgery {\it et al.}&\\ 
        Implementation of analog rotation gates && Gate synthesis with a number of $T$ gates distilled in magic state factories && Direct implementation with the transversal multi-rotation protocol&\\ 
        Parallelism of Non-Clifford gates && Parallelism of $T$ gate strictly depends on the supply rate of magic state factories. && Any set of spatially separate rotation gates can be executed simultaneously.  &\\ 
        Error mitigation && Not necessarily needed. && 
        \begin{tabular}{p{5.8cm}}
             (Stochastic error) We use the probabilistic coherent error cancellation to cancel the off-diagonal part of stochastic errors, and the usual probabilistic error cancellation to mitigate the remaining part of them.  \\ 
             (Control error) We use the randomized transversal rotation method to cancel systematic control errors.
        \end{tabular}  &\\
        Limitation on gate counts && In principle, any number of gates can be executed as long as we allow to consume any amount of physical qubits and execution time. && Total analog rotation angle that can be executed without excessive mitigation cost is restricted by a universal bound (Eq.~\eqref{eq: universal bound}) that depends on the physical error rate. &\\
        Target algorithms  && Any quantum algorithms available within hardware resources  && Analog rotation gate-based quantum algorithms, such as VQAs and Trotterization &\\
        \hline\hline
    \end{tabular}
    \label{tab:compare STAR}
\end{table*}
}

\subsubsection{Resource state preparation protocol}
\label{sec:Akahoshi injection}

The key technology of the STAR architecture is a space-time efficient preparation protocol for a resource state $\ket{m_\theta}_L$.
To develop the protocol, the authors of Ref.~\cite{Akahoshi2023} employed the $[[4, 1, 1, 2]]$-quantum subsystem code~\cite{Bacon2006} (for details see Ref.~\cite{Akahoshi2023} or Appendix.~\ref{Appendix:Akahoshi injection}).
In their protocol, they carefully prepared a resource state on the subsystem code, and then extended it to a surface code with some larger code distance via a patch deformation technique~\cite{Horsman2012,Litinski2019}.
Then, by teleporting the prepared resource state, they implemented a noisy analog rotation channel described as follows:
\begin{equation}
\begin{aligned}
\mathcal{N}_{\theta}^{\text{org}}:\ \hat{\rho} \ \to \ \mathcal{N}_{\theta}^{\text{org}}(\hat{\rho}) =&\  \mathcal{E}^{\text{org}} \circ \mathcal{R}_{\theta}(\hat{\rho})
\end{aligned}
\end{equation}
where $\mathcal{R}_{\theta}$ represents an ideal logical rotation gate with the target angle $\theta$,
\begin{equation}
\label{eq:ideal rotation channel}
    \mathcal{R}_{\theta}:\ \hat{\rho} \ \to \ \mathcal{R}_{\theta}(\hat{\rho}) = \hat{R}_{z,L}(\theta) \hat{\rho} \hat{R}_{z,L}^\dagger(\theta),
\end{equation}
and $\mathcal{E}^{\text{org}}$ denotes a stochastic Pauli-$Z$ error channel:
\begin{equation}
\begin{aligned}
    \mathcal{E}^{\text{org}}:\ \hat{\rho} \ \to \ \mathcal{E}^{\text{org}}(\hat{\rho}) = (1-P_L^{\text{org}})\cdot\hat{\rho}+ P_L^{\text{org}}\cdot \hat{Z}\hat{\rho} \hat{Z}. 
\end{aligned}
\end{equation}
Here, we introduce the label ``org" denoting  ``original" to distinguish their protocol from ours.
According to the theoretical and numerical calculation in Ref.~\cite{Akahoshi2023}, the logical error rate is determined as $P_L^{\text{org}}=\frac{2}{15}p_{\text{ph}} + \mathcal{O} (p^2_{\text{ph}})$ using a circuit-level noise model.
This is much better than the error rates reported in previous works~\cite{Li2015magic,Lao2022}.
As shown in the next section, we can improve the value of $P_L^{\text{org}}$ from $\frac{2}{15}p_{\text{ph}}$ to $\frac{1}{15}p_{\text{ph}}$ by modifying the protocol slightly.

Compared to the usual FTQC approach, it is notable that their protocol does not necessarily require the presence of an ancillary patch region dedicated to resource state preparation, unlike usual distillation techniques~\cite{Fowler2012,Gidney2019,Litinski2019magic}. 
The protocol requires only a single logical patch, and successful preparation is realized with a high probability, provided that the code distance and physical error rate are reasonably small.
Therefore, even for resource state preparation, we can exploit ancillary patch regions provided for logical operations, without implementing an additional spatial overhead. This also allows us to execute multiple rotation gates by running the preparation protocol at several ancilla patches in parallel.

\subsubsection{Error mitigation}

The residual error of the prepared resource state causes a small noise in the resulting rotation gate.
These noises can be mitigated using a standard error mitigation technique like PEC method.
In the case of the STAR architecture, we readily find that the PEC imposes an additional sampling overhead of $\gamma^2 \simeq e^{8N_{\text{rot}}P_{z,L}}$, where $N_{\text{rot}}$ is the number of analog rotations in the overall circuit.
Thus, we must maintain the gate number $N_{\text{rot}}$ within the order of $\order{1/P_{z,L}}$ to circumvent the exponential delay of quantum computation.
This is a clear bound that fairly limits the utility of the STAR architecture.

\subsection{Scope of applications}

Here, we briefly discuss the type of QEC codes to be assumed as a scope of applications of the STAR architecture.
As outlined above, in the original proposal in Ref.~\cite{Akahoshi2023}, the injection protocol strongly depends on the techniques of code deformation and the locality of surface codes.
By contrast, our injection protocol proposed below is based on a totally different technique, and it can generate a resource state with much higher fidelity, even
on any stabilizer codes, which include not only planar surface codes discussed in Ref~\cite{Akahoshi2023} but also more exotic QEC codes such as quantum low-density parity-check (LDPC) codes~\cite{Breuckmann2021} and latest quantum concatenated codes~\cite{Gidney2023yoked, Pattison2023hierarchical, Yamasaki2024constant,Yoshida2024concatenate, Goto2024}.
This update extends the scope of applications of the STAR architecture to more general fault-tolerant architectures that may suit quantum computing platforms such as superconducting circuits~\cite{Huang2020superconducting}, neutral atoms~\cite{Henriet2020neutral,Bluvstein2023}, trapped ions~\cite{Bruzewicz2019trapped,Silva2024}, photons~\cite{Slussarenko2019photonic}, and quantum dots~\cite{Zhang2018}.

In this work, to put these developments in perspective, we reinterpret the definition of the STAR architecture more broadly.
Specifically, we refer to the class of quantum computing architectures based on the following design principles as ``space-time efficient analog rotation quantum computing (STAR) architecture":
\begin{itemize}
    \item {\bf Partial fault-tolerance}: Quantum information is encoded on some error-correcting codes, and arbitrary Clifford operations are performed on it in a fault-tolerant manner.
    \item {\bf Noisy analog rotation gates}: Analog rotation gates are implemented using a non-fault-tolerant resource state preparation protocol followed by the gate teleportation of the prepared states. 
    \item {\bf Clifford + $\phi$ gate set}: Most logical operations are performed by synthesizing a gate set composed of the Clifford gates and analog rotation gates.
    \item {\bf Error mitigation}: Some error mitigation strategies are employed to suppress quantum errors occurring in analog rotation gates.
\end{itemize}
In Table.~\ref{tab:compare STAR}, we highlight the key features of the STAR architecture updated in this study, comparing it with typical FTQC architectures.
In particular, an essential difference from full-FTQC architecture is that the STAR architecture does not require any special equipment and footprint to implement non-Clifford gates such as the magic state factory and code switching to high-dimensional QEC codes in the conventional full-FTQC architecture.
This will remarkably broaden the scope of research on QEC codes 
toward the early-stage applications of quantum devices as a partially fault-tolerant quantum computer.

\section{State preparation protocol for small-angle rotation gates}
\label{sec:preparation protocol}

In this section, we formulate a novel resource state preparation protocol, dubbed {\it transversal multi-rotation protocol}, for implementing logical rotation gates with an arbitrary small angle. This protocol enables us to prepare a resource state $\ket{m_{\theta_*}}$ for logical $R_{z,L}(\theta_*)$ gates with a notably small infidelity of $\theta_*^{2(1-1/k)} P_{\text{ud}}$ in the small-angle limit ($\theta_* \ll 1$). 
Here, $P_{\text{ud}}\simeq \frac{k}{15} p_{\text{ph}}+\mathcal{O}({p_{\text{ph}}^2})$ represents the total error rate undetectable in the error detection process, and $k\ (\geq 2)$ is an integer appearing in our protocol, which is proportional to the code distance $d$. Notably, our protocol necessitates only one logical patch and an average execution time of approximately a single clock ($d$ code cycle) to prepare a resource state successfully. 
This represents considerable efficiency in terms of space and time compared with the conventional approach that utilizes lengthy Solovay-Kitaev decomposition~\cite{Kitaev1997_Review,Dawson2005,Ross2016} and costly magic state distillation~\cite{Fowler2012,Gidney2019,Litinski2019magic}.

In materials simulation with Trotter–Suzuki decomposition~\cite{Trotter1959,Suzuki1990,Suzuki1991}, we run a long sequence of rotation gates that have fairly small angles to ensure the accuracy of decomposition. In this situation, our protocol has a significant advantage, as the fidelity of the prepared resource states improves with decreasing the rotation angles.
A similar discussion will be held for variational quantum eigensolver with ansatzes such as unitary-coupled cluster ansatz~\cite{Bartlett1989,Hoffmann1988,Romero2018} and variational Hamiltonian ansatz~\cite{Wecker2015,Park2024}.
Moreover, several studies~\cite{Haug2021,Zhang2022escaping,Wang2023trainability,Park2024} revealed that keeping variational angles small often helps avoid the barren plateau problems, implying that our framework may be suitable not only for materials simulation but also for more general VQA tasks such as quantum machine learning.
We will discuss these promising applications in detail in Sec.~\ref{sec:application}.

\subsection{Choi \textbf{\textit{et al.}} protocol}
\label{sec:Choi}

The idea of our preparation protocol is closely related to the technique shown in Sec.~\ref{sec:Preliminary} and a key technique reported by Choi \textit{et al.}~\cite{Choi2023}. In what follows, we give a brief explanation regarding the latter idea.

The Choi \textit{et al.} protocol starts with a fault-tolerantly initialized Clifford state. Here let us assume that we prepare a logical state $\ket{+}_L$ of an arbitrary $[[n,1,d]]$ error-correcting code that has a logical-$Z$ operator in a form,
\begin{equation}
\label{eq:logical Z}
    \hat{Z}_L \equiv \prod_{i\in Q_z} \hat{Z}_i,
\end{equation}
where $\hat{Z}_i$ is a Pauli-$Z$ operator acting on the $i$-labeled
physical qubit and $Q_z$ is the set of qubits that define the support of $\hat{Z}_L$.
There is always some degree of arbitrariness in the choice of the set $Q_z$, and we only consider the case where $|Q_z|=d$. 
Any stabilizer codes can have a logical-$Z$ operator in the above form by tuning the physical Pauli frame of each qubit.

\begin{figure}[tb]
    \centering
    \hspace{-0.3cm}
    \includegraphics[width=8.7cm]{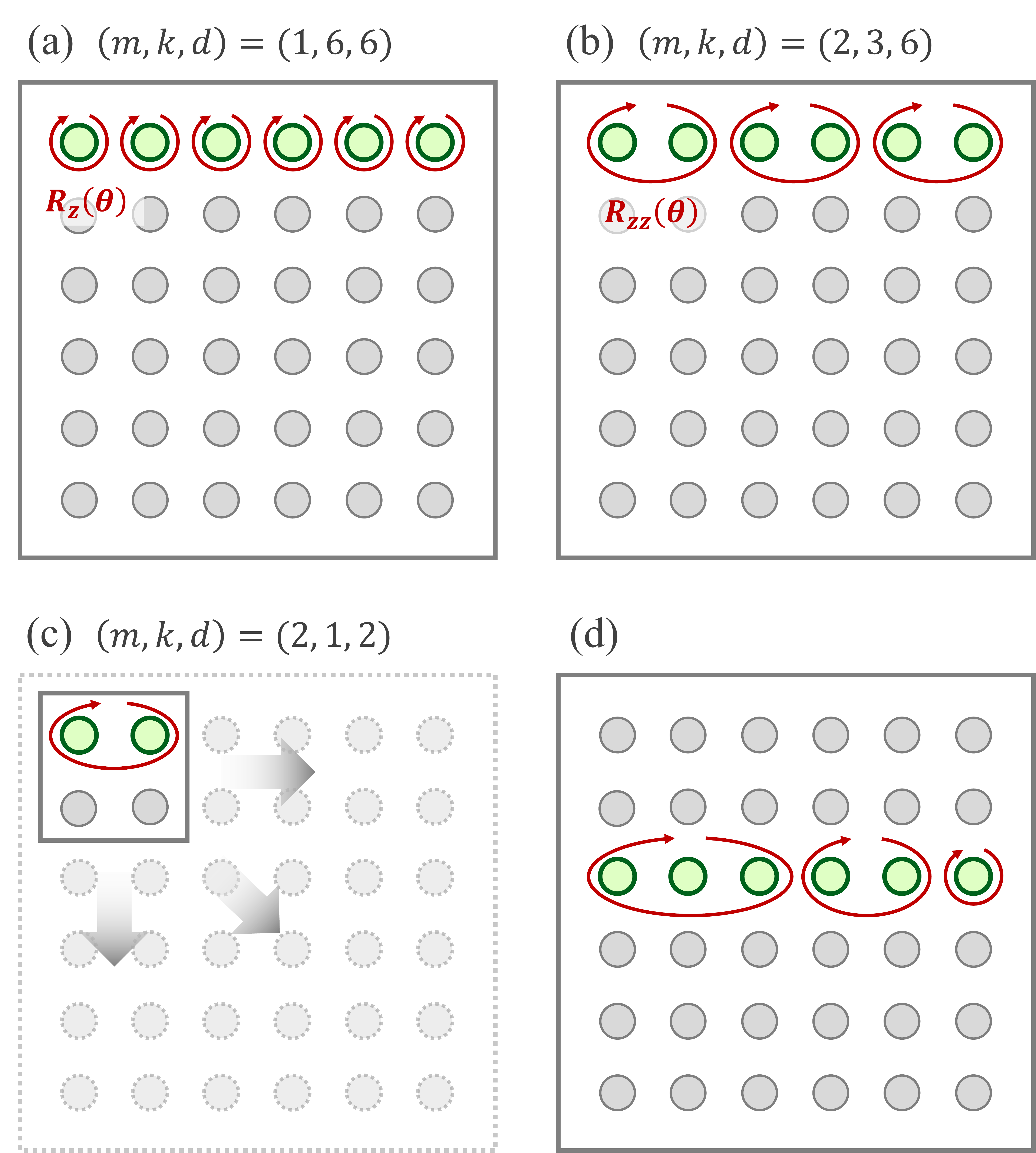}
    \caption{Schematic of transversal multi-rotation protocol on a rotated surface code with the code distance $d=6$. Every circle represents physical qubits that make up the surface code. The qubit set $Q_z$ for logical-$Z$ operation is colored in green. {\bf (a)} The case of $(m,k,d)=(1,6,6)$. This coincides with the Choi {\it et al.} protocol. {\bf (b)} The case of $(m,k,d)=(2,3,6)$. {\bf (c)} The case of $(m,k,d)=(2,1,2)$. This case is essentially equivalent to the protocol proposed in Ref.~\cite{Akahoshi2023}, where we expand the logical patch from $d=2$ to an arbitrary code distance right after post-selection. {\bf (d)} A more generic case of our protocol.  Generically, we can select another qubit set as a support of logical-$Z$ operation, and the transversal rotation gate can include multi-Pauli rotation gates with different weight. In such a case, we can no longer specify the protocol with only three parameters $(m,k,d)$.}
    \label{fig:transversal muti-rotation protocol}
\end{figure}

In general, a transversal rotation gate around the $Z$ axis on the set $Q_z$ (Fig.~\ref{fig:transversal muti-rotation protocol} (a)) can be decomposed into the following form:

\begin{equation}
\label{eq:Transversal_rotation}
\begin{aligned}
    \prod_{i\in Q_z} \hat{R}_{z,i}(\theta) &= \prod_{i\in Q_z} \left[ \cos\theta\cdot \hat{I}_i +i \sin\theta\cdot \hat{Z}_i \right]\\
    &= \sum_{n_z=0}^d  i^{n_z}\sin^{n_z}\theta  \cos^{d-n_z} \theta
    \sum_{wt(\hat{P})=n_z} \hat{P}, 
\end{aligned}
\end{equation}
where $\hat{R}_{z,i}(\theta)=e^{i\theta\hat{Z}_i}$ is a rotation gate acting on the $i$-th physical qubit~\cite{comment}, $wt(\hat{P})$ denotes the number of Pauli operators acting non-trivially in a Pauli string $\hat{P}$, and the sum in the last line is over all Pauli $Z$-strings that satisfy $wt(\hat{P})=n_z$. 
When $wt(\hat{P})=d$, the Pauli $Z$-string $\hat{P}$ coincides with the logical-$Z$ operator in Eq.~\eqref{eq:logical Z}.
Otherwise, each Pauli $Z$-string operator functions as an effective error that projects the input logical state out of the logical space.

Next, let us apply this transversal rotation gate to a prepared logical state $\ket{+}_L$.
Considering the condition of Eq.~\eqref{eq:logical Z}, we obtain a kind of noisy logical state,

\begin{widetext}
\begin{equation}
\label{eq: Choi protocol}
\begin{aligned}
    \prod_{i\in Q_z} \hat{R}_{z,i}(\theta) \ket{+}_L
    &\ =\  \cos^d \theta\ket{+}_L  + i^{d}\sin^d\theta  \ket{-}_L
     \ +\  (Z\text{-error terms})\\
    &\ =\ \left\{
        \begin{array}{ll}
            \sqrt{p_{\text{ideal}}} \cdot e^{-i\pi/4} \hat{R}_{x,L}(-\pi/4) \ket{m_{(-1)^j\theta_*}}_L \ +\  (Z\text{-error terms}) & \quad (d=2j) \\
            \sqrt{p_{\text{ideal}}} \cdot \ket{m_{(-1)^j\theta_*}}_L  \ +\  (Z\text{-error terms}) & \quad (d=2j+1)
        \end{array}
        \right.
\end{aligned}
\end{equation}
\end{widetext}
where $\ket{m_\theta}_L$ is the resource state introduced in Eq.~\eqref{eq:resource_state}.
The parameters $\theta_*$ and $p_{\text{ideal}}$ denote the logical rotation angle of the resource state and ideal success rate of Choi {\it et al.} protocol explained later, respectively.
The last term on the right side denotes a set of terms that correspond to noisy states that Pauli strings $\hat{P}$ other than $\hat{I}_L$ and $\hat{Z}_L$ act on and therefore are outside the logical space.
These error states can be removed by performing syndrome measurements twice to suppress measurement errors and by post-selecting the case where all measurement outcomes are $+1$. In particular, when we neglect physical noises, the probability that post-selection succeeds is given by
\begin{equation}
\label{eq:ideal success rate}
    p_{\text{ideal}}(\theta,d)\  \equiv\  \sin^{2d}\theta+\cos^{2d}\theta \ \simeq\  1- d \theta^2
    +\ \order{\theta^4}
\end{equation}
and, the rotation angle of the resource state is 
\begin{equation}
\label{eq:angle relation}
    \theta_*(\theta,d)\  \equiv\  \sin^{-1}\left(
        \frac{1}{\sqrt{p_{\text{ideal}}}} \sin^d\theta\right)
         \ \simeq \  \theta^d
    +\ \order{\theta^{d+2}}.
\end{equation}
Here, in the last equalities, we assume the case where the input angle $\theta$ is sufficiently small ($\theta \ll 1$).
If necessary, we can remove the extra sign of rotation angle or overall extra gate $\hat{R}_{x,L}(-\pi/4)$ by performing appropriate Clifford operations.

In summary, Choi {\it et al}. protocol enables the preparation of a resource state $\ket{m_{\theta_*}}_L$ to implement the rotation gate $\hat{R}_{z,L}(\theta_*)$ by following three procedures: (i) prepare a logical state $\ket{+}_L$, (ii) apply transversal Pauli-$Z$ rotation gate in Eq.~\eqref{eq:Transversal_rotation}, (iii) perform syndrome measurements twice and post-select the case that all the outcomes are $+1$. Otherwise, the state is discarded, and the procedures are repeated.

In actual devices, each gate operation in these procedures always accompanies an inevitable error, leading to the finite infidelity of output states.
While most of these errors are detected in the procedure (iii), a small part of them end up passing the post-selection process accidentally.
For example, when a single Pauli-$Z$ error arises on $Q_L$ right before and after the transversal gate in Eq.~\eqref{eq:Transversal_rotation}, the coefficients for $\ket{\pm}_L$ in Eq.~\eqref{eq: Choi protocol} are modified from $\cos^d\theta, i^d\sin^d\theta$ to $i\sin\theta \cos^{d-1}\theta, i^{d-1}\sin^{d-1}\theta \cos\theta$ respectively.
This type of the error state is indistinguishable from the ideal state via syndrome measurement, and thus contributes to a reduction in the fidelity of the output state.
In conclusion, the leading term of the state infidelity $1-F^{\text{c}}$ is approximately estimated as (for details of derivation, refer to the discussion around Eq.~\eqref{eq:state fidelity of multi-rotation protocol})
\begin{equation}
\label{eq: fidelity of Choi}
\begin{aligned}
    1-F^{\text{c}} \ &\simeq\  P^{\text{c}}_{\text{ud}}\cdot \sin^{2(d-1)}\theta \cos^{-1}\theta \\
    &\simeq\  P^{\text{c}}_{\text{ud}}\cdot\theta^{2(d-1)}\\
    &\simeq P^{\text{c}}_{\text{ud}}\cdot\theta_*^{2(1-1/d)},
\end{aligned}
\end{equation}
where $P^{\text{c}}_{\text{ud}}$ is the total error probability of undetectable errors occurring in the Choi {\it et al}. protocol, and $F^{\text{c}} \equiv \ev{\hat{\rho}^{\text{c}}_{\text{out}}}{m_{\theta_*}}_L$ is the state fidelity between the ideal resource state $\ket{m_{\theta_*}}_L$ and output state $\hat{\rho}^{\text{c}}_{\text{out}}$ of the protocol.
Therefore, using the Choi {\it et al}. protocol, we can prepare a resource state for a analog rotation gate with almost quadratically small infidelity as the rotation angle $\theta$ becomes small. 
The value of $P^{\text{c}}_{\text{ud}}$ can be determined by numerical simulations discussed in Appendix~\ref{appendix:numerical simulation}.
We will present numerical results for a simple example later (Fig.~\ref{fig:fidelity}).

\subsection{Transversal multi-rotation protocol}
\label{sec:Formulation of transversal multi-rotation protocol}

As discussed in the previous section, resource states prepared by Choi \textit{et al.} protocol are easily disturbed by a single Pauli-Z error. The origin of such undetectable errors includes idling errors occuring across the protocol, two-qubit gate errors in first-round syndrome measurement, and single-qubit rotation gate errors in the transversal rotation gate operation.
These errors prevent us from suppressing the total undetectable error rate $P_{\text{ud}}^{\text{c}}$.

In addition, the success rate of the Choi \textit{et al.} protocol rapidly decreases as the code distance $d$ increases, since the number of detectable error patterns becomes larger in proportion to the size of $Q_z$. In fact, as shown later, the original Choi \textit{et al.} protocol hardly passes the post-selection process when the code distance $d$ has a moderate value needed for the realistic setup of materials simulations.
Furthermore, the total error rate $P_{\text{ud}}$ increases in proportion to the code distance $d$.
This suggests that a moderately small code distance $d$ is preferable for preparing resource states with high fidelity and success rate. 
However, if we set the code distance too small, a non-negligible logical error inevitably occurs during subsequent procedures such as gate-teleportation or code deformation after the post-selection process in the Choi \textit{et al.} protocol. 


\subsubsection{Formulation}
\label{sec:formulation of injection protocol}

To alleviate the aforementioned issues, we develop a more elaborate protocol to prepare a resource state $\ket{m_\theta}$ with a higher fidelity and higher success rate.
The first idea of our protocol is to harness a transversal multi-Pauli rotation gate, instead of a transversal single Pauli-$Z$ rotation gate. Therefore, we call our protocol the {\it transversal multi-rotation protocol}.
For example, considering the case of two-qubit $ZZ$ rotation (Fig.~\ref{fig:transversal muti-rotation protocol} (b)), we apply the following gate on the logical state $\ket{+}$:
\begin{equation}
\label{eq:transversal multi rotation}
\begin{aligned}
    \prod_{i=1}^k \hat{R}_{zz,i}(\theta) &= \prod_{i=1}^{k} \left[ \cos\theta\cdot \hat{I}_{2i}\hat{I}_{2i+1} +i \sin\theta\cdot \hat{Z}_{2i}\hat{Z}_{2i+1} \right]\\
    &= \sum_{n_z=0}^k  i^{n_z}\sin^{n_z}\theta  \cos^{d-n_z} \theta
    \sum_{wt(\hat{P}_k)=2n_z} \hat{P}_k, 
\end{aligned}
\end{equation}
where we label each qubit in $Q_z$ with $i\in\{1,2,\cdots,d\}$ and assume that $d=2k$ for simplicity. An essential difference from the previous section is that all the Pauli strings in the last line have a weight of two or more.
Compared to Eq.~\eqref{eq:Transversal_rotation}, this operation has a great advantage in terms of noise resilience. This is because any single Pauli error $\hat{\sigma}$ $(=\hat{X},\hat{Y},\hat{Z})$ arising before and after the operation can be detected via the subsequent syndrome measurements as any operator in the form of $\hat{\sigma}\hat{P}_k$
never becomes a logical operator.

More generally, we can readily extend the above argument to the case where the transversal gate $\prod_{i=1}^d \hat{R}_{zz\cdots z,i}(\theta)$ consists of $m$-weight multi-$Z$ rotation $\hat{R}_{zz\cdots z,i}(\theta)$ for any positive integer $m$.
Obviously, this protocol for $m=1$ coincides with the Choi {\it et al.} protocol (Fig.~\ref{fig:transversal muti-rotation protocol} (a)).
Meanwhile, the case where $m=2$ and $d=2$ is essentially equivalent with the protocol explained in Sec.~\ref{sec:Akahoshi injection} if we execute the subsequent patch expansion process properly (Fig.~\ref{fig:transversal muti-rotation protocol} (c)).
The subtle differences between the two protocols is whether we start with the surface code or [4,1,1,2] subsystem code with $d=2$.

\begin{figure*}[tb]
    \centering
    \includegraphics[width=17cm]{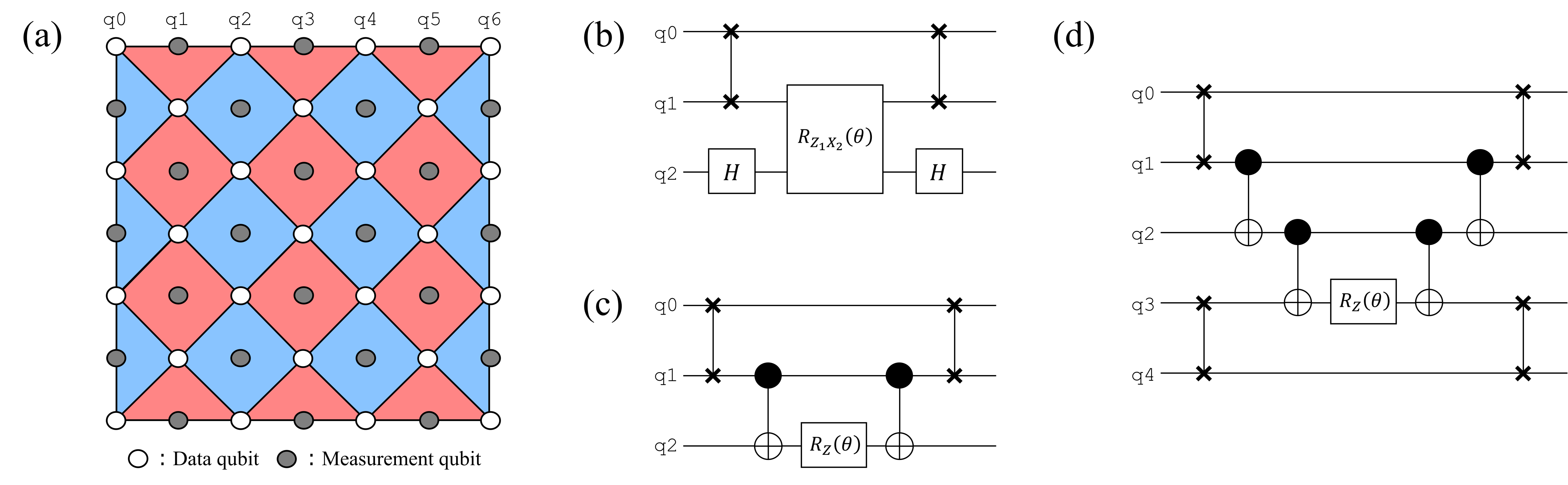}
    \caption{Implementation of a high-weight Pauli-$Z$ rotation gate under the constraints of nearest-neighbor connectivity. 
    {\bf (a)}~Qubit assignment to implement the unrotated planar surface code with the code distance $d=6$. Red (blue) panel denotes the $X$ ($Z$) stabilizer for the code.
    {\bf (b)} Implementation of a two-weight Pauli-$Z$ rotation gate $\hat{R}_{Z_0Z_2}(\theta)$ with nearest-neighbor CNOT/SWAP gates and the cross-resonance gate $\hat{R}_{Z_1X_2}(\theta)$. {\bf (c)} Another implementation of $\hat{R}_{Z_0Z_2}(\theta)$ with a single-qubit $Z$-rotation gate $\hat{R}_Z(\theta)$, instead of the cross-resonance gate. {\bf (d)} Implementation of a three-weight Pauli-$Z$ rotation gate $\hat{R}_{Z_0Z_2Z_4}(\theta)$ with a single-qubit $Z$-rotation gate. In our numerical simulation, we assume the method (c) and (d) to implement a two-weight and three-weigh Pauli-$Z$ rotation gates, respectively.}
    \label{fig:implementation of rotation gate}
\end{figure*}

On an actual device, we need to consider how to implement such a high-weight multi-$Z$ rotation gate under the limited gate set and connectivity of the hardware. 
For example, assuming superconducting devices with nearest-neighbor connectivity, we can implement high-weight Pauli-$Z$ rotation gates by combining several nearest-neighbor CNOT/SWAP gates with cross-resonance gates~\cite{Rigetti2010,Chow2011} or single-qubit $Z$-rotation gate (Fig.~\ref{fig:implementation of rotation gate}). 
In particular, for the latter case, it is well-known that a single-qubit $Z$-rotation gate can be implemented with extremely high precision by utilizing the virtual-$Z$ gate scheme~\cite{Mckay2017}.

In what follows, we focus on the case where $d=mk$ for simple description; thus, we specify each protocol with three parameters $(m,k,d)$.
However, it is easily confirmed that the following analyses and formulas hold for a more general case, 
where we utilize generic transversal multi-$Z$ rotation that comprises of multi-$Z$ rotation gates with different weights.
In such a case, we must reinterpret the factor $k$ as the number of multi-$Z$ rotation gates performed transversally.

\begin{figure}
    \centering
    \hspace{-1.2cm}
    \includegraphics[width=8.3cm]{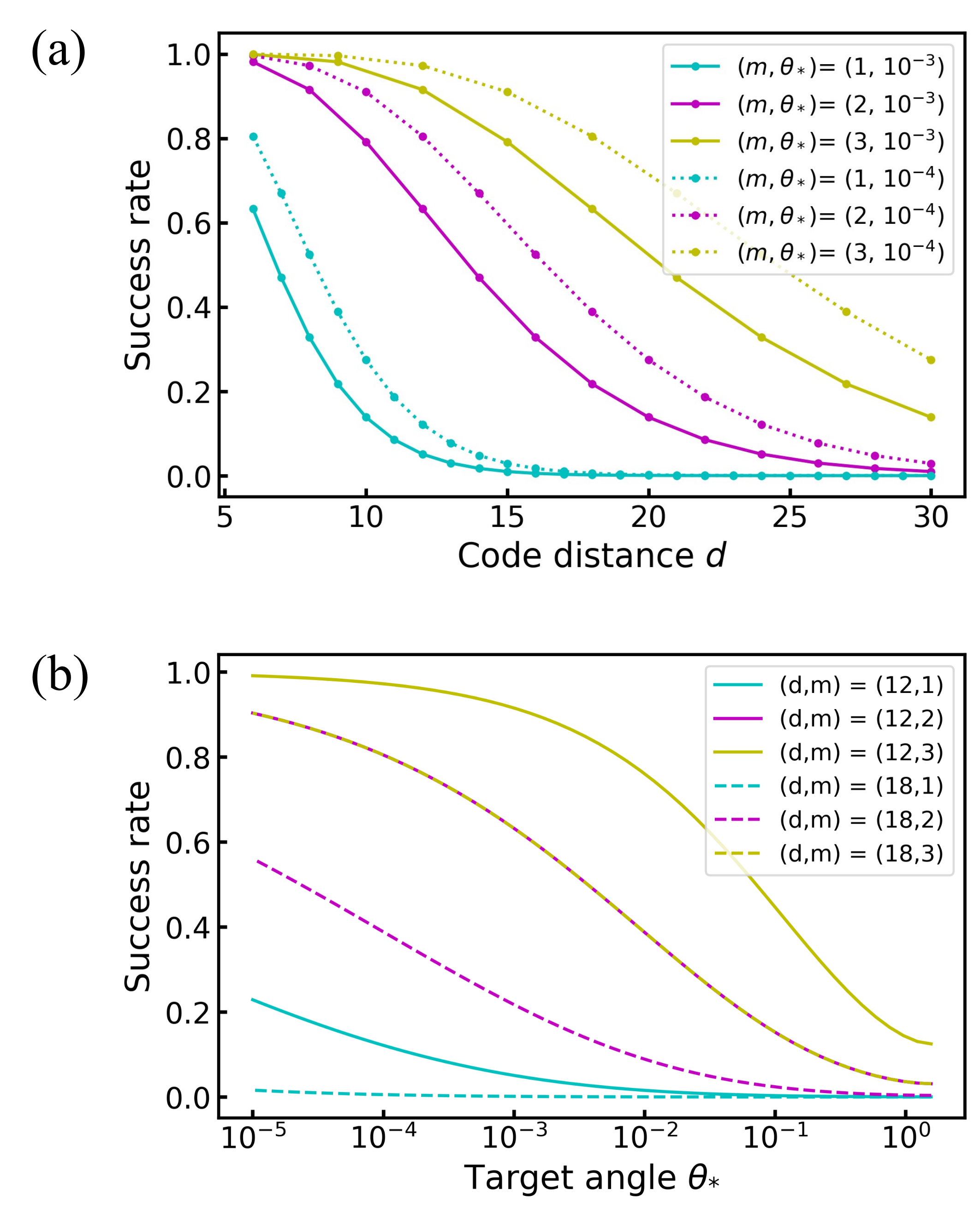}
    \caption{Success rate of resource state preparation via transversal multi-rotation protocol in the ideal limit ($p_{\text{ph}}=0$): {\bf (a)} dependence on code distances $d$ and {\bf (b)} on target angles $\theta_*$.  We show the data for various values of code distance $d$, rotation-weight $m$, and target angles $\theta_*$. Clearly, our protocol ($m=2,3$) has a much higher success rate than the Choi {\it et al.} protocol ($m=1$).} 
    \label{fig:ideal success probability}
\end{figure}

Let us now resume a more quantitative analysis of our protocol.
First, consider the ideal case where we can perform the transversal multi-rotation gate (Eq.~\eqref{eq:transversal multi rotation}) and subsequent stabilizer measurements without noise ($p_{\text{ph}}=0$).
In this case, following the discussion in Sec.~\ref{sec:Choi}, we find that the transversal multi-rotation protocol generate a resource state $\ket{m_{\theta_*(\theta,k)}}_L$ with an ideal success rate $p_{\text{ideal}}(\theta,k)$ using Eq.~\eqref{eq:ideal success rate} and Eq.~\eqref{eq:angle relation}.
This result suggests that, in the ideal limit, we can improve the success rate by increasing the value of weight $m$, since the function $p_{\text{ideal}}(\theta,k)$ monotonically increases as $k$ $(=d/m)$ decreases. 
This can be confirmed quantitatively by referring to Fig.~\ref{fig:ideal success probability}, which plots $p_{\text{ideal}}(\theta,k)$ for several parameters.
This figure shows that the Choi {et al.} protocol ($m=1$) hardly succeeds even without physical errors when the code distance is over around $10$. Meanwhile, our protocol for $m=2$ and $m=3$ possess a success rate several times higher for the same parameters $(d, \theta_*)$.

On the other hand, considering the case where a single undetectable error occurs, the output angle becomes an incorrect value $\theta_{\text{error}}$.
Such an undetectable error originates from $ZZ$-errors arising in the execution of $\hat{R}_{zz,i}(\theta)$,
while other single errors in the entire circuit, including syndrome measurement circuits, can be detected via syndrome measurements, as already mentioned.
For example, assuming that we implement $\hat{R}_{zz,i}(\theta)$ with a native two-qubit gate such as the cross-resonant gate (Fig.~\ref{fig:implementation of rotation gate}(b)), we readily find that the total error probability $P_{\text{ud}}$ amounts to $\frac{k}{15}p_{\text{ph}}$ under circuit-level noise model.
Meanwhile, assuming that we implement $\hat{R}_{zz,i}(\theta)$ with the circuit shown in Fig.~\ref{fig:implementation of rotation gate}(c) and that the single-qubit $Z$-rotation gate can be perfectly performed with the virtual-$Z$ scheme, $P_{\text{ud}}$ amounts to $\frac{2k}{15}p_{\text{ph}}$.
These estimates can also be verified by numerical calculations discussed in Appendix~\ref{appendix:numerical simulation}.


Taking these modifications from Sec.~\ref{sec:Choi} into account, the density matrix of the output state is obtained as
\begin{equation}
\label{eq:final output state}
\begin{aligned}
 \hat{\rho}_{\text{out}}\simeq \  &\frac{1}{p_{\text{suc}}} \Big[ p_{\text{ideal}}(1-Q)\cdot \ket{m_{\theta_*}}\bra{m_{\theta_*}}_L  \\
    & +\ p_{\text{error}}P_{\text{ud}}\cdot \ket{m_{\theta_{\text{error}}}}\bra{m_{\theta_{\text{error}}}}_L\Big] + {\mathcal{O}(p_{\text{ph}}^2)}\\
\end{aligned}
\end{equation}
after we modify the state with Clifford operations properly. 
Here, $Q$ $(\propto p_{\text{ph}})$ is the probability of discarding the output state even though it equals the target state slightly modified by physical errors.
The normalized factor $p_{\text{suc}}=p_{\text{ideal}}(1-Q)+p_{\text{error}}P_{\text{ud}}$ corresponds to the success rate of our protocol under physical errors if we neglect the contributions of $\mathcal{O}(p_{\text{ph}}^2)$.
The second term in the square bracket corresponds to the undetectable error state with a probability amplitude
\begin{equation}
\begin{aligned}
    p_{\text{error}}(\theta,k)& \  \equiv\  \sin^{2}\theta\cos^{2}\theta(\sin^{2k-4}\theta+\cos^{2k-4}\theta)\\
    &\  \simeq \ \theta^2  +\ \order{\theta^4},
\end{aligned}
\end{equation}
and an incorrect rotation angle
\begin{equation}
\label{eq:error angle}
\begin{aligned}
    \theta_{\text{error}}(\theta,k)&  \  \equiv\  -\sin^{-1}\left(
        \frac{1}{\sqrt{p_{\text{error}}}} \sin^{k-1}\theta \cos\theta\right)\\
        & \ \simeq \  -\theta^{k-2}
    +\ \order{\theta^{k}}.
\end{aligned}
\end{equation}
Furthermore, the state infidelity is calculated as 
\begin{equation}
\label{eq:state fidelity of multi-rotation protocol}
\begin{aligned}
    1-F &=1-\ev{\hat{\rho}_{\text{out}}}{m_{\theta_*}}_L\\
    &= 1- \frac{p_{\text{ideal}}}{p_{\text{suc}}}(1-Q)
    - \frac{p_{\text{error}}}{p_{\text{suc}}}P_{\text{ud}}
    |\braket{m_{\theta_*}}{m_{\theta_{\text{error}}}}_L|^2\\
    &\simeq P_{\text{ud}} \left(\frac{p_{\text{error}}}{p_{\text{ideal}}}\right)  (1- |\braket{m_{\theta_*}}{m_{\theta_{\text{error}}}}_L|^2)+ {\mathcal{O}(p_{\text{ph}}^2)}\\
    & = P_{\text{ud}} \left(\frac{p_{\text{error}}}{p_{\text{ideal}}}\right) \sin^2(\Delta_{\theta_*}) + {\mathcal{O}(p_{\text{ph}}^2)},
\end{aligned}
\end{equation}
where we introduce the over-rotation angle $\Delta_{\theta_*} \equiv \theta_{\text{error}}- \theta_*$.
In particular, if we focus on the leading term in the limit $\theta\to 0$, we obtain the asymptotic behavior $1-F\simeq P_{\text{ud}} \cdot\theta^{2(k-1)} \simeq P_{\text{ud}} \cdot\theta_*^{2(1-1/k)}$, which is interpreted as an extension of Eq.~\eqref{eq: fidelity of Choi} to the case of $m \geq 2$.


\begin{figure}
    \centering
  \begin{tabular}{l}
{\large (a)} \\
   \includegraphics[width=8.3cm]{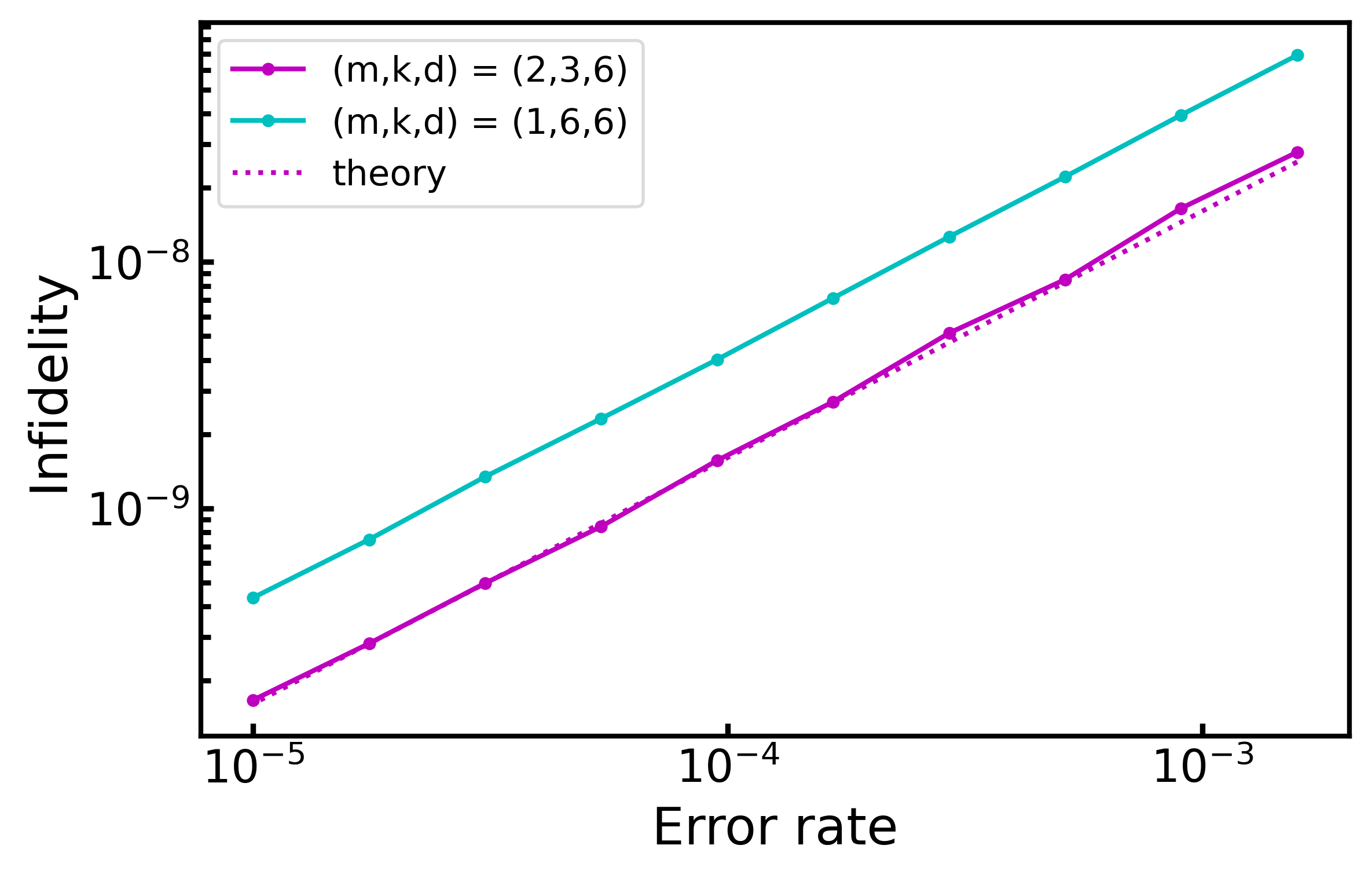}\\
{\large (b)} \\
    \includegraphics[width=8.3cm]{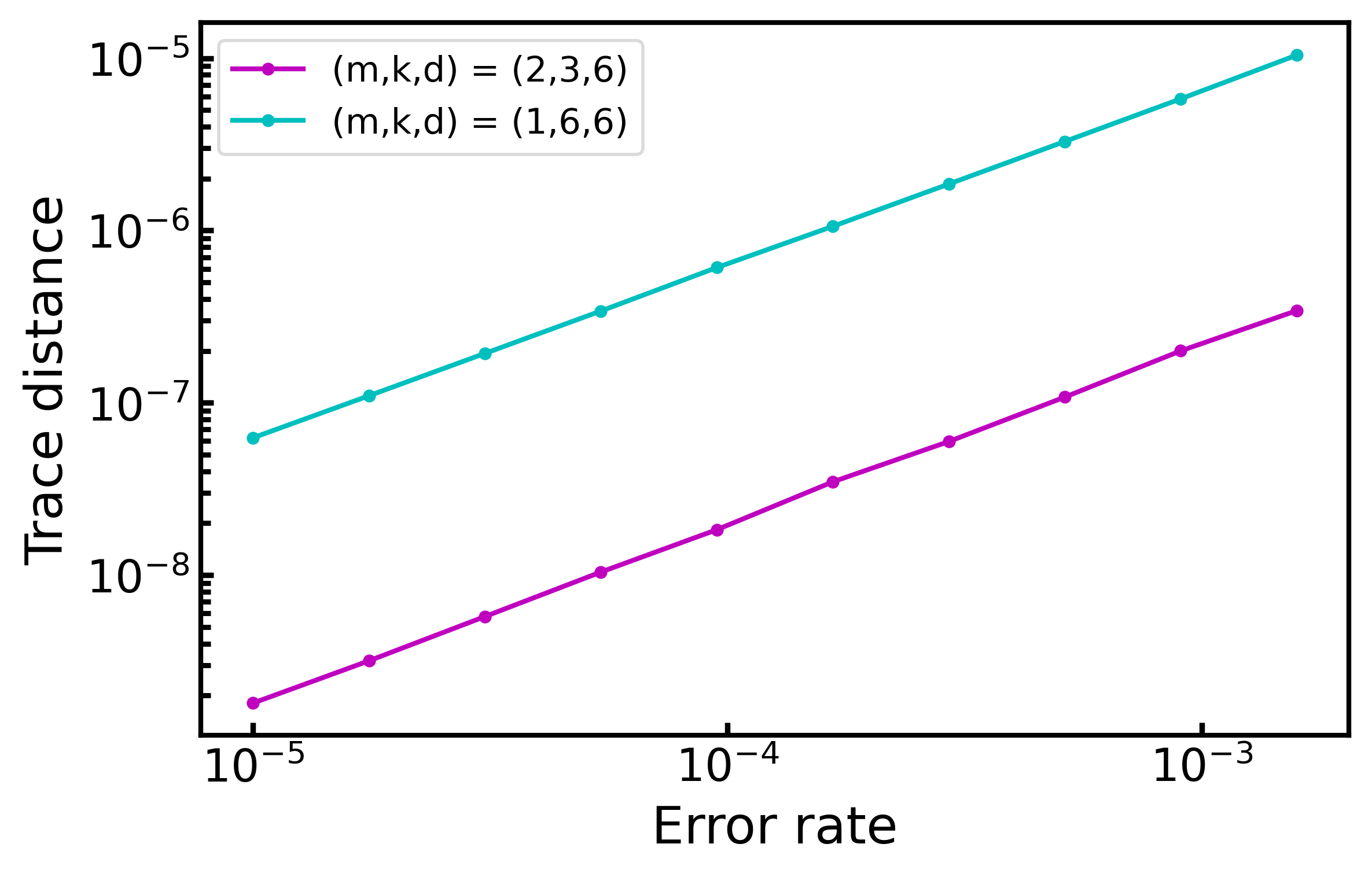}
  \end{tabular}
    \caption{Error analysis of Choi {\it et al.} protocol and transversal multi-rotation protocol with the measure of {\bf (a)} the state infidelity and {\bf (b)} the trace distance. We assume that the logical qubit is encoded on the unrotated surface code and set the target angle to $\theta_*=10^{-3}$. The plotted values are obtained by calculating $2\times 10^7$ samples. For the case of $m=2$, we implement the two-weight Pauli-$Z$ rotation gate using the circuit shown in Fig.~\ref{fig:implementation of rotation gate}(c). The dotted line in (a) represents the theoretical line obtained in Eq.~\eqref{eq:state fidelity of multi-rotation protocol} with $P_{\text{ud}}=2k/15$.}
    \label{fig:fidelity}
\end{figure}

Finally, we compare our protocol with the Choi {\it et al.} protocol in more detail, in terms of the quality of prepared resource states.
In Fig.~\ref{fig:fidelity}(a), we show the result of numerical calculations of state infidelity, assuming the case of $(m,k,d)=(2,3,6)$ and $(1,6,6)$ (see Appendix~\ref{appendix:numerical simulation} for details of the numerical calculations).
It suggests that our protocol ($m=2$) can achieves a smaller infidelity than the Choi {\it et al.} protocol ($m=1$).
This originates from the fact that our protocol reduces the total undetectable error rate $P_{\text{ud}}$ more efficiently.
However, in terms of state infidelity, the Choi {\it et al.} protocol has a slightly better scaling with respect to the target angle $\theta_*$. This is because the state infidelity scales as $\theta_*^{2(1-1/d)}$ in the Choi {\it et al.} protocol, while it scales as $\theta_*^{2(1-1/k)}$ ($k<d$) in our protocol.

Here we should note that the state infidelity cannot assess the effect of the off-diagonal error correctly.
In fact, as mentioned in Sec.~\ref{sec:error model}, rotation gates implemented via the gate-teleportation of the prepared resource state show a worst-case error rate proportional to $\order{|\theta_*|p_{\text{ph}}}$, rather than to $\order{|\theta_*|^{2(1-1/k)}p_{\text{ph}}}$.
To address this issue, we next consider the trace distance~\cite{Nielsen2000}. We can calculate the leading term for the output state in Eq.~\eqref{eq:final output state} as follows:
\begin{equation}
\label{eq: Trace distance of resource state}
\begin{aligned}
    D_{\text{tr}}&(\rho_{\text{out}},\ket{m_{\theta_*}}\bra{m_{\theta_*}}_L)\\
    &= \ \frac12 \abs{\rho_{\text{out}} - \ket{m_{\theta_*}}\bra{m_{\theta_*}}_L}\\
    & \simeq \ \frac{p_{\text{error}}P_{\text{ud}}}{2p_{\text{ideal}}} \big| \ket{m_{\theta_{\text{error}}}}\bra{m_{\theta_{\text{error}}}}_L- \ket{m_{\theta_*}}\bra{m_{\theta_*}}_L\big|\\
    &= P_{\text{ud}} \left(\frac{p_{\text{error}}}{p_{\text{ideal}}}\right) \sin(\Delta_{\theta_*}), 
\end{aligned}
\end{equation}
where the term of $\mathcal{O}(p_{\text{ph}}^2)$ is neglected in the second equation.
Similar to the analysis after Eq.~\eqref{eq:state fidelity of multi-rotation protocol}, we obtain the asymptotic behavior in a small-angle limit as $D_{\text{tr}}(\rho_{\text{out}},\ket{m_{\theta_*}}\bra{m_{\theta_*}}_L)\simeq P_{\text{ud}} \cdot\theta^{k} \simeq P_{\text{ud}} \cdot\theta_*$. Notably, this formula is independent of the value of $k$; hence the performance of protocols is determined only by the value of $P_{\text{ud}}$.
This suggests that our protocol for $m\geq 2$ always outperforms the Choi {\it et al.} protocol (m=1) in terms of trace distance, as well as its success rate.
In Fig.~\ref{fig:fidelity}(b), we show the numerical results for the trace distance. 
This result suggests that our protocol can achieve a trace distance that is at least an order of magnitude smaller than that of the Choi {\it et al.} protocol.

\begin{algorithm}[tb]
\SetKwComment{Comment}{/* }{ */}
\SetKwInOut{Input}{Input}
\SetKwInOut{Output}{Output}
\caption{Transversal multi-Pauli rotation protocol with the optimal post-selection}
\label{alg:Transversal multi-Pauli rotation protocol}
\Input{
\begin{itemize}
\renewcommand{\labelitemi}{}
    \item $\mathcal{S} \gets$ Stabilizer group of the stabilizer code that we use
    \item $\mathcal{S}_{PS}(\subset \mathcal{S})\gets$ Stabilizers in the post-selection regime
    \item $\theta_* \gets$ Target rotation angle
    \item $m \gets$ Weight of multi-Pauli rotation
    \item $Q_z \gets$ Qubit set that form a support of $\hat{Z}_L$
\end{itemize}
}
\Output{
\begin{itemize}
\renewcommand{\labelitemi}{}
    \item $\ket{m_{\theta_*}} \gets$ Resource state for implementing $\hat{R}_z(\theta)$
\end{itemize}
}
\BlankLine
Set all the data physical qubits in $\ket{+}$ state\;
Measure the stabilizer set $\mathcal{S}$ to generate $\ket{+}_L$ state\;
\If{There are unexpected syndromes in $\mathcal{S}_{PS}$}{\Return Failure}
\BlankLine
Apply a transversal multi-Pauli rotation gate on $Q_z$ with a Pauli weight $m$ and a physical angle $\theta$ that satisfies Eq.~\eqref{eq:angle relation}\;
\BlankLine
\For{$i\leftarrow 1$ \KwTo $2$}{
Measure the stabilizer set $\mathcal{S}$\;
\If{There are unexpected syndromes in $\mathcal{S}_{PS}$}{\Return Failure}
}
\Return Success
\end{algorithm}

\subsubsection{Optimal post-selection}

Next let us discuss optimising the post-selection of the prepared state in syndrome measurement processes. 
In the original proposal in Ref.~\cite{Choi2023}, the authors assumed that states with error syndromes should be rejected to remove detected errors.
Although this approach suppresses adverse effects of all detectable errors confidently, it 
leads to a non-negligible failure rate of the preparation protocol, $p_{\text{fail}} \equiv 1-p_{\text{suc}}$, which is roughly proportional to the number of physical qubits comprising logical codes when the error rate is sufficiently small ($p_{\text{ph}}\ll 1$).
This indicates that the failure rate scales as $p_{\text{fail}}\propto d^2$ when using the planar surface code.
In a typical situation, where we try to execute a quantum circuit of moderate size and the code distance becomes around several tens, such a rapid increase in the failure rate 
can be an obvious disadvantage that delays the execution time of analog rotation gates.

In what follows, we formulate a more flexible approach to remove the adverse effects of detectable errors.
The essential idea is to adopt a hybrid approach that optimally combines post-selections and quantum error corrections.
Namely, we carefully reject only the states with error syndromes that implies the appearance of incorrect resource states and then correct the remaining errors in the states that pass the post-selection process.

\begin{figure}
    \centering
    \includegraphics[width=8cm]{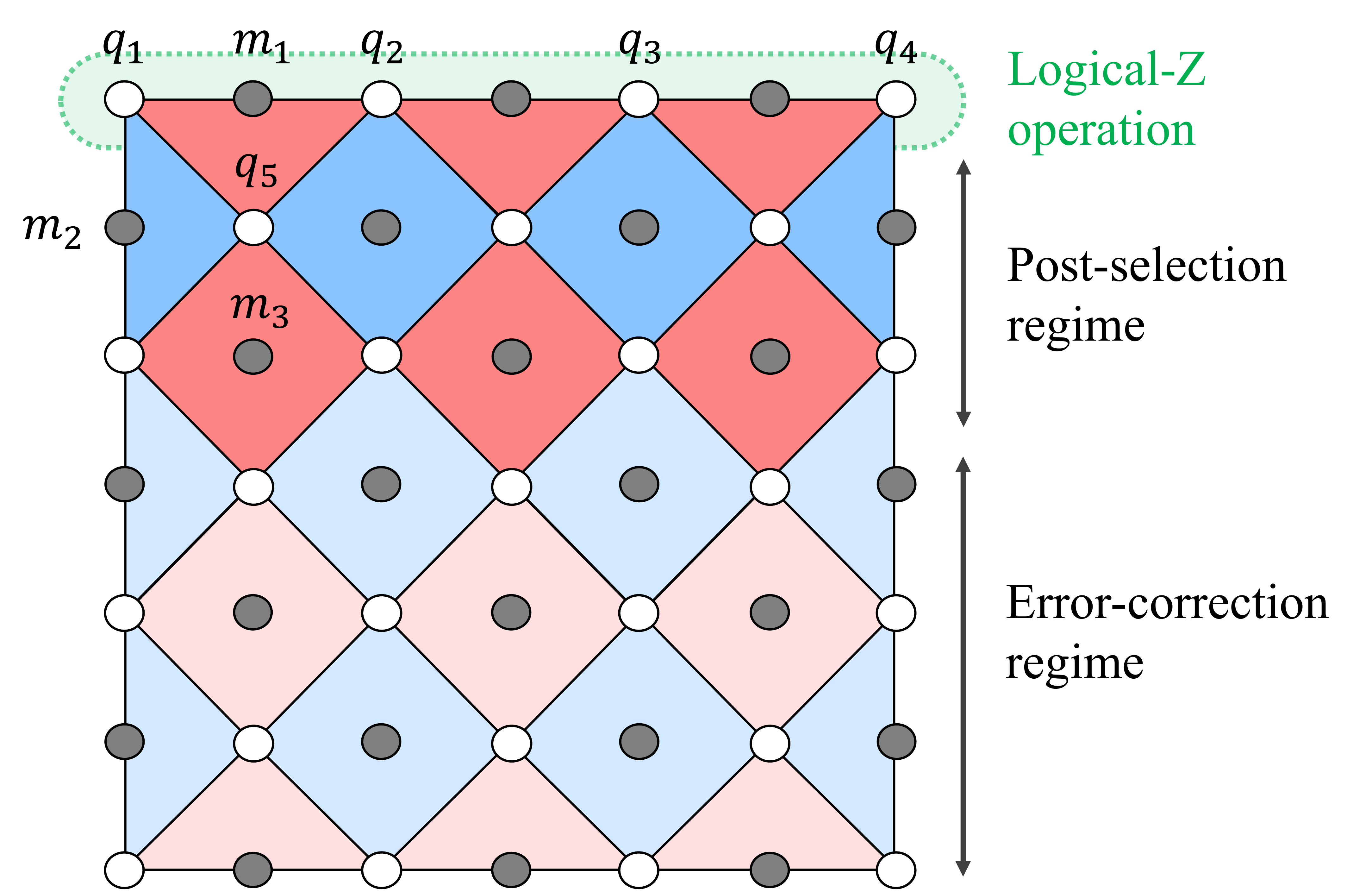}
    \caption{Segmentation of stabilizers for optimal post-selection on an unrotated surface code. $Z$ ($X$)-stabilizers are colored in blue (red). In our post-selection scheme, we discard the output state if we obtain unexpected syndromes from stabilizers with dark color throughout the two rounds of syndrome measurements after the transversal rotation gate. Then, if we pass the post-selection processes, we correct any errors after performing $(d-2)$-times syndrome measurements to suppress measurement errors.}
    \label{fig:post-selection}
\end{figure}

For instance, we focus on distinguishing the two quantum states that appear in Eq.~\eqref{eq: Choi protocol},
\begin{equation}
\begin{aligned}
    \ket{m_{\theta_*}}_L&\ \propto \ \cos^d \theta\ket{+}_L  + i^{d}\sin^d\theta  \ket{-}_L, \\
    \hat{Z}_1\ket{m_{\theta_{\text{error}}}}_L&\ \propto\ 
    i\sin \theta \cos^{d-1} \theta \hat{Z}_1 \ket{+}_L\\
    & \ \ \ \ \ \ \ + i^{d-1}\sin^{d-1}\theta \cos \theta \cdot \hat{Z}_1\ket{-}_L.
\end{aligned}
\end{equation}
The latter state is one of the origins of the leading error in our protocol.
Because these states belong to different orthogonal stabilizer subspaces, the stabilizer measurement process projects the superposition state in Eq.~\eqref{eq: Choi protocol} into an orthogonal state including $\ket{m_{\theta_*}}_L$ or $\hat{Z}_1\ket{m_{\theta_{\text{error}}}}_L$ probabilistically.
If we assume an ideal stabilizer measurement, we can distinguish $\ket{m_{\theta_*}}_L$ and $\hat{Z}_1\ket{m_{\theta_{\text{error}}}}_L$ based on whether the measurement qubit $m_1$ in Fig.~\ref{fig:post-selection} returns an unexpected error syndrome in the first round or not.

Realistically, errors in stabilizer measurements make the analysis more complex.
For example, if a measurement error occurs at qubit $m_1$ in the first round, we fail to distinguish the two states. To avoid this scenario, we have to discard the case where an error syndrome occurs at the qubit $m_1$ in the second round.
As a more complex example, there is a case where a correlated error, such as $\hat{Z}\otimes \hat{X}$ or $\hat{Z}\otimes \hat{Z}$, occurs in the CNOT gate between $m_1$ and $q_5$.
This type of errors leads to an error syndrome at $m_2$ or $m_3$ in the first or second round, depending on the ordering of the CNOT gates.

These analyses can readily be extended to the problems of distinguishing $\ket{m_{\theta_*}}_L$ from $\hat{Z}_i\ket{m_{\theta_{\text{error}}}}_L$ and to the case of $m\neq 1$.
In conclusion, it is sufficient to reject only states with error syndromes in a specific regime (post-selection regime) shown in Fig.~\ref{fig:post-selection} during the $\ket{+}_L$ state preparation process and the first and second stabilizer measurement processes.
Because the width of the post-selection regime remains unchanged as the code distance increases, the failure rate of this post-selection approach scales linearly with the code distance $d$.
This is in contrast to the fact that the failure rate of the original approach scales quadratically with $d$, as mentioned before.

\begin{figure}  
    \centering
    \hspace{-0.5cm}
    \includegraphics[width=8.2cm]{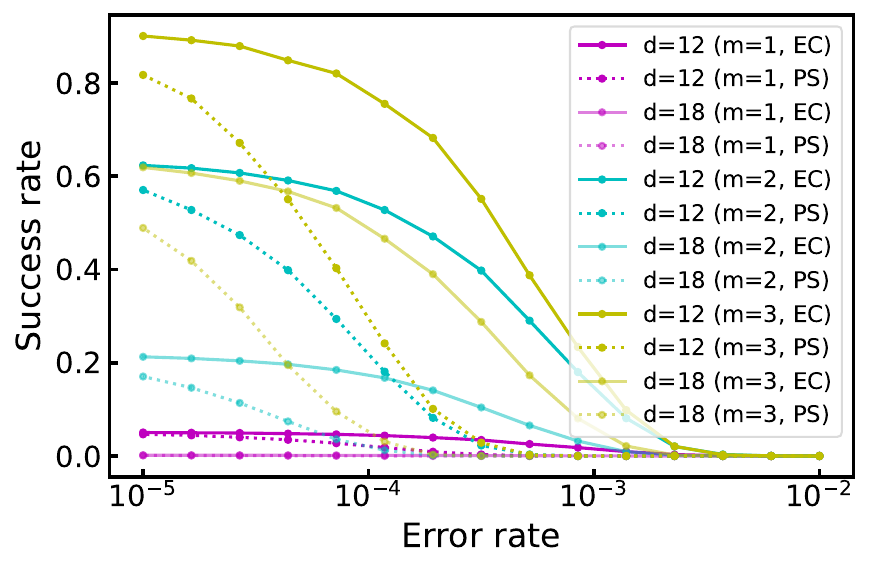}
    \caption{Success rate of the proposed state preparation protocol against various error rates $p_{\text{ph}}$. We fix the target angle with $\theta_*=10^{-3}$ and plot curves for several weights ($m=1,2,3$) and code distances ($d=12,18$) based on two types of post-selection approach. ``PS" denotes the original post-selection approach discussed in Ref.~\cite{Choi2023}, where we reject states with any error syndromes. Meanwhile, ``EC" denotes the hybrid approach proposed in this work, where we limit the post-selection regime to the band-shaped area in Fig.~\ref{fig:post-selection}.}
    \label{fig:success_rate_single_trial}
\end{figure}

The procedures in our protocol are summarized in Algorithm~\ref{alg:Transversal multi-Pauli rotation protocol}. 
Based on these procedures, we perform a numerical simulation of our protocol, and obtain the success rate under a finite error rate $p_{\text{ph}}$, as shown in Fig.~\ref{fig:success_rate_single_trial}.
This result clearly shows that, under a moderate error rate ($p_{\text{ph}}\gtrsim 10^{-3}$), the proposed hybrid approach (EC) achieves a success rate several orders of magnitude higher than the original post-selection approach (PS) in Ref.~\cite{Choi2023}.

\subsubsection{Other remarks}

\begin{figure}[tb]
    \centering
    \hspace{-0.5cm}
    \includegraphics[width=8.2cm]{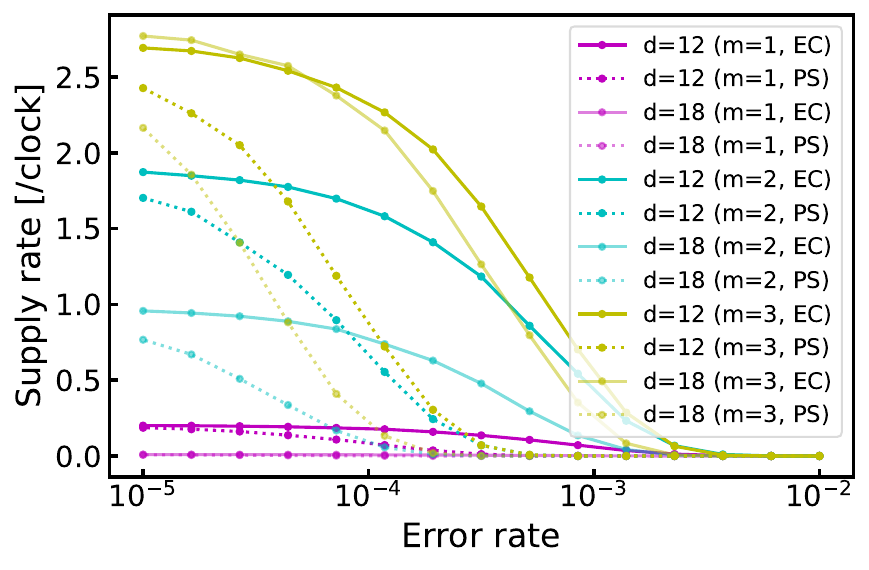}
    \caption{Supply rate of resource states per single clock for each preparation protocols. We fix the target angle with $\theta_*=10^{-3}$ and plot curves for several weights ($m=1,2,3$) and code distances ($d=12,18$) based on two types of post-selection approach, as in Fig.~\ref{fig:success_rate_single_trial}.}
    \label{fig:success_rate_in_clock}
\end{figure}

Finally, we give a few remaining remarks on our preparation protocol.
The first remark is that, in our protocol,  we can conduct multiple trials of state preparation during a clock (= $d$ code cycle) until we succeed.
Our preparation protocol includes $\ket{+}$ state preparation, transversal multi-Pauli rotation with SWAP gates, and two rounds of stabilizer measurements.
Because these operations take around four code cycles for $m\leq 3$, we can make $d/4$ trials in a clock to prepare a resource state.
In Fig.~\ref{fig:success_rate_in_clock}, we show the supply rate of resource states per single clock for our preparation protocol.
This result suggests that, for $p_{\text{ph}}=10^{-4}$, we prepare one or more resource states only in a single clock with a single code patch. Even for $p_{\text{ph}}=10^{-3}$, we maintain the supply rate at one by assigning two or three code patches for resource state preparation.

The second is that our protocol need not be run on a code patch with the same code distance as that of the data code patches.
This is because, by performing patch deformation~\cite{Horsman2012, Litinski2019} simultaneously with the second stabilizer measurement, the code distance of surface codes can be expanded from smaller to larger.
To maintain low infidelity of a prepared resource state, 
the initial value of the code distance should be large enough to prevent idling logical errors from limiting the achievable value of infidelity.

Finally, we note that our protocol is also applicable to any stabilizer code other than the planar surface code. 
Once the logical-$Z$ gate is defined as a direct product of the Pauli operator on a qubit set $Q_z$, we can generate a resource state with a transversal rotation gate on $Q_z$. In particular, on neutral atom or trapped ion devices, it would be possible to carry out a transverse multi-Pauli gate with a high weight value $m$ without suffering from a connectivity limitation.

\section{Stochastic error mitigation}
\label{sec:Stochastic error mitigation}

In this section, we will explain our strategy to mitigate the adverse effects of stochastic errors in our framework.
First, we clarify the error channel for the noisy rotation gate produced by our state preparation protocol. 
Then, we remark that these errors rapidly accumulate until we succeed in the RUS procedure for successful gate teleportation.
This is because, every time we fail a trial of gate teleportation, we have to double the rotation angle $\theta_*$; additionally the worst-case error rate linearly increases with the target angle $\theta_*$.

Next, we illustrate a method to cancel the coherent (off-diagonal) part of the stochastic error by applying inverse rotation probabilistically.
It reduces the worst-case error rate of our noisy rotation gate from $\order{|\theta_*| p_{\text{ph}}}$ into $\order{|\theta_*|^{2-2/d} p_{\text{ph}}}$ without incurring additional measurement costs.
We refer to this method as the {\it probabilistic coherent error cancellation}.
Unfortunately, it is also shown that the accumulation of stochastic errors reaches $\order{|\theta_*| p_{\text{ph}}}$ on average, rather than $\order{|\theta_*|^{2-2/d} p_{\text{ph}}}$, even with the cancellation method.
To alleviate the error accumulation, we propose an optimized approach to properly switch two different preparation protocols depending on the target angle in the RUS process. This can roughly halves the errors that accumulate in the RUS process, even though the averaged error rate remains in the order of $\order{|\theta_*| p_{\text{ph}}}$.

Finally, we formulate how to mitigate the remaining stochastic errors using an usual PEC method~\cite{Temme2017,Endo2018}.
In particular, we present a clear formula that relates the error mitigation cost in our framework with the total analog angles rotated through the entire circuit.

In the following, we will omit the subscript "$L$" for the description of states and gates for the sake of simplicity.

\subsection{Error channel model for our protocol}
\label{sec:error model}

First, before discussing how to mitigate stochastic errors, we clarify the error model for our noisy rotation channel.
By preparing a resource state $\ket{m_{\theta_*}}$ via our preparation protocol, we can implement a noisy logical Pauli-$Z$ rotation gate via the gate teleportation.
According to Eq.~\eqref{eq:final output state}, its noisy gate is described by the following quantum channel when the gate teleportation succeeds in the first trial:
\begin{equation}
\label{eq:noisy rotation channel in the first trial}
\begin{aligned}
    \mathcal{N}_{\theta_*}:\ \hat{\rho} \ \to \ \mathcal{N}_{\theta_*}(\hat{\rho}) =&\  \mathcal{E}_{\theta_*} \circ \mathcal{R}_{\theta_*}(\rho)\\
    =& \ (1-P_L) \cdot \mathcal{R}_{\theta_*}(\rho) \\
    & \qquad + P_L \cdot \mathcal{R}_{\theta_{\text{error}}}(\rho) + \order{|\theta_*|^2 p_{\text{ph}}^2},
\end{aligned}
\end{equation}
where $\mathcal{R}_{\theta_*}$ is defined in Eq.~\eqref{eq:ideal rotation channel} as an ideal logical rotation channel with the target angle $\theta_*$, and $\mathcal{E}_{\theta_*}$ denotes the following stochastic over-rotation channel with an angle $\Delta_{\theta_*}$:
\begin{equation}
\label{eq:our error channel}
\begin{aligned}
    \mathcal{E}_{\theta_*}:\ \hat{\rho} \ \to \ \mathcal{E}_{\theta_*}(\hat{\rho}) &= (1-P_L)\cdot\hat{\rho}\\
    & \  + P_L\cdot \mathcal{R}_{\Delta_{\theta_*}}(\hat{\rho})
+ \order{|\theta_*|^2 p_{\text{ph}}^2}. 
\end{aligned}
\end{equation}
Here we define the logical error rate $P_L(\theta_*)\equiv p_{\text{error}}P_{\text{ud}}/p_{\text{suc}}\simeq \theta_*^{2/k} P_{\text{ud}}$ and the over-rotation angle $\Delta_{\theta_*}\equiv \theta_{\text{error}}- \theta_*\simeq -\theta_*^{1-2/k}$ with the error angle $\theta_{\text{error}}$ in Eq.~\eqref{eq:error angle}.
We remind the reader that $k$ is the parameter that counts the number of multi-Pauli rotation gates used in the transversal rotation gate for our preparation protocol (see Sec.~\ref{sec:Formulation of transversal multi-rotation protocol}).

To evaluate the quality of quantum channels with an arbitary error channel $\mathcal{E}$,
it is convenient to introduce two common metrics of gate errors: average error rate $\varepsilon_{\text{av}}(\mathcal{E})$ and worst-case error rate $\varepsilon_\diamond(\mathcal{E})$.
In particular, according to Eq.~\eqref{eq: error rate formula} in Appendix~\ref{Appendix:definition of error rate}, these metrics for the error channel $\mathcal{E}_{\theta_*}$ in Eq.~\eqref{eq:our error channel} are calculated as follows:
\begin{equation}
\label{eq:average error rate of rotation gate}
    \varepsilon_{\text{av}}(\mathcal{E}_{\theta_*})=\frac23 P_L \Delta_{\theta_*}^2 \ \simeq\  \frac23\theta_*^{2(1-1/k)} P_{\text{ud}},
\end{equation}
\begin{equation}
\label{eq:worst-case error rate of rotation gate}
    \varepsilon_{\diamond}(\mathcal{E}_{\theta_*})=
    P_L|\Delta_{\theta_*}|\sqrt{1+\Delta_{\theta_*}^2 }\ \simeq\  |\theta_*| P_{\text{ud}}.
\end{equation}
In particular, in the small-angle limit ($\theta_*\to 0$), the average error rate has the same value as the resource state infidelity in Eq.~\eqref{eq:state fidelity of multi-rotation protocol}, except for an extra factor $2/3$. This is a natural consequence of the similarity between the definitions of state infidelity and average error rate.
Meanwhile, the worst-case error rate show different scaling with respect to the target angle $\theta_*$ in the same limit, implying more severe assessment of the error channel.
These differences arise because the coherent error term ($\propto \hat{Z}\hat{\rho}-\hat{\rho} \hat{Z}$) plays the most dominant role in determining the worst-case error rate.

In what follows, we focus on the analysis with the worst-case error rate as it naturally provides a clear upper bound of the error rate for entire quantum circuits via the chaining property (Eq.~\eqref{eq:chaining property}).

\subsection{Accumulation and cancellation of errors in repeat-until-success process}

Next, let us discuss how the stochastic error in Eq.~\eqref{eq:our error channel} accumulates during the RUS process for the gate-teleportation.
As explained in Sec.~\ref{sec:RUS}, in the RUS process, we must double the rotation angle every time we fail to teleport the rotation gate. Therefore, we naively imagine that the exponential accumulation of errors will occur when we repeatedly fail the gate teleportation, since the channel's error rate $\varepsilon_{\diamond}(\mathcal{E}_{\theta_*})$ is proportional to the target angle $\theta_*$. 

As a starting point, consider the case where we success the gate teleportation on the $K$-th trials ($K \geq 2$). In this case, we can describes the resulting rotation channel explicitly as 
\begin{equation}
\label{eq:Kth trial channel}
\begin{aligned}
    \mathcal{N}_{\theta_*}^K(\hat{\rho})\ 
    &\equiv\ \mathcal{N}_{2^{K-1}\theta_*}\circ \mathcal{N}_{-2^{K-2}\theta_*}\circ\cdots \circ \mathcal{N}_{-2\theta_*} \circ \mathcal{N}_{-\theta_*}(\hat{\rho})\\
    &=\  \mathcal{E}_{\theta_*}^K \circ \mathcal{R}_{\theta_*}(\hat{\rho}),    
\end{aligned}
\end{equation}
where $\mathcal{E}_{\theta_*}^K$ is an effective error channel that describes the accumulated error through the $K$ trials of RUS:
\begin{widetext}
\begin{equation}
\label{eq: gate teleportation channel}
\begin{aligned}
    \mathcal{E}_{\theta_*}^K(\hat{\rho})\ &= \ \mathcal{E}_{2^{K-1}\theta_*}\circ \mathcal{E}_{-2^{K-2}\theta_*}\circ\cdots \circ \mathcal{E}_{-2\theta_*} \circ \mathcal{E}_{-\theta_*}(\hat{\rho}v) \\
    &= \ \left(1-P_L(2^{K-1}\theta_*)
    -\sum_{n=0}^{K-2}P_L(-2^n\theta_*) \right)\hat{\rho}
    + P_L(2^{K-1}\theta_*)\hat{R}_z(\Delta_{2^{K-1}\theta_*})\hat{\rho} \hat{R}_z^\dagger(\Delta_{2^{K-1}\theta_*})  \\
     & \qquad\qquad\qquad+
    \sum_{n=0}^{K-2} P_L(-2^n\theta_*)\cdot \hat{R}_z(\Delta_{-2^n\theta_*})\hat{\rho} \hat{R}_z^\dagger(\Delta_{-2^n\theta_*}) 
     + \order{|\theta_*|^2 p_{\text{ph}}^2}\\
     & = (1-x_{\theta_*}^K) \hat{\rho} +  iy_{\theta_*}^K(\hat{Z}\hat{\rho} -\hat{\rho} \hat{Z}) +  x_{\theta_*}^K \hat{Z}\hat{\rho} \hat{Z} + \order{|\theta_*|^2 p_{\text{ph}}^2},
\end{aligned}
\end{equation}
where we omit the subscript ``$L$" for a logical rotation gate $\hat{R}_{z,L}(\theta)$, and introduce the following two parameters for later convenience:
\begin{equation}
\label{eq: factors of RUS channel}
\begin{aligned}
    x_{\theta_*}^K &= P_L(2^{K-1}\theta_*)\cdot \sin^2 (\Delta_{2^{K-1}\theta_*}) + \sum_{n=0}^{K-2}P_L(-2^n\theta_*)\cdot \sin^2 (\Delta_{-2^n\theta_*}),\\
    y_{\theta_*}^K &= \frac12 P_L(2^{K-1}\theta_*)\cdot \sin (2\Delta_{2^{K-1}\theta_*}) + \frac12 \sum_{n=0}^{K-2}P_L(-2^n\theta_*)\cdot \sin (2\Delta_{-2^n\theta_*}).
\end{aligned}
\end{equation}
\end{widetext}
When the rotation angle exceeds $\pi/8$ in the RUS process, reducing the angle to the smallest value obtained by employing a logical $S$ gate is preferable for minimising the error rate, which is proportional to the target angle.
More specifically, once the rotation angle $2^n\theta_*$ satisfies $ 2^n\theta_*> \pi/8\geq 2^{n-1}\theta_*$ at the $n$-th RUS trial, we replace the channel $\mathcal{N}_{2^{n}\theta_*}$ in Eq.~\eqref{eq:Kth trial channel} with $\mathcal{N}_{\Lambda(2^{n}\theta_*)}\circ\mathcal{R}_{\pi/4}$ to reduce the rotation angle as small as possible.
Here we introduce a wrapping function $\Lambda(x)\equiv |x-\pi/4|$, which leads to the relation of $\Lambda(2^{n}\theta_*)<\pi/8$.
By applying the similar procedure to the subsequent trials, we can always keep the analog rotation angles to be less than $\pi/8$.
In what follows, we assume that the quantum channels $\mathcal{N}_{\theta_*}^K$ and $\mathcal{E}_{\theta_*}^K$ implicitly include such modifications when the parameter $K$ satisfies $2^{K-1}>\pi/8$.


In an actual RUS process, we complete it on the $K$-th trials with probability $2^{-K}$.
Therefore, by averaging the channel $\mathcal{N}_{\theta_*}^K$ over any possible $K$, we yield the explicit form of an analog rotation channel $\tilde{\mathcal{N}}_{\theta_*}$ that is finally obtained after the RUS process as follows:
\begin{equation}
\label{eq:Rotation channel after RUS}
    \tilde{\mathcal{N}}_{\theta_*}(\rho) \equiv \sum_{K=1}^{\infty} \left(\frac12 \right)^K \mathcal{N}_{\theta_*}^K(\rho)   
    = \tilde{\mathcal{E}}_{\theta_*} \circ  \mathcal{R}_{\theta_*}(\rho),
\end{equation}
where $\tilde{\mathcal{E}}_{\theta_*}$ denotes an effective error channel for $\tilde{\mathcal{N}}_{\theta_*}$, which is represented as
\begin{equation}
    \tilde{\mathcal{E}}_{\theta_*}(\rho)\  \equiv\  \sum_{K=1}^{\infty} \left(\frac12 \right)^K \mathcal{E}_{\theta_*}^K(\rho).
\end{equation}
We provide a detailed analysis of the worst-case error rate of $\tilde{\mathcal{E}}_{\theta_*}$ in Appendix.~\ref{Appendix:RUS}. While the calculation is intricate, it is technically straightforward.
Fortunately, the obtained conclusion is that the error rate $\varepsilon_\diamond(\tilde{\mathcal{E}}_{\theta_*})$ is given almost in the order of $\order{|\theta_*| p_{\text{ph}}}$,
while error accumulation due to the RUS process yields a moderately large prefactor that is dependent on $\theta_*$ logarithmically.
Reducing this overhead is the primary focus of our subsequent discussions.

\subsection{Probabilistic coherent error cancellation}

As mentioned in Sec.~\ref{sec:error model}, the coherent (off-diagonal) term in Eq.~\eqref{eq:our error channel} plays a dominant role in determining the worst-case error rate of our rotation channels.
To address this issue, we now formulate a post-processing method to cancel the coherent term in the channel $\mathcal{E}_{\theta_*}$, which we refer to as the {\it probabilistic coherent error cancellation}.
To this end, we consider the following post-processing quantum channel,
\begin{equation}
\begin{aligned}
    \mathcal{C}_{\theta_*}(\hat{\rho}) &\equiv (1-P_L)\cdot\hat{\rho} + P_L \cdot \tilde{\mathcal{N}}_{-\Delta_{\theta_*}}(\hat{\rho})  \\
    & \simeq (1-P_L)\cdot\hat{\rho} + P_L \cdot \mathcal{R}_{-\Delta_{\theta_*}}(\hat{\rho})
    + \order{|\theta_*| p_{\text{ph}}^2}.
\end{aligned}
\end{equation}
This channel is easily implemented by applying the analog rotation gate in Eq.~\eqref{eq:Rotation channel after RUS} with a target angle $-\Delta_{\theta_*}$ and probability $P_L$.
In the last equality, we used the fact that $P_L \simeq \order{|\theta_*|^{2/k}p_{\text{ph}}}$ and $\varepsilon_\diamond(\tilde{\mathcal{E}}_{-\Delta_{\theta_*}})\simeq \order{|\theta_*|^{1-2/k} p_{\text{ph}}}$.

By applying the channel $\mathcal{C}_{\theta_*}$ after the noisy rotation channel $\mathcal{N}_{\theta_*}$, we obtain a composed quantum channel as follows:
\begin{equation}
    \mathcal{N}^{c}_{\theta_*}(\hat{\rho})\ \equiv\  \mathcal{C}_{\theta_*} \circ \mathcal{N}_{\theta_*}(\hat{\rho}) \ =\  \mathcal{E}_{\theta_*}^{c} \circ \mathcal{R}_{\theta_*}(\hat{\rho}),
\end{equation}
where an effective error channel for $\mathcal{N}^{c}_{\theta_*}$ is defined as $\mathcal{E}_{\theta_*}^{c} \equiv \mathcal{C}_{\theta_*} \circ \mathcal{E}_{\theta_*}$.
Then, we decompose the error channel in straightforward manner as
\begin{equation}
\label{eq:canceled rotation gate}
\begin{aligned}
    \mathcal{E}_{\theta_*}^{c}(\hat{\rho})\ &\simeq \ 
    (1-P_L)^2 \hat{\rho} + P_L^2 \cdot \mathcal{R}_{-\Delta_{\theta_*}}\circ \mathcal{R}_{\Delta_{\theta_*}}(\hat{\rho})\\
    & \qquad + P_L (1-P_L) \left(\mathcal{R}_{\Delta_{\theta_*}}(\hat{\rho})+ \mathcal{R}_{-\Delta_{\theta_*}}(\hat{\rho})\right)\\
    & \qquad + \order{|\theta_*| p_{\text{ph}}^2}\\
    & \simeq [1-2P_L\sin^2(\Delta_{\theta_*})]\cdot \hat{\rho}  \\
    & \qquad + 2P_L \sin^2(\Delta_{\theta_*})\cdot \hat{Z}\hat{\rho} \hat{Z} + \order{|\theta_*| p_{\text{ph}}^2},
\end{aligned}
\end{equation}
where we used the following identity in the last equation:
\begin{equation}
\begin{aligned}
    \mathcal{R}_{\Delta_{\theta_*}}(\hat{\rho}) &+ \mathcal{R}_{-\Delta_{\theta_*}}(\hat{\rho})\\
    &= 2\left[
    \cos^2(\Delta_{\theta_*})\cdot\hat{\rho} + \sin^2(\Delta_{\theta_*})\cdot \zhat\hat{\rho} \zhat
    \right].
\end{aligned}    
\end{equation}
Then, by applying Eq.~\eqref{eq: error rate formula}, the worst-case error rate of $\mathcal{E}_{\theta_*}^{c}$ is evaluated as 
\begin{equation}
\label{eq:improved scailing}
    \varepsilon_\diamond(\mathcal{E}_{\theta_*}^{c}) = 2P_L \sin^2(\Delta_{\theta_*}) 
    \simeq 2\theta_*^{2(1-1/k)} P_{\text{ud}},
\end{equation}
where $k$ is the parameter describing a transversal multi-Pauli rotation gate introduced in Sec.~\ref{sec:Formulation of transversal multi-rotation protocol}.
This means that the proposed post-processing channel $\mathcal{C}_{\theta_*}$ helps reduce the worst-case error rate of our noisy rotation channel from $\order{\theta_* p_{\text{ph}}}$ to $\order{|\theta_*|^{2(1-1/k)} p_{\text{ph}}}$.
This scaling is comparable with the one in Eq.~\eqref{eq:average error rate of rotation gate}, rather than that in Eq.~\eqref{eq:worst-case error rate of rotation gate}.
Furthermore, the proposed error cancellation scheme does not require additional measurement costs, unlike usual PEC techniques~\cite{Temme2017,Endo2018}. In addition, its time overhead is almost negligible because we rarely apply non-identical operations in the post-processing channel $\mathcal{C}_{\theta_*}$.

Next, reconsider the RUS process using the improved rotation channel $\mathcal{N}^{c}_{\theta_*}$.
By repeating the similar discussion as in the previous subsection, we obtain the following analog rotation channel, instead of $ \tilde{\mathcal{N}}_{\theta_*}$ in Eq.~\eqref{eq:Rotation channel after RUS}:
\begin{equation}
    \tilde{\mathcal{N}}^c_{\theta_*}(\hat{\rho}) \equiv \sum_{K=1}^{\infty} \left(\frac12 \right)^K \mathcal{N}_{\theta_*}^{c,K}(\hat{\rho})   
    = \tilde{\mathcal{E}}^c_{\theta_*} \circ  \mathcal{R}_{\theta_*}(\hat{\rho}),
\end{equation}
where we introduce several notations as follows:
\begin{equation}
\begin{aligned}
    \mathcal{N}_{\theta_*}^{c,K}(\hat{\rho})\ 
    &\equiv\ \mathcal{N}^c_{2^{K-1}\theta_*}\circ \mathcal{N}^c_{-2^{K-2}\theta_*}\circ\cdots \circ \mathcal{N}^c_{-2\theta_*} \circ \mathcal{N}^c_{-\theta_*}(\hat{\rho})\\
    &=\  \mathcal{E}_{\theta_*}^{c,K} \circ \mathcal{R}_{\theta_*}(\hat{\rho}), \\   
    \tilde{\mathcal{E}}^{c,K}_{\theta_*}(\hat{\rho})\  
    &\equiv \ \mathcal{E}^c_{2^{K-1}\theta_*}\circ \mathcal{E}^c_{-2^{K-2}\theta_*}\circ\cdots \circ \mathcal{E}^c_{-2\theta_*} \circ \mathcal{E}^c_{-\theta_*}  \\
    \tilde{\mathcal{E}}^c_{\theta_*}(\hat{\rho})\  &\equiv\  \sum_{K=1}^{\infty} \left(\frac12 \right)^K \mathcal{E}_{\theta_*}^{c,K}(\hat{\rho}).
\end{aligned}
\end{equation}
Then, using the formula in Eq.~\eqref{eq:canceled rotation gate}, we yield the explicit form of $\tilde{\mathcal{E}}^c_{\theta_*}$ as
\begin{equation}
\label{eq:error model for the final rotation channel}
    \tilde{\mathcal{E}}^c_{\theta_*} \ =\ 
    (1-\tilde{P}_L)\cdot \hat{\rho} + \tilde{P}_L\cdot \hat{Z}\hat{\rho} \hat{Z},
\end{equation}
Here $\tilde{P}_L(\theta_*)$ is the effective error rate of $\mathcal{N}_{\theta_*}^{c,K}$, and it is explicitly determined as
\begin{equation}
\label{eq:final error rate}
    \tilde{P}_L(\theta_*) = \sum_{K=1}^{\infty} \left(\frac12 \right)^K \sum_{n=1}^{K} \varepsilon_\diamond(\mathcal{E}_{2^{n-1}\theta_*}^{c}) 
\end{equation}
As shown in Eq.~\eqref{eq:improved scailing}, the error rate $\varepsilon_\diamond(\mathcal{E}_{2^{n-1}\theta_*}^{c})$ scales as $\order{|\theta_*|^{2(1-1/k)} p_{\text{ph}}}$ when the parameter $n$ is sufficiently small.
However, when the parameter $K$ satisfies $2^{K-1}\theta_*\gtrsim1$, the corresponding accumulated errors $\sum_{n=1}^{K} \varepsilon_\diamond(\mathcal{E}_{2^{n-1}\theta_*}^{c})$ reaches $\order{ p_{\text{ph}}}$.
This is the most dominant factor in determining the value of $\tilde{P}_L(\theta_*)$.
Because the above case appears in the RUS process with the probability of $2^{-K}\simeq \order{|\theta_*|}$, $\tilde{P}_L(\theta_*)$ scales linearly with $|\theta_*|$ and $p_{\text{ph}}$, even with the probabilistic coherent error cancellation.

\begin{figure}
    \centering
    \includegraphics[width=8cm]{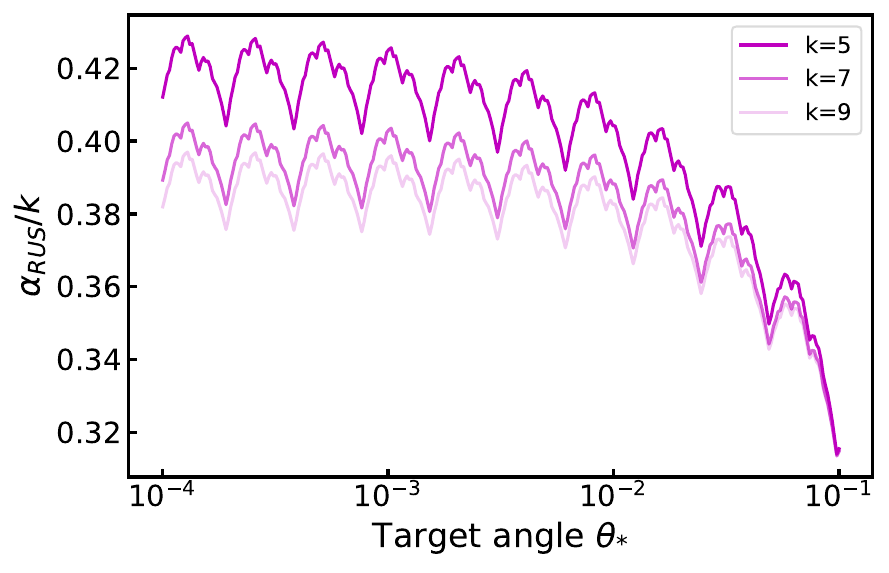}
    \caption{Numerical results for the prefactor $\alpha_{\text{RUS}}$ in Eq.~\eqref{eq:alpha_RUS}. Because it scales almost linearly with $k$, we plot the curves of $\alpha_{\text{RUS}}/k$.}
    \label{fig:RUS_factor}
\end{figure}

Finally, for later convenience, we introduce a new prefactor $\alpha_{\text{RUS}}$ to satisfy the relation 
\begin{equation}
\label{eq:alpha_RUS}
    \tilde{P}_L(\theta_*)=\alpha_{\text{RUS}}|\theta_*|p_{\text{ph}}.
\end{equation}
Because $P_{\text{ud}}$ scales linearly with $k$, the prefactor $\alpha_{\text{RUS}}$ behaves almost linearly with $k$.
Fig.~\ref{fig:RUS_factor} depicts the numerical results of $\alpha_{\text{RUS}}$ by directly calculating Eq.~\eqref{eq:final error rate}. In the calculation, we use the formula of $P_{\text{ud}}=\frac{k}{15}p_{\text{ph}}$.
Because the wrapping function $\Lambda(x)$ makes the series of angles complicated, $\alpha_{\text{RUS}}$ show a complex dependence on the target angle $\theta_*$. As a result, we find that $\alpha_{\text{RUS}}/k$ has a value of approximately 0.4 in the small-angle region ($\theta_*\ll1$), regardless of the value of $k$.

\subsection{Protocol switching in the RUS process}
\label{sec:Protocol switching}

In the previous discussions, we have formulated the RUS process and the resulting logical error rate of analog rotation gates, based on the transversal multi-rotation protocol.
In this subsection, we briefly show that we can further reduce the error rate by properly switching the preparation protocol during the RUS process.

\begin{figure}
    \centering
    \includegraphics[width=8cm]{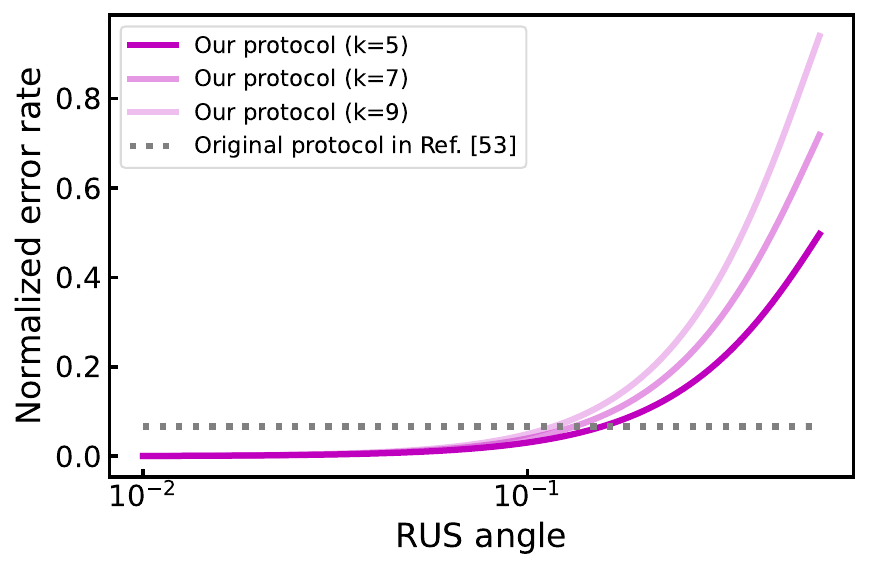}
    \caption{Comparison between the (normalized) error rates of the transversal multi-rotation protocol in Eq.~\eqref{eq:improved scailing} and the original protocol proposed in Ref.~\cite{Akahoshi2023}. The former becomes larger than the latter when the RUS angle exceeds about $10^{-1}$. In this plot, we normalize the error rate by the physical error rate $p_{\text{ph}}$}
    \label{fig:comparison_error_rate}
\end{figure}

Our idea is simple. According to Sec.~\ref{sec:Akahoshi injection} and Sec.~\ref{sec:Formulation of transversal multi-rotation protocol}, by utilizing the original preparation protocol in Ref.~\cite{Akahoshi2023}, we can generate an analog rotation gate with an error rate of $\frac{1}{15}p_{\text{ph}}$.
As shown in Fig.~\ref{fig:comparison_error_rate}, this error rate becomes superior to that obtained using our protocol, when the RUS process fails repeatedly and the RUS angle becomes fairly large.
Therefore, it is possible to improve the logical error rate of analog rotation channels by switching the preparation protocol from ours to the original one, once the error rate in Eq.~\eqref{eq:improved scailing} exceeds the value of $\frac{1}{15}p_{\text{ph}}$.

\begin{figure}
    \centering
    \includegraphics[width=8cm]{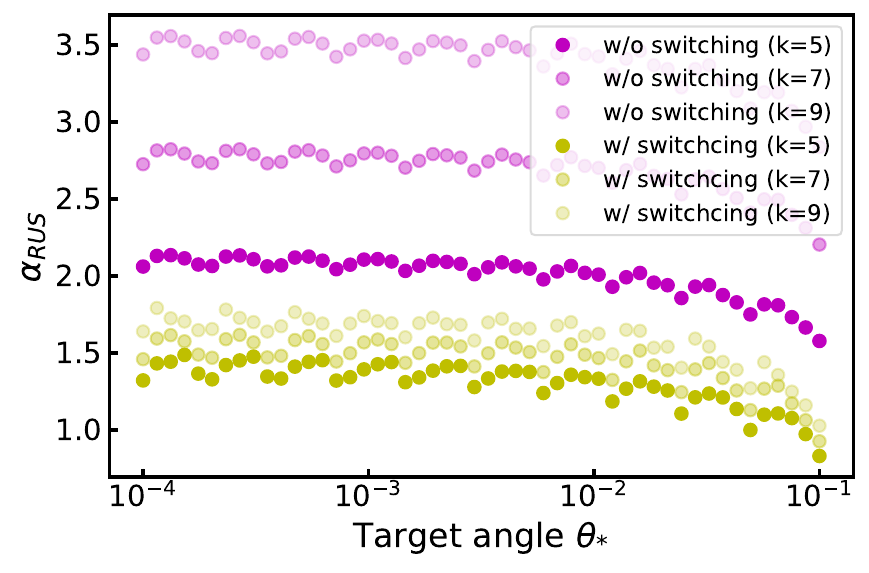}
    \caption{Numerical results for the prefactor $\alpha_{\text{RUS}}$ with and without the protocol switching. Unlike in Fig.~\ref{fig:RUS_factor}, here we does not normalize $\alpha_{\text{RUS}}$ by $k$.}
    \label{fig:hybrid RUS}
\end{figure}

Fig.~\ref{fig:hybrid RUS} demonstrates the numerical results of $\alpha_{\text{RUS}}$ with and without protocol switching.
The figure illustrates that the value of $\alpha_{\text{RUS}}$ can be reduced by approximately half by using the proposed scheme.
Furthermore, $\alpha_{\text{RUS}}$ becomes almost independent of $k$ under the protocol switching, 
while $\alpha_{\text{RUS}}$ orginally scales linearly with $k$.
This is because the error rate of the original protocol, which is independent of $k$, becomes  the primary factor for determining the total error rate when the protocol switching is applied.
Thus, in the following sections, we assume that the prefactor $\alpha_{\text{RUS}}$ has a value of around $1.5$, regardless of the value of $k$.

\subsection{Probabilistic error cancellation and total
mitigation cost}

Finally, we discuss how the standard PEC method~\cite{Temme2017,Endo2018} is utilized to mitigate the residual errors in the noisy rotation channel $\tilde{\mathcal{N}}^c_{\theta_*}$.

First, the error in the channel is described with a simple Pauli-$Z$ error channel $\tilde{\mathcal{E}}^c_{\theta_*}$ presented in Eq.~\eqref{eq:error model for the final rotation channel}.
The inverse map for $\tilde{\mathcal{E}}^c_{\theta_*}$ is explicitly constructed as 
\begin{equation}
\label{eq:inverse map}
    \tilde{\mathcal{E}}^{c,-1}_{\theta_*} \ =\ 
    \frac{1-\tilde{P}_L}{1-2\tilde{P}_L}\rhohat 
    -\frac{\tilde{P}_L}{1-2\tilde{P}_L}\zhat\rhohat \zhat.
\end{equation}
In other words, by applying the inverse map after our rotation channel $\tilde{\mathcal{N}}^c_{\theta_*}$, we can realize an ideal rotation gate with angle $\theta_*$ as follows:
\begin{equation}
\label{eq:stochastic error cancellation}
    \mathcal{R}_{\theta_*}=\  \tilde{\mathcal{E}}^{c,-1}_{\theta_*}
    \circ\  \tilde{\mathcal{N}}^{c}_{\theta_*}
    = \gamma \left((1-\tilde{P}_L) \cdot\tilde{\mathcal{N}}^{c}_{\theta_*}
    -\tilde{P}_L\cdot \mathcal{Z} \circ \tilde{\mathcal{N}}^{c}_{\theta_*}\right),
\end{equation}
where $\gamma = (1-2\tilde{P}_L)^{-1}$ and $\mathcal{Z}$ is a Pauli-$Z$ gate channel.

In general, the inverse map in Eq.~\eqref{eq:inverse map} cannot be implemented only by sampling some unitary gates probabilistically, since it includes a negative probability.
However, from Eq.~\eqref{eq:stochastic error cancellation}, the noise-free expectation value of any observable $\hat{O}$ can be decomposed as
\begin{equation}
\label{eq: expectation value}
    \ev*{\hat{O}}_{\mathcal{R}_{\theta_*}}\  \simeq \ \gamma\left((1-\tilde{P}_L) \ev*{\hat{O}}_{\tilde{\mathcal{N}}^{c}_{\theta_*}}  - \tilde{P}_L\ev*{\hat{O}}_{\mathcal{Z} \circ \tilde{\mathcal{N}}^{c}_{\theta_*}}\right),
\end{equation}
where $\ev*{\hat{O}}_{\mathcal{N}}=\tr[\hat{O}\mathcal{N}(\rhohat)]$ is the expectation value of $\hat{O}$ after applying a channel $\mathcal{N}$ to a target state.  Thus, 
by performing the Monte Carlo sampling of correcting operations $\mathcal{Z}$ depending on the weight of their coefficients, we can estimate the noise-free expectation value of any observable $O$.
Now it is crucially important that the variance of the estimator in Eq.~\eqref{eq: expectation value} is amplified by the factor $\gamma^2$ from the one without PEC.
Consequently, in the PEC, we requires $\gamma^2$ times more samples to suppress the amplified statistical errors sufficiently.

More generally, assume the situation where we perform a sequence of noisy rotation channel with angles $\{\theta_{*,i}\}_{i=1,\cdots,N_g}$. 
In such a situation, the overall factor denoting the total mitigation cost is evaluated with $\gamma_i\equiv(1-2\tilde{P}_L(\theta_{*,i}))^{-1}$ as 
\begin{equation}
\label{eq:mitigation cost in generic case}
\begin{aligned}
    \gamma_{\text{total}}^2 \ &\equiv\ \prod_{i=1}^{N_g} \gamma_i^2 \ =\  \prod_{i=1}^{N_g} \left(
    \frac{1}{1-2\tilde{P}_L(\theta_{*,i})}
    \right)^2\\
    &\simeq \ e^{4P_{\text{total}}}
    \ =\ e^{4\alpha_{\text{RUS}}\theta_{\text{total}} p_{\text{ph}}},
\end{aligned}
\end{equation}
where we assume that $\alpha_{\text{RUS}}$ is approximately independent of $\theta_*$ and introduce the total error rate and the total rotation angle as follows:
\begin{equation}
\label{eq: definition of total error and total rotation}
\begin{aligned}
    P_{\text{total}}\equiv \sum_{i=1}^{N_g} \tilde{P}_L(\theta_{*,i}) = \alpha_{\text{RUS}}\theta_{\text{total}} p_{\text{ph}},\ \ 
    \theta_{\text{total}} \equiv  
    \sum_{i=1}^{N_g} |\theta_{*,i}|.
\end{aligned}
\end{equation}
This result is fairly remarkable in the sense that the total error mitigation cost depends on the total rotation angle $\theta_*$, rather than on the total number of rotation gates $N_g$.
As shown in Sec.~\ref{sec:application}, this leads to the properties that, for materials simulation, the tractable problem size in our framework is determined by the 1-norm of the target Hamiltonian, rather than by the number of the terms.

Finally we remark that a recent work~\cite{Tsubouchi2024} has proposed a cost-optimal error mitigation method aimed at mitigating logical
errors in non-Clifford operations with minimal sampling
overhead.
The work suggests that the sampling overhead in Eq.~\eqref{eq:mitigation cost in generic case} could be reduced to $e^{2P_{\text{total}}}$ in some highly structured circuits, such as Trotter simulation circuits, by converting noise to global white noise.
By using this technique, we might be able to double the size of circuits executable on our framework.

\section{Control error cancellation}
\label{sec:Control error suppression}

In the previous sections, we implicitly assume that error channels that describe physical native operations are described well with stochastic Pauli error channels, such as the depolarizing error channels presented in Eq.~\eqref{eq:single depolarizing error} and Eq.~\eqref{eq:double depolarizing error}.
However, in actual devices, we never can neglect the presence of control errors in physical gates due to calibration errors, misalignment of the quantization axes, cross-talk errors, and so on.

In this section, we consider the effect of control errors on our state preparation protocols and methods to suppress them.
First, we show that over-rotations around $Z$-axis inevitably shift the output logical angle of the transversal multi-rotation protocol from the target angle.
We then develop a novel error cancellation scheme, dubbed as {\it randomized transversal rotation}. We show that, 
assuming a reasonable error model, 
leading coherent errors arising in the transversal multi-rotation protocol can be canceled by randomly switching the direction of each physical rotation.

\subsection{Control error model}

\begin{figure}
    \centering
    \includegraphics[width=8.7cm]{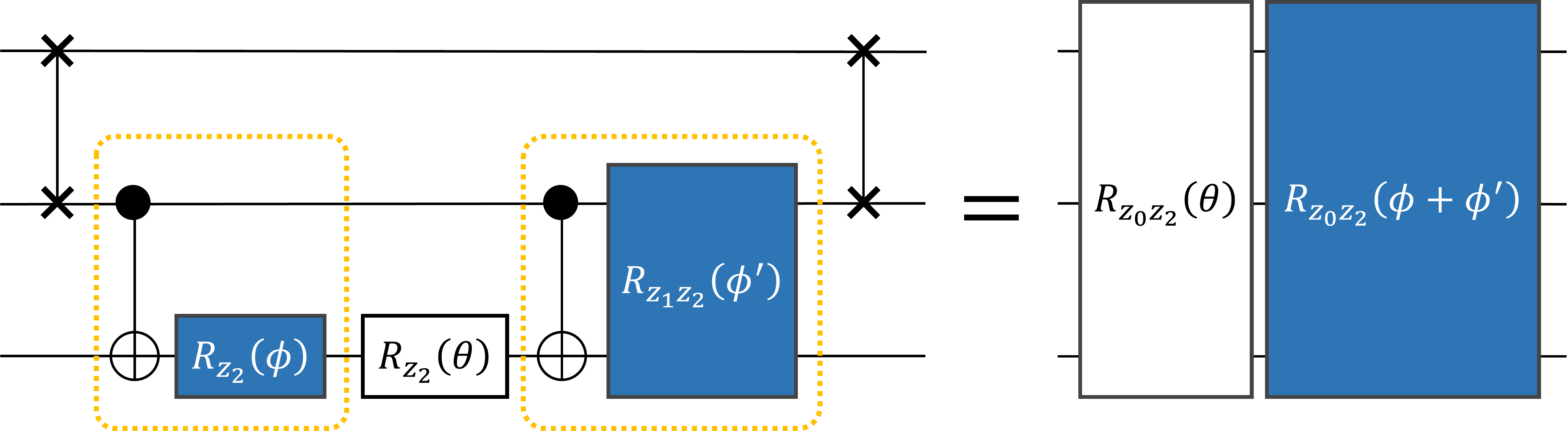}
    \caption{Over-rotation due to control errors in the CNOT gates before and after the virtual-Z rotation gate $R_{z_2}(\theta)$. Another type of multi-Pauli rotation errors accompanied with the CNOT gates can be detected via syndrome measurements in transversal multi-rotation protocol.}
    \label{fig:control error in CNOT}
\end{figure}

To demonstrate our idea, we start with a simplified model for control errors.
Specifically, we assume a unitary error model where rotation angles in the transversal multi-rotation gate (Eq.~\eqref{eq:transversal multi rotation}) are slightly shifted by over-rotation errors as follows:
\begin{equation}
\label{eq:over-rotation model}
    \prod_{i=1}^k \hat{R}_{zz,i}(\theta)\ \ \to \ \ \prod_{i=1}^k \hat{R}_{zz,i}(\theta+\phi_i).
\end{equation}
Here we assume the case of $(m, k)=(2, d/2)$ as an example, and $\phi_i$ denotes a small shift angle in the $i$-th rotation gate due to control errors such as an imperfect gate calibration.
Note that standard benchmarking methods~\cite{Knill2008benchmarking,Eisert2020} for evaluating gate error rates $p_{\text{ph}}$ measure combined effects of stochastic errors and control errors.
Consequently, in our notations, it is natural to assume that the above parameters satisfy at least
$|\phi_i|^2 \lesssim p_{\text{ph}}$.

In what follows, we only assume the case where we implement $\hat{R}_{zz}(\theta)$ in Eq.~\eqref{eq:over-rotation model} using the circuit in Fig.~\ref{fig:implementation of rotation gate} (c).
In this setup, we can implement the single-qubit rotation gate $\hat{R}_{z}(\theta)$ in the circuit with an incredibly high precision by utilizing the virtual-$Z$ gate scheme~\cite{Mckay2017}.
This implies that we should regard the control errors of the CNOT gates before and after $\hat{R}_{z}(\theta)$ in Fig.~\ref{fig:implementation of rotation gate} (c) as the primary sources of over-rotation errors.
Fig.~\ref{fig:control error in CNOT} shows which type of control error contributes to the angle shift $\phi_i$ in Eq.~\eqref{eq:over-rotation model}.
These errors seem to lack physically plausible causal mechanisms; therefore, the magnitude of the errors is expected to be relatively small.
Furthermore, it is noteworthy that, in our setup, the angle shift $\phi_i$ will not depend on the value of the target angle $\theta$, since the control errors in the CNOT gates arise independently of how the angle $\theta$ is chosen and the unitary gates that contribute to the over-rotation commutes with the single-qubit rotation gate with the angle $\theta$.
Hence, in the following, we suppose that shift angles $\{\phi_i\}$ in Eq.~\eqref{eq:over-rotation model} are constant with respect to $\theta$.

Finally, we comment on the possibility of more general types of errors.
As an illustration, let us assume a situation where each rotation gate in the transversal multi-Pauli rotation is modified by a weak $XX$-rotation gate:
\begin{equation}
    \prod_{i=1}^k \hat{R}_{zz,i}(\theta)\ \ \to \ \ \prod_{i=1}^k \hat{R}_{xx,i}(\phi_i)\hat{R}_{zz,i}(\theta),
\end{equation}
In this case, the erroneous unitary gate can be decomposed as $\hat{R}_{xx,i}(\phi_i)= \cos(\phi_i)\cdot \hat{I}\hat{I} + i \sin(\phi_i)\cdot \hat{X}\hat{X}$, and each term on the right-hand side maps the input state to a syndrome space with different stabilizer eigenvalues.
In particular, as $\hat{X}\hat{X}$ term causes unexpected error syndromes, we can detect and remove it via the syndrome measurements in our preparation protocol by the same arguments as in Sec.~\ref{sec:formulation of injection protocol}.
These analyses are readily extended to more generic error models, which may include not only over-rotation errors but also phase-misalignment and cross-talk errors.
For these reasons, we assume that the error mechanism presented in Eq.~\eqref{eq:over-rotation model} is the only source of the logical angle error that occurs even in the order of $\order{\phi_i}$.

\subsection{Effect of over-rotation error on prepared resource state}

Next, let us consider how the over-rotation errors in Eq.~\eqref{eq:over-rotation model} modify the output state of our preparation protocol.
By performing transversal multi-rotation protocol under this type of error, we obtain an imperfect resource state with a shifted logical angle
\begin{equation}
\label{eq: shifted angle}
\begin{aligned}
    \tilde{\theta}_*(\theta,k, \{\phi_i\})\  &\equiv\  \sin^{-1}\left(
    \frac{1}{\sqrt{\tilde{p}_{\text{ideal}}}} \prod_{i=1}^k \sin(\theta+\phi_i)\right),
\end{aligned}
\end{equation}
where $\tilde{p}_{\text{ideal}}$ is the success rate when no errors occur other than over-rotation errors in Eq.~\eqref{eq:over-rotation model}, 
\begin{equation}
    \tilde{p}_{\text{ideal}}(\theta,d,\{\phi_i\})\  \equiv\  \prod_{i=1}^k \sin^2(\theta+\phi_i)+\prod_{i=1}^k \cos^2(\theta+\phi_i). 
\end{equation}
For example, if we assume an ideal case where $\phi_i \ll \theta \ll 1$, the above equation leads to
\begin{equation}
\label{eq:bare relative error}
\begin{aligned}
    \tilde{\theta}_*(\theta,k, \{\phi_i\})
    &  \ \simeq \  \prod_{i=1}^k (\theta+\phi_i)^d\\
    &\ \simeq \ \theta_*(\theta,d)\cdot\left(1+\sum_i \phi_i/\theta\right),
\end{aligned}
\end{equation}
where $\theta_*(\theta,k)$ is the target logical angle in Eq.~\eqref{eq:angle relation}.
This result implies that, in our preparation protocol, 
the relative error to the logical target angle $\epsilon_{\text{rel}}\equiv|\tilde{\theta}_*-\theta_*|/\theta_*$ is determined by the cumulative sum of the relative errors $\phi_i/\theta$ to the physical angles $\theta$.

In principle, we can estimate the magnitude of the relative error with very high precision, by executing the QPE algorithm for the rotation gate $\hat{R}_z(\tilde{\theta}_*)$.
Therefore, in reality, it may be helpful to calibrate the transversal rotation gate for  reducing the estimated relative errors as much as possible before executing large-scale practical tasks such as materials simulation.
Such a calibration will not take so much time, since the quantum circuit for the estimation is far shallow compared to the practical tasks discussed in Sec.~\ref{sec:application} and it could be performed for several ancillary patches simultaneously.
In what follows, we consider how to suppress over-rotation errors that remain even after such calibration.

\subsection{Randomized transversal rotation}

Next, we introduce random flips in the direction of transversal rotation. 
Namely, we consider the following form of transversal rotation, instead of that in Eq.~\eqref{eq:over-rotation model}:
\begin{equation}
    \prod_{i=1}^k \hat{R}_{zz,i}(\theta+\phi_i)
    \ \ \to \ \ 
    \prod_{i=1}^k \hat{R}_{zz,i}((-1)^{n_i}\theta+\phi_i),
\end{equation}
where $n_i$ $(=0,1)$ is a parameter that determines whether we flip the direction of the $i$-th rotation gate or not.
By employing this gate, our preparation protocol yields a resource state with a shifted logical angle 
\begin{equation}
\begin{aligned}
    \sin^{-1}&\left(
    \frac{1}{\sqrt{\tilde{p}_{\text{ideal}}}} \prod_{i=1}^k \sin((-1)^{n_i}\theta+\phi_i)\right)\\
    &=\ (-1)^{\sum_in_i}\sin^{-1}\left(
    \frac{1}{\sqrt{\tilde{p}_{\text{ideal}}}} \prod_{i=1}^k \sin(\theta+(-1)^{n_i}\phi_i)\right)\\
    &=\ (-1)^{\sum_in_i}\cdot \tilde{\theta}_*(\theta,k, \{(-1)^{n_i}\phi_i\}).
\end{aligned}
\end{equation}
When $\sum_in_i\equiv 1$ (mod 2), by applying a logical-$X$ gate to the output state, we obtain a resource state with a logical angle of $\tilde{\theta}_*(\theta,k, \{(-1)^{n_i}\phi_i\})$.
This means that, by reversing the sign of the rotation angle $\theta$, we can effectively reverse the sign of the over-rotation angle $\phi_i$, assuming that $\phi_i$ does not depend on $\theta$.
Realistically, we need not apply a logical-$X$ gate to remove the factor $(-1)^{\sum_in_i}$, since we can treat it by modifying the feedback operation in the gate teleportation.

By sampling a resource state with a logical angle $\tilde{\theta}_*(\theta,k, \{(-1)^{n_i}\phi_i\})$ randomly, we yield the following quantum channel after teleporting the state:
\begin{widetext}
\begin{equation}
\begin{aligned}
    \mathcal{R}^{\text{ave}}(\hat{\rho}) &\equiv \frac{1}{2^k}\sum_{\{n_i\}}\hat{R}_z(\tilde{\theta}_*(\{n_i\})) \hat{\rho} \hat{R}_z^\dagger(\tilde{\theta}_*(\{n_i\})) \\
     &=\left(\frac{1}{2^k}\sum_{\{n_i\}}\cos^2(\tilde{\theta}_*(\{n_i\})) \right) \hat{\rho} \  +  \ i\left(\frac{1}{2^k}\sum_{\{n_i\}}\sin(\tilde{\theta}_*(\{n_i\})) \cos(\tilde{\theta}_*(\{n_i\}))\right) (\hat{Z}\hat{\rho}-\hat{\rho}\hat{Z}) \\
    &\qquad\qquad+\left(\frac{1}{2^k}\sum_{\{n_i\}}\sin^2(\tilde{\theta}_*(\{n_i\})) \right)\hat{Z}\hat{\rho}\hat{Z}\\
    & = \hat{R}_z(\tilde{\theta}_*^{\text{ave}}) \hat{\rho} \hat{R}_z^\dagger(\tilde{\theta}_*^{\text{ave}}) + \order{\theta_*^2\phi_i^2},
\end{aligned}
\end{equation}
\end{widetext}
where we denote $\tilde{\theta}_*(\theta,k, \{(-1)^{n_i}\phi_i\})$ as $\tilde{\theta}_*(\{n_i\})$ briefly and introduce the averaged rotation angle over all possible flips as follows:
\begin{equation}
    \tilde{\theta}_*^{\text{ave}}(\theta,k, \{\phi_i\})
    \equiv \frac12 \arcsin\left(\frac{1}{2^k}\sum_{\{n_i\}}\sin(2\tilde{\theta}_*(\{n_i\})) \right).
\end{equation}
Then, we expand the function $\sin(2\tilde{\theta}_*(\theta,k, \{\phi_i\}))$ with respect to the parameters $\{\phi_i\}$ as
\begin{equation}
\begin{aligned}
    \sin(2\tilde{\theta}_*(\theta,k, \{\phi_i\}))
    &= \sin(2\theta_*(\theta,k)) + \sum_i C_i \phi_i \\
    & \qquad\qquad 
    +\sum_{i\leq j} C_{ij} \phi_i \phi_j +\cdots,
\end{aligned}
\end{equation}
where $C_i$ and $C_{ij}$ are the expansion coefficients with respect to $\phi_i$.
Using this notation, the averaged rotation angle can be rewritten as 
\begin{equation}
    \sin(2\tilde{\theta}_*^{\text{ave}}(\theta,k, \{\phi_i\})) = \sin(2\theta_*(\theta,k))  + \sum_i C_{ii} \phi_i^2 + \order{\phi_i^3}.
\end{equation}
We can calculate the explicit form of $C_{ii}$, and obtain a simple form $C_{ii}(\theta,k) = 2\theta_*(\theta,k)(1 + \order{\theta})$ in small angle regime ($\theta\ll1 $).
Therefore, if we assume an ideal case where $\phi_i \ll \theta \ll 1$, the averaged rotation angle scales as 
\begin{equation}
\label{eq:relative error with randomization}
    \tilde{\theta}_*^{\text{ave}}(\theta,k, \{\phi_i\}) \simeq \theta_*(\theta,k)\left(1+\sum_i\phi_i^2  \right).
\end{equation}
Compared to Eq.~\eqref{eq:bare relative error}, this result suggests that our randomized method suppresses the relative error to the logical angle $\theta_*$ from $|\sum_i \phi_i/\theta|$ to $\sum_i \phi_i^2/2$.

\begin{figure}[tb]
  \centering
  \begin{tabular}{l}
{\large (a)} \\
   \includegraphics[width=8.3cm]{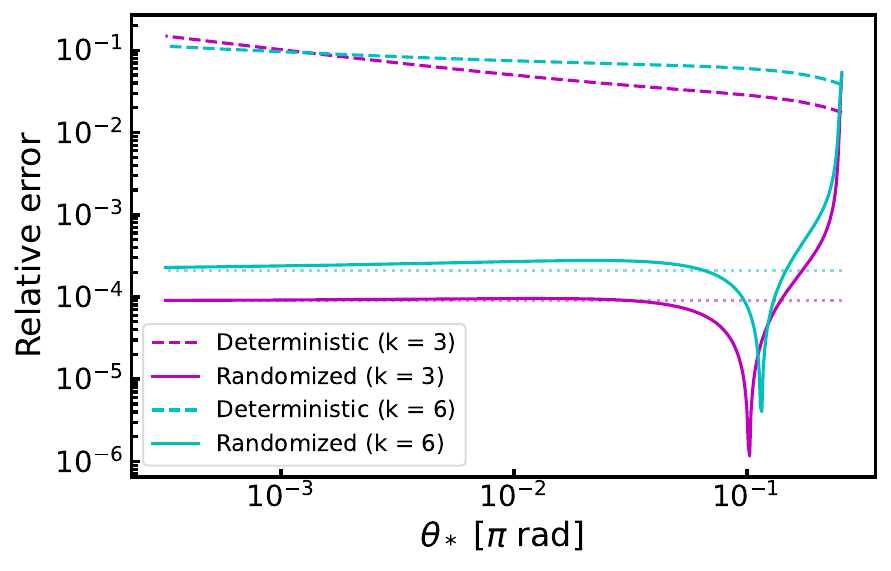}\\
{\large (b)} \\
    \includegraphics[width=8.3cm]{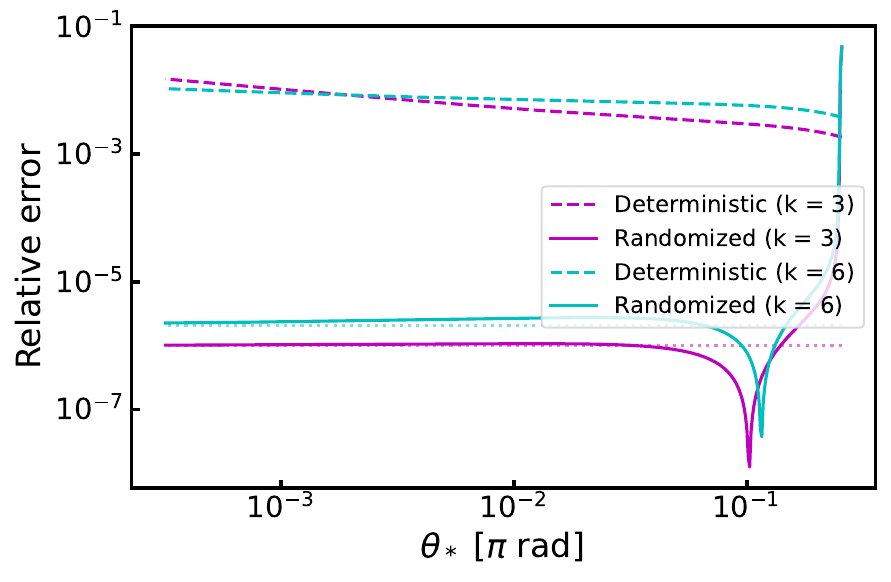}
  \end{tabular}
\caption{Relative error of the logical angle to the target angle $\theta_*$ with (solid line) and without (dashed line) the application of our randomized methods. We set the values of physical over-rotation angles $\{\phi_i\}$ by sampling from a uniform distribution in the range $[0,\phi_{\text{max}}]$ with (a) $\phi_{\text{max}}=10^{-2} $ [rad] and (b) $\phi_{\text{max}}=10^{-3} $ [rad]. Plotted values are obtained by averaging the results over 100 samplings. Dotted line denotes the theoretical value expected from Eq.~\eqref{eq:relative error with randomization}. The cusps in the curves are due to the change in sign of the relative error.}
\label{fig:relative error}
\end{figure}

In Fig.~\ref{fig:relative error}, we numerically compare the relative errors to the target angle $\theta_*$ with and without our randomized method. 
The plot suggests that, in a small angle regime ($\theta_*\ll 1$), the relative error approaches the constant value $\sum_i \phi_i^2/2$ when we use our randomized method. This is consistent with the asymptotic behavior in Eq.\eqref{eq:relative error with randomization}.

\section{Promising applications}
\label{sec:application}

Next, we explore some of the promising applications that fully harness the potential of our framework.
As demonstrated in Eq.~\eqref{eq:mitigation cost in generic case}, quantum circuits executed on our framework must satisfy
the following universal bound on the total accumulation error $P_{\text{total}}$ to keep the additional measurement cost for PEC at a moderate value:
\begin{equation}
\label{eq: universal bound}
   P_{\text{total}}\ =\  \alpha_{\text{RUS}}\theta_{\text{total}} p_{\text{ph}} \ \lesssim \ 1,
\end{equation}
where $\theta_{\text{total}}$ denotes the total angle rotated across the entire circuit in a single shot.
$\alpha_{\text{RUS}}$ is the prefactor that quantifies the error accumulation due to the RUS process for gate teleportation, and it is evaluated as $\alpha_{\text{RUS}}\sim 1.5$ with the protocol switching.
In Sec.~\ref{sec:Stochastic error mitigation}, we have shown that the multiplicative factor that determines the measurement cost of PEC scales as $e^{4P_{\text{total}}}$ approximately.
For example, when $P_{\text{total}}=1$, we need to perform $e^4\simeq 55$ times more measurement shots to mitigate the errors in the estimates of expectation values.
This constraint severely limits the range of possible applications of our architecture.

For this reason, it is desirable that quantum algorithms executed on our framework are not only significantly more complex than those feasible on NISQ devices, but are also composed of rotation gates with as small rotation angles as possible.
In addition, quantum circuits with high parallelism are more desirable, as we can execute multiple rotation gates in parallel without any additional spatial cost.
In what follows, we present some promising scenarios that can harness the great potential of our framework even under these requirements.

\subsection{Quantum simulation with the Trotter circuit}

The simplest and most promising application of our framework is the simulation of quantum many-body dynamics using the Trotterization techniques~\cite{Lloyd1996,Beverland2022assessing}.
It is well-known as a valuable approach for elucidating various nonequilibrium phenomena~\cite{Fauseweh2024review}, including chemical reactions~\cite{Lidar1999, Kassal2008} and laser-induced many-body physics~\cite{Magann2021, Chan2023grid, Kohler1995review,Cavalleri2018review, Ghimire2019review}, and for evaluating
theoretical tools such as the many-body Green's function~\cite{Wecker2015_2, Bauer2016, Kreula2016, Kanasugi2023}.
For example, the authors of Ref.~\cite{Childs2018speedup} previously proposed that simulating disordered quantum spin models can be an early practical application of quantum computers, as it requires significantly fewer resources than other classically infeasible problems such as prime factoring and quantum chemistry. Their estimates revail that the Trotterization approach is the most preferable of several leading algorithms if empirical error estimates of the Trotter error suffice.

Trotter circuits can also be adapted to estimate the eigenvalues of many-body Hamiltonians with QPE~\cite{Abrams1999,Aspuru-Guzik2005,Lin2022,Ding2023QCELS}, and to prepare the ground state with the adiabatic (or imaginary) time evolution~\cite{Aharonov2003,Wecker2015_2, Motta2020QITE, Lin2021, Kosugi2022} or the quantum eigenvalue transformation of unitary matrices
(QETU)~\cite{Dong2022preparation}.
Furthermore, beyond quantum many-body simulations, Trotter simulation holds promise for applications in areas such as combinatorial optimization~\cite{Kadowaki1998} and Markov chain Monte Carlo sampling~\cite{Layden2023}, and simulation of differential equations~\cite{Lloyd2020_2,Dodin2021plasma}.

In the following, as two important examples, we provide brief resource analyses of the Trotter simulation of quantum many-body dynamics, and the eigenvalue estimations with the QPE in our framework. For QPE, we present a more detailed estimate of its space-time cost in Sec.~\ref{sec:resource estimation}.

\subsubsection{Trotter simulation of quantum many-body dynamics}
\label{sec:Dynamics simulation}

The Trotter circuit consists of a long sequence of analog rotation gates with fairly small angles; typically, we can execute several rotation gates in parallel. 
These features meet our requirements to fully exploit the potential of our framework. 
For example, using the second-order Trotter decomposition, we decompose the time-evolution operator $e^{-iT\hat{\mathcal{H}}}$ as follows:
\begin{equation}
\label{eq: Trotter circuit}
\begin{aligned}
    e^{-iT\hat{\mathcal{H}}} &\simeq \ \left(\prod_{i=1}^L e^{-i\left(\frac{a_iT}{2N}\right)\hat{P}_i}\prod_{i=L}^1 e^{-i\left(\frac{a_iT}{2N}\right)\hat{P}_i}\right)^N\\
    & = \ \left(\prod_{i=1}^L \hat{R}_{P_i}(\theta_i)\prod_{i=L}^1 \hat{R}_{P_i}(\theta_i)
    \right)^N
\end{aligned}
\end{equation}
where we define the number of Trotter steps as $N$ and each rotation angle as $\theta_i=-a_iT/2N$. We assume that the Hamiltonian is specified as a linear combination of the Pauli string operators $\hat{P}_i$: 
\begin{equation}
\label{eq:Hamiltonian}
    \hat{\mathcal{H}}=\sum_{i=1}^L a_i\hat{P}_i.
\end{equation}

Applying Eq.~\eqref{eq: universal bound} to the Trotter circuit, we obtain the upper bound on the runtime $T$,
\begin{equation}
\label{eq: trotter time bound}
       T \ \lesssim \  \frac{1}{\alpha_{\text{RUS}}\lambda p_{\text{ph}}}.
\end{equation}
Here we use the fact that the total angle in the circuit is calculated as  $\theta_{\text{total}}=\lambda T$, where $\lambda\equiv\norm{\hat{\mathcal{H}}}_1= \sum_{i=1}^L |a_i|$ is the 1-norm of the Hamiltonian $\hat{\mathcal{H}}$.
As long as the evolution time $T$ satisfies this inequality, we can accurately perform Trotter simulation with an admissible error mitigation cost of the order $\order{1}$.
Here it is noteworthy that Eq.~\eqref{eq: trotter time bound} does not depend on the number of Trotter steps $N$ because the error rate of each analog rotation gate is proportional to $1/N$.

As an intriguing example, let us consider a one-dimensional Heisenberg model disordered by a random magnetic field. 
Its Hamiltonian is described as
\begin{equation}
    \hat{\mathcal{H}}=  \sum_j (\hat{X}_j\hat{X}_{j+1}+\hat{Y}_j\hat{Y}_{j+1}+\hat{Z}_j\hat{Z}_{j+1})
    +\sum_j h_j \hat{Z}_j,
\end{equation}
where we assume a periodic condition, and random magnetic field $h_j\in[-h, h]$ is chosen uniformly and randomly.
This model is known to exhibit intriguing dynamical behaviors, such as many-body localization~\cite{Nandkishore2015review}; therefore, it has been extensively studied in the condensed matter community to explore the nature of self-thermalization in closed quantum systems~\cite{Nandkishore2015review, Pal2010,Luitz2015}.
Despite these efforts, very little is known about the transition between the thermal and localized phases; furthermore, the most extensive numerical study has been limited to at most 22 spins~\cite{Luitz2015} due to the difficulty of simulating quantum systems classically. To address this issue, Ref.~\cite{Childs2018speedup} investigated the resource for simulating the above model on a quantum computer, and suggested that this task could be an early promising application for demonstrating practical quantum speedups.


The averaged 1-norm of the above Hamiltonian is evaluated as $\lambda=(3+h/2)N_{\text{site}}$, where $N_{\text{site}}$ is the site number.
Assuming that $\alpha_{\text{RUS}}\simeq 1.5$ and $h=1$, Eq.~\eqref{eq: trotter time bound} yields the upper bound
\begin{equation}
    T \ \lesssim \  \frac{4}{21N_{\text{site}}p_{\text{ph}}}.
\end{equation}
Then, assuming $p_{\text{ph}}=10^{-4}$ and $N_{\text{site}}=100$, this bound suggests that we can simulate the many-body dynamics with an arbitrary precision, up to the maximum runtime of $T_{\text{max}}\sim 19$ without paying excessive error mitigation costs.
Such long-time many-body dynamics are significantly more complex than those recently demonstrated on a NISQ device in Ref.~\cite{Kim2023evidence}. These dynamics will be intractable on a classical computer without introducing some bold approximations. We expect that Trotter simulation discussed here, combined with well-established techniques for analyzing real-time quantum dynamics~\cite{Serbyn2014,Schreiber2015,Smith2016,Brydges2019entanglement}, will offer new insights into the nature of nonequilibrium quantum many-body phenomena.

Finally, it should be noted that the 1-norm $\lambda$ does not necessarily increase linearly with the number of Hamiltonian terms, unlike in the case of the above spin model. In electronic structure problems in quantum chemistry, molecular Hamiltonians contain a huge number of weak long-range Coulomb interaction terms. In such a case, 
the complexity of the Hamiltonian is not directly related to the strictness of the evolution time bound in Eq.~\eqref{eq: trotter time bound}, while it is closely related to the execution time.

\subsubsection{Eigenvalue estimation with QPE}
\label{sec:appliccation of QPE}

Next, consider QPE with the Trotter decomposition. Estimating the phase to an accuracy $\epsilon$ requires the maximum evolution time $T_{\text{max}}$ to be at least $\pi/\epsilon$ for typical QPE algorithms such as the so-called textbook-type QPE~\cite{Nielsen2000}, even if the overlap between the initial state and the ground state, $\eta$, is sufficiently large.
Circuits with such a large depth are undesirable for our framework.

Meanwhile, in the last few years, several studies have developed other types of QPE methods suitable for early-FTQC devices, where $T_{\text{max}}$ is suppressed to be relatively smaller and only one ancillary qubit is required~\cite{Lin2022,Wan2022randomized,Wang2023gaussian,Wang2023rejection,Ding2023QCELS,Ding2023simultaneous,Ding2023robust, Ding2024filter,Ni2023robust, Li2023multiple}.
In particular, Refs.~\cite{Ding2023QCELS,Ding2023simultaneous, Ni2023robust,Li2023multiple} proposed quantum (multiple-)phase estimation algorithms that achieve the Heisenberg-limited scaling in the total runtime.
Importantly, in their algorithm, the maximum runtime $T_{\text{max}}$ scales as $\delta/\epsilon$, where the prefactor $\delta$ vanishes as the initial overlap $\eta$ approaches one.
Other works also mentioned that one can estimate eigenvalues with $T_{\text{max}}=\tilde{\mathcal{O}}(\Delta^{-1}\log(1/\epsilon))$ assuming a lower bound on the spectral gap     $\Delta$~\cite{Ding2023QCELS,Ding2023simultaneous,Ding2023robust, Ding2024filter,Wang2023gaussian,Wang2023rejection}.  
Furthermore, several works suggested that some of these algorithms are much more noise-resilient than conventional QPE algorithms~\cite{Kshirsagar2022,Ding2023robust}.

Most of these latest algorithms are based on an iterative execution of the Hadamard test circuit with different runtimes smaller than $T_{\text{max}}$ and provide an optimal classical post-processing of the outcomes. 
Since we can effectively cut the runtime in half using a familiar technique in Appendix.~\ref{Appendix: Hadamard test},  
the universal bound in Eq.~\eqref{eq: universal bound} yields 
\begin{equation}
\label{eq:mitigation cost in QPE}
    P_{\text{total}}\ \equiv \ \alpha_{\text{RUS}} \cdot \lambda \cdot \frac{T_{\text{max}}}{2}\cdot p_{\text{ph}} \ \lesssim \ 1,
\end{equation}
where we use the fact that the total angle satisfies 
$\theta_{\text{total}}\leq \lambda T_{\text{max}} /2$ for each Hadamard test circuit.
Then, assuming the typical scaling $T_{\text{max}}\sim \delta/\epsilon$, this leads to a bound on the Hamiltonian 1-norm
\begin{equation}
   \lambda \ \lesssim \  \frac{2\epsilon}{\alpha_{\text{RUS}}\delta p_{\text{ph}}}.
\end{equation}
For example, assuming the overlap $\eta$ is sufficiently large, the prefactor $\delta$ is heuristically determined as $\delta\simeq 0.06$ for the QPE algorithms proposed in Refs.~\cite{Ding2023QCELS,Ding2023simultaneous}.
Therefore, assuming $p_{\text{ph}}=10^{-4}$, we obtain $\lambda \lesssim 2.2\times 10^2$ for chemical accuracy $\epsilon=10^{-3}$.
This upper bound is much larger than the 1-norms of small molecules, such as H${}_2$O and NH${}_3$, obtained in Ref.~\cite{Loaiza2023}.
Meanwhile it is several times smaller than those of highly correlated molecules such as the FeMo cofactor of nitrogenase (important in nitrogen fixation)~\cite{Reiher2017} and ruthenium metal complexes (important in carbon dioxide capture)~\cite{Burg2021} reported in Ref.~\cite{Koridon2021} (for example, the 1-norm for FeMo  cofactor is estimated as $\lambda=1511$ Hartree using the bounds in Ref.~\cite{Koridon2021}).
Furthermore, in Sec.~\ref{sec:resource estimation}, we show that our framework allows us to analyse the 2D Hubbard model with an execution time that is significantly smaller than that of the classical approach using tensor network techniques.
These observations suggest that our framework has a great potential in eigenvalue spectrum estimation for solids and chemical molecules beyond classically tractable problem sizes.

\subsection{Application of near-term algorithms}

Another promising application of our architecture might be found in near-term algorithms, including VQAs~\cite{Cerezo2021review} and other modern proposals~\cite{Huang2020measurement,Huggins2022,Xu2023QCQMC,Layden2023,Kanno2023qsci,Robledo2024QSCI}.
These algorithms have various potential applications such as quantum simulation of materials~\cite{Tilly2022}, combinatorial optimization~\cite{Blekos2024review}, finance~\cite{Herman2022review}, and quantum machine learning~\cite{Zeguendry2023, Wang2024}. Typically, their circuits consist of a sequence of parallelizable analog rotation gates and simple Clifford gates, which is often easier to implement than that of long-term algorithms.
These properties are preferable to the near-term application of our framework.
In addition, the VQAs are relatively resilient to unitary errors such as over-rotations, because these errors only serve to shift the location of the optimal value for the variational optimization~\cite{McClean2016}. This point is also beneficial for the STAR architecture, where small coherent errors inevitably remain in analog rotations even after performing the error suppression discussed in Sec.~\ref{sec:Control error suppression}.

These near-term algorithms often suffer from their poor scalability in terms of the number of measurements~\cite{Wecker2015,Elfving2020,Gonthier2022,Tilly2022}, vanishing gradient of cost functions~\cite{McClean2018,Wang2021,Cerezo2022,Ragone2023,Larocca2024review}, NP-hardness of variational optimization~\cite{Bittel2021NP-Hard}, and exponentially growing measurement cost for error mitigation~\cite{Takagi2022,Takagi2022_2,Tsubouchi2022}. 
However, several recent proposals in this context suggest that we can partially alleviate these problems by making good use of Hamiltonian partitioning~\cite{Tilly2022, McClean2016, Gokhale2019, Huggins2021measurement}, classical shadow~\cite{Abbas2023backpropagation,Boyd2022,Nakaji2023shadow,Huang2024QCQMC,Boyd2024QSE}, the locality of target systems~\cite{Mizuta2022,Kanasugi2023,Kanasugi2024LSVQC}, and tensor network techniques~\cite{Rudolph2023tensor,Watanabe2023tensor}.
Furthermore, on our framework, we will be able to avoid the excessive time overhead for error mitigation and greatly reduce the measurement cost for estimating the expectation value of some observables by using long-term algorithms for optimal quantum estimation~\cite{Knill2007,Huggins2022optimal}.

In what follows, we briefly discuss some promising applications of near-term algorithms on our framework.

\subsubsection{Guideline for designing the VQAs}

First, let us consider a technique for reducing errors in the VQAs. For example, suppose a variational circuit is constructed in the form
\begin{equation}
    \hat{V}(\{\theta_i\})=\prod_i C_i \hat{R}_z(\theta_i),
\end{equation}
where $C_i$ is an arbitrary Clifford gate, which is performed without errors in the STAR architecture.
The sum of the variational parameters $\{\theta_i\}$ should satisfy the following upper bound from Eq.~\eqref{eq: universal bound}:
\begin{equation}
     \sum_i|\theta_i| \ \lesssim \ \frac{3}{2p_{\text{ph}}}.
\end{equation}
Here we use the fact that $\alpha_{\text{RUS}}\lesssim 1.5$ for arbitrary angles.
One possible approach to suppress the accumulation of errors in the VQA is to introduce a penalty term in the cost function in the optimization process. That is, if the original cost function is given as $L(\{\theta_i\})$, we can modify it into 
\begin{equation}
    L_{\text{new}}(\{\theta_i\}) = L(\{\theta_i\}) + \beta \sum_i |\theta_i|,
\end{equation}
where $\beta$ is a parameter that tunes the weight of the penalty term. Although such a modification may spoil the expressibility of variational circuits slightly, it is crucial for decreasing the total error in the circuits, enabling the execution of much deeper and more complex circuits in our frameworks. 
Moreover, recent several works~\cite{Haug2021,Zhang2022escaping,Wang2023trainability,Park2024} point out that keeping the variational parameters small often helps to avoid the barren plateau problems.
These perspectives might become a new guideline for designing variational circuits and cost functions in the VQAs for our framework.

\subsubsection{Variational quantum compilation}

Next, we briefly discuss a possible approach to suppress the total rotation angle of the entire circuit while increasing the parallelism of rotation gates, or to transform any shallow quantum circuits into the Clifford+$\phi$ gate set.
The key idea is to employ variational quantum compilation (VQC)~\cite{Khatri2019,Sharma2020} or its cousins~\cite{Heya2018,Bilek2022,Jones2022,He2021, Huang2024}, which are 
known as powerful tools to compress quantum circuits into variational circuit ansatzes with a simple gate architecture, based on a hybrid quantum-classical approach. For example, VQC and its local variant~\cite{Mizuta2022,Kanasugi2024LSVQC} was previously applied to approximate time evolution operators with a compressed circuit depth~\cite{Cirstoiu2020,Mizuta2022} and to evaluate eigenvalue spectra of many-body systems through the Green function technique~\cite{Kanasugi2023, Kanasugi2024LSVQC}. 
Similar interesting work can also be found in Ref.~\cite{Morisaki2023}. In the work, by applying the automatic
quantum circuit encoding algorithm~\cite{Shirakawa2021}, the authors attempt to compile the PREPARE circuit for the LCU protocol~\cite{Childs2012} with a variational quantum circuit without ancillary qubits.
By utilizing these techniques, we expect that we can recompile the circuit for various quantum algorithms into a more shallow and simpler one comprising the Clifford+$\phi$ gate set.
Such an approach will further broaden the applicability of our framework in the future.

\subsubsection{Possible application of the quantum-selected configuration interaction}

Finally, let us consider the applicability of other modern near-term algorithms~\cite{Huang2020measurement,Huggins2022,Xu2023QCQMC,Layden2023,Kanno2023qsci,Robledo2024QSCI} as an alternative approach beyond VQAs.
These algorithms are usually designed to overcome some of the difficulties we face in VQAs; however, it still seems challenging to achieve quantum advantages with these algorithms on NISQ devices. In the following, we focus on the quantum-selected configuration interaction (QSCI)~\cite{Kanno2023qsci} as an intriguing  example.

QSCI is a class of hybrid quantum-classical algorithms for calculating the ground- and excited-state energies of many-body Hamiltonians on near-term quantum devices.
It is originally proposed in Ref.~\cite{Kanno2023qsci}, and recently, demonstrated experimentally on a 133-qubit Heron quantum processor~\cite{Robledo2024QSCI}.
In the algorithm, we prepare an approximate ground state on a quantum computer and then measure the state in the computational basis.
From the measurement outcomes, we can identify the electron configurations important for reproducing the ground state, leading to an effective Hamiltonian of the target systems.
In general, such a sampling of important bases is classically infeasible, thereby providing a potential quantum speedup in QSCI.
Finally, by diagonalizing the effective Hamiltonian on a classical computer, we can obtain the ground-state energy and corresponding eigenvector.
Notably, QSCI is demonstrated to be noise resilient numerically and experimentally~\cite{Kanno2023qsci, Robledo2024QSCI} and, in principle, it is free of the costly optimization of parametrized quantum circuits. Furthermore, QSCI can give a rigorous upper bounds on the ground-state energy even under any quantum errors. 
These properties may allow us to perform eigenvalue estimation for systems of classically intractable sizes, even in situations where we have limited quantum resources, smaller than those required for QPE.

The major challenge in QSCI is to produce an approximate ground state that sufficiently overlaps with the true ground state. For example, in Ref.~\cite{Robledo2024QSCI}, this procedure is carried out in a fairly simplified manner using a shallow variational unitary circuit pretrained on a classical computer, because the feasible circuit size is limited on a NISQ device.
However, such an approach is expected to fail to produce a good approximate ground state for most classically intractable systems, thus significantly undermining the quantum advantage of QSCI.

On the other hand, in our framework, we can utilize more elaborated approaches to prepare an approximate ground state, such as the adiabatic (or imaginary) time evolution~\cite{Aharonov2003,Wecker2015_2, Motta2020QITE, Lin2021, Kosugi2022} or the quantum eigenvalue transformation of unitary matrices
(QETU)~\cite{Dong2022preparation}.
For example, Ref.~\cite{Yoshioka2022hunting} suggests that the evolution time required for adiabatic preparation, $t_{\text{ASP}}$, is at most a few tens to achieve the infidelity of $0.1$ for typical spin models with a system size of $N\sim 100$. Similar results are obtained in Ref.~\cite{Sugisaki2022adiabatic} for small molecules such as N${}_2$. 
According to Sec.~\ref{sec:Dynamics simulation}, such a long-time simulation can be implemented in our framework, allowing us to fully exploit the potential quantum advantage of QSCI. 

However, in this case, it is problematic whether we can accomplish the sampling procedure for QSCI within a realistic time, because the preparation protocols discussed above take much more run time than that employed in Ref.~\cite{Robledo2024QSCI}. We will be able to alleviate this issue by employing techniques such as the parallel implementation of rotation gates and the variational quantum compilation of Trotter circuit~\cite{Mizuta2022,Kanasugi2023,Kanasugi2024LSVQC}. 
This issue will be discussed in the future work.

\subsection{Comparison with conventional approaches}

To close this section, we re-emphasize the 
pros and cons of our framework by comparing it with the conventional NISQ and FTQC approaches.
In usual NISQ approaches, we directly perform target quantum circuits at the physical level.
Physical gates usually work faster than their encoded counterpart in our framework, while they are strictly constrained by the connectivity of the hardware.
However, unlike in our framework, any gate operations in NISQ approach contribute to the decoherence of quantum states, regardless of whether it is the Clifford or non-Clifford gate.
This leads to a significantly rapid accumulation of errors, strongly restricting the problem size that can be handled on NISQ devices. 
Furthermore, in our framework, we can easily execute multi-Pauli rotation gates on spatially distant qubits with lattice surgery techniques~\cite{Litinski2019}.
In addition, as shown in Sec.~\ref{sec:Control error suppression}, our framework can keep the relative error to the target angle much smaller than that in the NISQ devices. These features are very desirable for use in various algorithms including the Trotter circuit simulation, implying the superiority of our framework over the NISQ approach.

Next, let us compare our framework with the conventional FTQC approach (see also TABLE.~\ref{tab:compare STAR}).
As already mentioned in Sec.~\ref{sec:Preliminary}, our framework can avoid tedious Solovay-Kitaev decomposition~\cite{Kitaev1997_Review,Dawson2005,Ross2016} and costly magic state distillation~\cite{Fowler2012,Gidney2019,Litinski2019magic}, as we directly prepare an analog resource state $\ket{m_\theta}$.
This is a major advantage in reducing the space-time cost for the execution of quantum circuits.
Furthermore, compared to the usual state preparation with distillation techniques, our preparation protocol has high parallelism, that is, we can generate multiple resource states simultaneously at each local code patch.
This enables us to execute multiple analog rotation gates in parallel, as long as there is no overlap in the routing areas required for the gate teleportation with lattice surgery.
By utilizing such high parallelism, our framework is expected to achieve a significant acceleration relative to the conventional FTQC approach, especially in the simulation of models with local interactions such as the Hubbard model. Typically, fermion-to-qubit mappings, such as the Jordan-Wigner transformation~\cite{Jordan1928}, yield non-local terms, making parallel implementation of rotation gates difficult. However, such issues will be 
alleviated by utilizing familiar techniques such as the fermionic swap network~\cite{Babbush2018fswap,Kivlichan2018}.

The parallel implementation of rotation gates in Trotter circuits is straightforward.
Whereas, in the Hadamard test circuit for the QPE, we need to perform a series of nonlocal controlled-rotation gates, rather than local rotation gates.
In particular, since every controlled-rotation gate connects to the same ancilla qubit, 
the parallel execution of the circuit seems to be difficult.
Fortunately, under some reasonable assumptions, we can easily avoid these difficulties by utilizing the so-called {\it control-free implementation} of controlled time evolution~\cite{Huggins2020control-free,Lu2021, Russo2021, Lin2022, Dong2022preparation}.
This technique will play a pivotal role in fully exploiting the high parallelism in our framework, as well as the fermionic swap network technique.

\section{Resource analysis for QPE}
\label{sec:resource estimation}

Estimating the ground state energy of Hamiltonian is a fundamentally important problem that underlies computational materials science~\cite{Bauer2020,Cao2019,McArdle2020}.
This task is often cited as a promising candidate for practical
quantum speedups, although there is some skepticism regarding the evidence~\cite{Lee2023evaluating}.
In this section, to demonstrate the usefulness of our framework, we provide a resource analysis in the QPE for typical many-body Hamiltonians such as the Hubbard model. In what follows, we only assume the case where logical qubits are encoded on the rotated surface code~\cite{Horsman2012} and any Clifford operation is executed using the lattice surgery techniques~\cite{Horsman2012,Litinski2019}.

\subsection{QCELS algorithm}

\begin{figure}
    \centering
    \fontsize{11pt}{11pt}\selectfont
    \mbox{
        \Qcircuit @C=1em @R=.7em {
        \lstick{\ket{+}}    & \ctrl{1}                    & \gate{\hat{W}} & \measureD{M_{X}}  \\
        \lstick{\ket{\psi_0}} & \gate{e^{-in\tau \hat{\mathcal{H}}}}     & \qw      & \qw   \\
        }
    }    
    \caption{Hadamard test circuit for QCELS algorithm. We set $\hat{W} = \hat{I}$ (or $ \hat{S}^{\dag}$) to estimate the real (or imaginary) part of $\bra{\psi_0} e^{-in\tau \hat{\mathcal{H}}} \ket{\psi_0}$. }
    \label{fig:circuit for QCELS}
\end{figure}
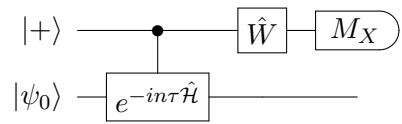

In this section, we explain a state-of-the-art QPE algorithm designed for early-FTQC devices, called the quantum complex exponential least squares (QCELS) algorithms~\cite{Ding2023QCELS,Ding2023simultaneous}. 
In this algorithm, we repeatedly run the Hadamard test circuit (Fig.~\ref{fig:circuit for QCELS}) for the unitary $U=e^{-i n \tau \hat{\mathcal{H}}}$ ($n = 0, \dots, K-1$) with an input state $\ket{\psi_0}$ and a small time evolution interval $\tau$. 
By running the circuit $N_s$ times, we prepare the following data set:
\begin{equation}
    \mathcal{D} = \{ (n\tau, Z_n) \}_{n=0}^{K-1},
\end{equation}
where $Z_n$ is an estimator for $\bra{\psi_0} e^{-i n\tau \hat{\mathcal{H}}} \ket{\psi_0}$:
\begin{equation}
    Z_n \equiv \frac{1}{N_s} \sum_{k=1}^{N_s} (X_{k,n}+iY_{k,n})\ \xrightarrow{N_s\to\infty} \ \bra{\psi_0} e^{-i n\tau \hat{\mathcal{H}}} \ket{\psi_0}.
\end{equation}
Here, $X_{k,n}$ and $Y_{k,n}$ are independent random variables obtained from the Hadamard test circuit with different $\hat{W}$. They provide estimates for the real and imaginary parts of $\bra{\psi_0} e^{-i n\tau \hat{\mathcal{H}}} \ket{\psi_0}$, respectively.

Next, using the data set $\mathcal{D}$, we define the mean-square error function as follows:
\begin{equation}
    L^{(0)}(r, \theta) = \frac{1}{N} \sum_{n=0}^{K-1} |Z_n - r e^{-i n\tau\theta}|^2. \label{eq:loss_func}
\end{equation}
For example, if the initial state $\ket{\psi_0}$ equals the ground state $\ket{\psi_{GS}}$, the estimator $Z_n$ approaches $e^{-in\tau E_0}$ in the limit $N_s\to\infty$. Here $E_0$ is the ground state energy. 
Therefore, we readily notice that, in such a limit, one of the global minimum $\left( r^*, \theta^* \right)$ of $L(r, \theta)$ coincides with $r_*=1$ and $\theta_*=E_0$. 
In QCELS, with a fixed number of samples $N_s$, we can estimate the ground state energy with higher precision as the maximum runtime $T_{\text{max}} = (K-1)\tau$ increases.
However, since the loss function $L(r, \theta)$ is periodic in the transformation $\theta\to \theta\pm 2\pi/\tau$, we need to remove this uncertainty by introducing some proper scheme.

To this end, the (multi-level) QCELS algorithm adopts a level-by-level approach to estimate the ground state energy $E_0$.
More specifically, we start with small time interval $\tau_0=\delta/K$ and estimate the eigenvalue using the time-series data $Z^{(0)}_n \equiv \bra{\psi_0} e^{-i n \tau_0 \hat{\mathcal{H}}} \ket{\psi_0}$ ($n = 0,1, \dots, K-1$).
Here $\delta$ is a prefactor that should be set properly.
This procedure gives a rough estimate of $E_0$.
Then, after doubling the time interval as $\tau_1=2\tau_0$, we repeat the same procedure and get a refined estimate for $E_0$.
Repeating these procedures up to $\tau_J$, we estimate $E_0$ with target precision $\epsilon$.
Here the integer $J$ is specified as $J(\epsilon)=\lceil \log_2(1/\epsilon) \rceil+1$.

In conclusion, estimating the ground state energy using QCELS algorithm requires the maximum runtime 
\begin{equation}
\label{eq:maximum runtime}
    T_{\text{max}}=K\tau_J=\delta/\epsilon,
\end{equation}
and the total runtime~\cite{Comment2}
\begin{equation}
\label{eq: total runtime}
    T_{\text{total}} = \sum_{j=1}^J \sum_{n=0}^{K-1}2 N_s n \tau_j \\
    =\sum_{j=1}^J K(K-1)N_s \tau_j,
\end{equation}
where $\tau_j=2^{j-1}\tau_0$.
Using the inequality $2/\epsilon \leq 2^J < 4/\epsilon$, we can relate $T_{\text{total}}$ with $T_{\text{max}}$ implicitly as follows:
\begin{equation}
    2(K-1)N_s T_{\text{max}} \ \leq \ T_{\text{total}} \ < \  4(K-1)N_s T_{\text{max}}.
\end{equation}
According to Ref.~\cite{Ding2023QCELS}, the prefactor $\delta$ is determined as $\delta\simeq 0.06$ if we assume the ground state energy estimation for the 8-site transversal field Ising model with $K=5$, $N_s=100$, and $\eta > 0.6$.
In general, the prefactor $\delta$ depends on the system size indirectly through factors such as the overlap $\eta$ and the spectrum gap.
However, in the following estimation, we will use the heuristic parameters presented above, as it is not our purpose to elucidate the intricate behavior of the prefactor in detail.

\subsection{Resource estimation: Theory}

Now we are ready to estimate the space-time cost for executing the QCELS algorithm for many-body Hamiltonians.
To achieve the desired level of precision $\epsilon_{\text{target}}$ in this task, it is necessary to clarify the required precision in each subroutine.
In our case, there are two sources of errors that affect our energy estimation: (i) Trotter error $\epsilon_{\text{Trotter}}$ and (ii) algorithmic error in the QCELS algorithm $\epsilon_{\text{QPE}}$.
In the worst scenario, these errors 
additively contribute to the total error~\cite{Reiher2017,Kivlichan2020improved}.
Therefore, to achieve a target precision of $\epsilon_{\text{target}}
$, we need to satisfy the following constraint:
\begin{equation}
\label{eq:constraint for precision}
    \epsilon_{\text{Trotter}} + \epsilon_{\text{QPE}} \leq \epsilon_{\text{target}}.
\end{equation}

\begin{figure*}
    \centering
    \large
    \mbox{
        \Qcircuit @C=1.0em @R=0.7em {
        & & & & \mbox{Repeat $N_{j,n}$ times\qquad\qquad\qquad\qquad} & & & &\\
        \lstick{\ket{+}} & \multigate{1}{\hat{R}_{Z \otimes P_1}(\theta_1)}  & \multigate{1}{\cdots}  &   \multigate{1}{\hat{R}_{Z \otimes P_L}(\theta_L)} & \multigate{1}{\hat{R}_{Z \otimes P_L}(\theta_L)}  & \multigate{1}{\cdots}  &   \multigate{1}{\hat{R}_{Z \otimes P_1}(\theta_1)}  & \gate{\hat{W}} & \measureD{M_{X}}  \\
        \lstick{\ket{\psi}} & \ghost{\hat{R}_{Z \otimes P_1}(\theta_1)} & \ghost{\cdots}  & \ghost{\hat{R}_{Z \otimes P_L}(\theta_L)} & \ghost{\hat{R}_{Z \otimes P_L}(\theta_L)} & \ghost{\cdots}  & \ghost{\hat{R}_{Z \otimes P_1}(\theta_1)}  & \qw      & \qw \gategroup{2}{2}{3}{7}{.7em}{--}  \\
        }
    }
    \vspace{0.2cm}
    \caption{Quantum circuit equivalent to that in Fig.~\ref{fig:Hadamard_test} under the approximation of the second-order Trotter decomposition. Here each rotation angle is determined as $\theta_i=a_i\Delta t/2$, where $a_i$ is the coefficient of the Hamiltonian in Eq.~\eqref{eq:Hamiltonian}, and $\hat{Z}\otimes \hat{P}_i$ represents a direct product of the Pauli-$Z$ operator on the ancilla qubit and a Pauli string operator included in the Hamiltonian.
    The circuit enclosed by a dotted line corresponds to a single Trotter step for $e^{-i(\hat{Z}\otimes \hat{\mathcal{H}})n\tau_j}$; thus we must repeat this circuit $N_{j,n}\equiv \lceil n\tau_j/2\Delta t \rceil$ times.
    We set $\hat{W} = \hat{I}$ (or $ \hat{S}^{\dag}$) to estimate the real (or imaginary) part of $\bra{\psi} e^{-in\tau \hat{\mathcal{H}}} \ket{\psi}$. }
    \label{fig:Trotter circuit for QCLES}
\end{figure*}
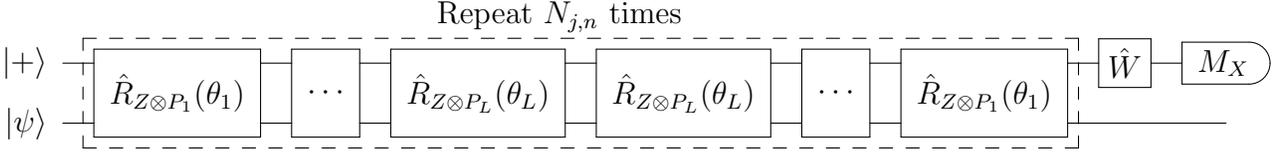

As discussed in the previous subsection, the algorithmic error in the QCELS algorithm is related to the maximum and total runtime via Eq.~\eqref{eq:maximum runtime}
and \eqref{eq: total runtime}, respectively.
Meanwhile, the Trotter error is related to the so-called Trotter error norm $W$~\cite{Kivlichan2020improved} as 
\begin{equation}
    W \Delta t^2 \leq \epsilon_{\text{Trotter}}, 
\end{equation}
when we assume the second-order Trotter decomposition in Eq.~\eqref{eq: Trotter circuit}.
Here $\Delta t \equiv T/N$ is the time interval in a single Trotter step.
Thus, in what follows, we will keep the time interval as $\Delta t = \sqrt{\epsilon_{\text{Trotter}}/W}$.
In conclusion, to achieve the precisions $\epsilon_{\text{Trotter}}$ and $\epsilon_{\text{QPE}}$, 
the total number of Trotter steps in the QCELS algorithm is given as
\begin{equation}
    N_{\text{total}} \equiv \frac{T_{\text{total}}}{2\Delta t} = \sum_{j=1}^{J} 2^{j-2}\delta(K-1)N_s \sqrt{\frac{W}{\epsilon_{\text{Trotter}}}},
\end{equation}
where $J=J(\epsilon_{\text{QPE}})=\lceil \log_2(1/\epsilon_{\text{QPE}}) \rceil+1$, and we leverage the fact that the controlled-$e^{-i\tau\hat{\mathcal{H}}}$ operation in the Hadamard test requires only half the number of Trotter steps needed for $e^{-i\tau\hat{\mathcal{H}}}$ (see Appendix.~\ref{Appendix: Hadamard test}).
Similarly, we determine the maximum number of Trotter steps required for a single shot of the Hadamard test, as
\begin{equation}
    N_{\text{max}} \equiv \frac{T_{\text{max}}}{2\Delta t} = \frac{\delta}{2\epsilon_{\text{QPE}}}\sqrt{\frac{W}{\epsilon_{\text{Trotter}}}}.
\end{equation}

The next step is to determine the optimal value of the code distance $d$ for the surface code. This is achieved by discussing how long the logical information should be protected from Clifford errors in the QCELS algorithm.
As discussed in detail in Appendix.~\ref{Appendix: Hadamard test}, 
we can execute the QCELS algorithm with the circuit in Fig.~\ref{fig:Trotter circuit for QCLES}, whose outcomes are equivalent to those of the Hadamard test circuit in Fig.~\ref{fig:circuit for QCELS}.
Clearly, this circuit only contains multi-Pauli rotation gates in the form of $\hat{R}_{Z \otimes P_i}(\theta_i)$, except for the final gate $\hat{W}$.
Here we define two types of time units for logical operations: (i) {\it code cycle} as the period for a single round of stabilizer measurements and (ii) {\it clock} as the period of $d$ code cycles.
As discussed in Ref.~\cite{Litinski2019}, if the resource state $\ket{m_{\theta_i}}$ is already prepared, we can execute the rotation gate $\hat{R}_{Z \otimes P_i}(\theta_i)$ at least within 9 clocks via lattice surgery techniques for the compact blocks.

\begin{figure}[b]
    \centering
    \includegraphics[width=8.7cm]{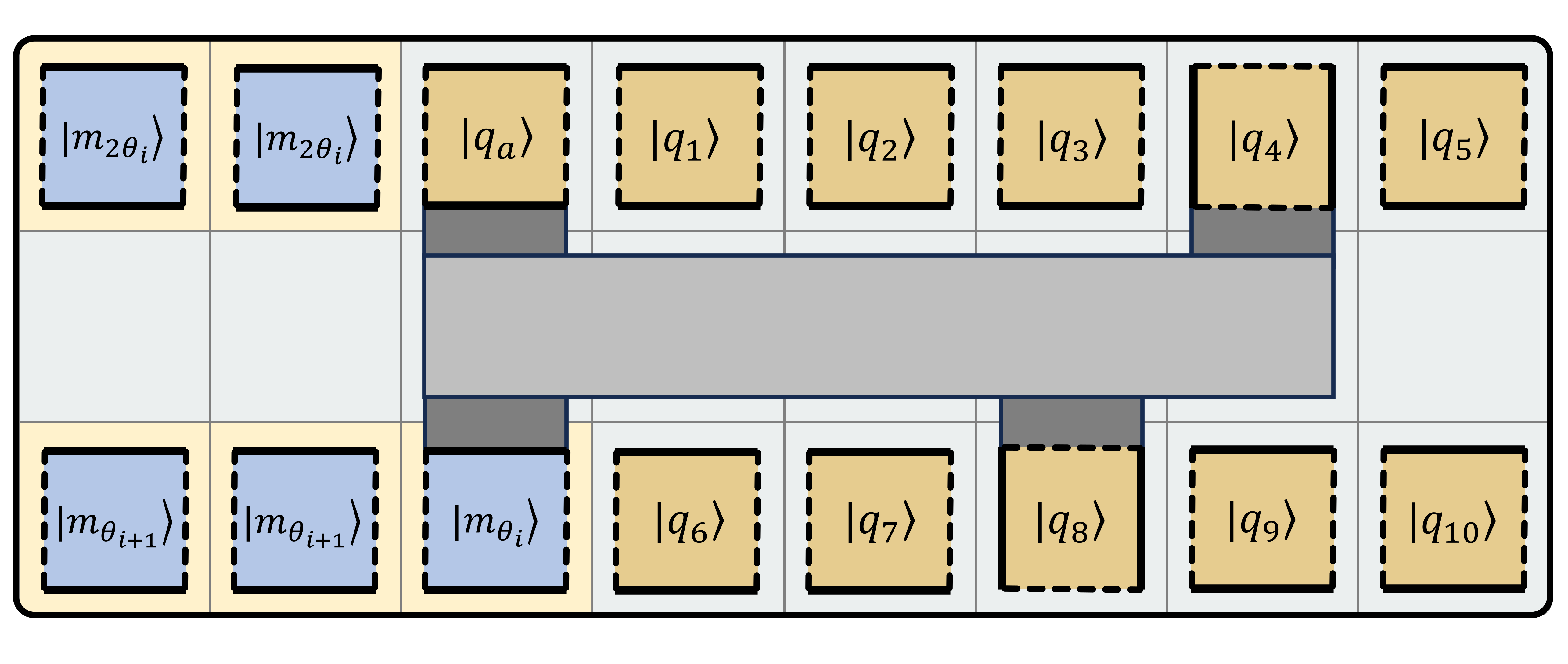}
    \caption{Patch arrangement considered to estimate the space-time cost for executing the QCELS algorithm. The light brown patches are allocated to encode logical data qubits. The ancilla region in the middle line is used for lattice surgery operations such as multi-Pauli measurements. The yellow ancilla region is used for preparing an ancilla state for executing a $\pi/4$-rotation gate or preparing resource states for executing an analog rotation gate. While executing the rotation gate $\hat{R}_{Z \otimes P_i}(\theta_i)$, we prepare resource states needed in the next step by using the remaining free patches.}
    \label{fig:footprint}
\end{figure}

In the following discussion, we assume the patch arrangement depicted in Fig.~\ref{fig:footprint}.
This arrangement is designed for the sequential execution of the circuit shown in Fig.~\ref{fig:Trotter circuit for QCLES}.
It is important to note that, in this work, we do not consider the possibility of executing multiple rotation gates in parallel to accelerate quantum computation, leaving this as an interesting future work.
In the arrangement in Fig.~\ref{fig:footprint}, we use the yellow ancilla region to prepare an ancilla state for executing an $\pi/4$ rotation gate or preparing resource states for executing an analog rotation gate. 
The number of patches assigned to the yellow region is determined to avoid the time delay due to the resource state preparation.
In the arrangement in Fig.~\ref{fig:footprint},
we can freely use four logical patches during at most 7 clocks after performing $\pi/4$ rotations during the lattice surgery procedure for $\hat{R}_{Z \otimes P_i}(\theta_i)$.
By allocating these areas to generate the two resource states $\ket{m_{\theta_{i+1}}}$ and $\ket{m_{2\theta_{i+1}}}$,
we are ready to perform the next rotation gate, 
regardless of whether the gate teleportation succeeds or fails.
Because the success rate in the resource state preparation is sufficiently high as shown in Fig.~\ref{fig:success_rate_in_clock}, 
we can neglect the latency time to prepare a resource state unless the code distance becomes too large.
For these reasons, in the following estimation of the execution time, we suppose that each rotation gate in Fig.~\ref{fig:Trotter circuit for QCLES} can be executed within the ideal number of clocks given in Ref.~\cite{Litinski2019}.

To summarize the above augments, we conclude that the QCELS algorithm requires executing the Hadamard test circuit with the depth of at most $4dLC_{\text{av}}N_{\text{max}}$ code cycles.
Here we define $C_{\text{av}}$ as the average number of clocks required for executing the rotation gate $\hat{R}_{Z \otimes P_i}(\theta_i)$ that appears in the circuit in Fig.~\ref{fig:Trotter circuit for QCLES}. The factor 4 originates from the fact that we need to repeat gate teleportation twice on average in the RUS process, and the circuit in Fig.~\ref{fig:Trotter circuit for QCLES} includes $2L$ rotation gates in a single Trotter step.
Meanwhile, using a minimum-weight perfect matching decoder under the circuit-level noise model, the logical error rate per code cycle is approximated for the  surface code as~\cite{Fowler2018}
\begin{equation}
    p_L(p_{\text{ph}},d) = 0.1\times(100p_{\text{ph}})^{(d+1)/2}.
\end{equation}
The optimal value of the code distance $d$ should be determined to satisfy $p_L(p_{\text{ph}},d)^{-1}\gg 4dLC_{\text{av}}N_{\text{max}}N_{\text{patch}}$,
so that errors in the Clifford operations are sufficiently suppressed.
Here $N_{\text{patch}}$ is the number of code patches in Fig.~\ref{fig:footprint} and is determined as $N_{\text{patch}}=3/2 \times (N_{\text{sys}}+6)$, where $N_{\text{sys}}$ is the number of logical qubits required to encode the target model.
In the above argument, we implicitly assume that idling errors on any code patches, including any ancilla patches, always contribute to logical errors.
This seems to be slightly pessimistic since most of the ancilla patch region does not work during the sequential execution of gates.
In this study, we will determine the code distance $d$ to satisfy
\begin{eqnarray}
\label{eq:code distance condition}
    p_L(p_{\text{ph}},d)^{-1}\  \geq \ 100 \times 4dLC_{\text{av}}N_{\text{max}}N_{\text{patch}}.
\end{eqnarray}
The factor $100$ is introduced to ensure that logical errors do not affect our estimate, which is more stringent than the condition assumed in a previous related work~\cite{Yoshioka2022hunting}.

Once the code distance $d$ is determined, we can estimate the space-time cost, namely the number of  physical qubits and the execution time, required for the QCELS algorithm.
For the spatial cost, it is easily estimated as $N_{\text{patch}}\times 2d^2$ physical qubits by assuming the patch alignment in Fig.~\ref{fig:footprint}. 
Meanwhile, for the time cost, we must clarify the physical time taken in a single code cycle.
Realistically, it is lower-bounded by the stabilizer measurement time and the decoding time for quantum error correction. 
On a current superconducting qubit chip~\cite{Arute2019,Google2023suppressing}, a single round of syndrome extraction takes less than 1 $\mu$s.
Meanwhile, for the decoding time, some of the latest algorithms can process syndrome data in less than 1 $\mu$s per round of syndrome extraction on a single CPU core~\cite{Higgott2023sparse} or on a low-end hardware platform such as FPGA or ASIC~\cite{Barber2023ASIC}, assuming a moderate value of the code distance.
For these reasons, we assume that a single code cycle takes 1 $\mu$s, as in previous related works~\cite{Yoshioka2022hunting,Babbush2018qubitization,Kivlichan2020improved}.
Then, the ideal execution time for our task is evaluated as $4dLC_{\text{av}}N_{\text{total}}$ $\mu$s.
However, in actual devices, we need to include the additional time overhead due to the sampling cost for error mitigation.
Following the same reasoning as in Eq.~\eqref{eq:mitigation cost in QPE}, we can estimate this multiplicative factor as $\gamma_\tau^2=e^{2 \alpha_{\text{RUS}}\lambda \tau p_{\text{ph}}}$ for the Hadamard test with time evolution $\tau$.
Accounting for these factors in Eq.~\eqref{eq: total runtime}, 
we arrive at the following formula for the total execution time of the QCELS algorithm on early-FTQC devices:
\begin{equation}
\label{eq:total execution time in physical time unit}
\begin{aligned}
    \mathcal{T}_{\text{total}} &=  4dLC_{\text{av}} \sum_{j=1}^J \sum_{n=0}^{K-1} \gamma^2_{n\tau_j} N_s n\tau_j \sqrt{\frac{W}{\epsilon_{\text{Trotter}}}} \ [\mu s]
\end{aligned}
\end{equation}


\subsection{Resource estimation: Example}

\begin{table*}[tb]
    \centering
    \begin{tabular}{|wc{2.3cm}|wc{2.3cm}|wc{1.9cm}|wc{1.9cm}|wc{1.9cm}|wc{1.9cm}|wc{1.9cm}|wc{1.9cm}|}
    \hline
        \multicolumn2{|c|}{Problem size} & \multicolumn2{|c|}{Code distance} & \multicolumn{2}{|c|}{Physical qubits} &   \multicolumn{2}{|c|}{Execution time (hours)} \\ \hline
        Lattice size & Data qubits & $p_{\text{ph}}=10^{-3}$ & $p_{\text{ph}}=10^{-4}$ &  $p_{\text{ph}}=10^{-3}$ & $p_{\text{ph}}=10^{-4}$ & $p_{\text{ph}}=10^{-3}$ & $p_{\text{ph}}=10^{-4}$\\ \hline
        $6\times 6$ & 73 & 21 & 11 & 1.03e+05 & 2.83e+04 & 1.18e+03 & 7.88e+01 \\ \hline
        $8\times 8$ & 129 & 21  & 11  & 1.77e+05 & 4.86e+04 & 2.74e+04 & 2.15e+02\\ \hline
        $10\times 10$ & 201 & 23 & 11 & 3.27e+05 & 7.48e+04 & 1.40e+06 & 5.08e+02\\ \hline
    \end{tabular}
    \caption{Space-time cost for estimating the ground-state energy
of the 2D Hubbard model with a single QPU. Here we set the target precision and the Hamiltonian parameters as $\epsilon=0.01$ and $U/t=4$, respectively. The column ``Data qubits" denotes the number of logical qubits required for executing the circuit in Fig.~\ref{fig:Trotter circuit for QCLES}. The column ``Physical qubits" denotes the total number of physical qubits required for our task, including all components to implement the floor plan in Fig.~\ref{fig:footprint}. The execution time includes the sampling overhead for error mitigation.}
    \label{tab:resource estimation}
\end{table*}

\begin{table}[tb]
    \centering
    \begin{tabular}{|wc{2.1cm}|wc{2.1cm}|wc{1.8cm}|wc{1.8cm}|}
    \hline
    \multicolumn2{|c|}{Problem size} & \multicolumn2{|c|}{Execution time (sec)}\\
    \hline
    Lattice size & Data qubits & $p_{\text{ph}}=10^{-3}$ & $p_{\text{ph}}=10^{-4}$ \\
    \hline
    $6\times 6$ & 73 & 5.72e+02 & 3.00e+02 \\
    \hline
    $8\times 8$ & 129 & 1.36e+03 & 7.10e+02\\
    \hline
    $10\times 10$ & 201 & 2.88e+03 & 1.38e+03 \\
    \hline
    \end{tabular}
    \caption{The minimum execution time that can be achieved for ground state energy estimation of the 2D Hubbard model assuming a fully parallel computation with a large number of QPUs. Please note that, unlike Table.~\ref{tab:resource estimation}, the unit of time is given in seconds.}
    \label{tab:resource estimation with parallel QPUs}
\end{table}

As a simple demonstration, we present a concrete value of the execution time and the number of qubits that is required for estimating the ground state energy of the 2D Hubbard model.
The 2D Hubbard model~\cite{Hubbard1964} is one of the most familiar models in condensed matter physics, which captures the physics of strongly correlated electron systems.
Despite its simplicity, this model exhibits amazingly rich phases, such as anti-ferromagnetism and the Mott insulator, and it is also regarded as a simplified model for high-temperature superconductors~\cite{Arovas2022hubbard}.
In the context of quantum computation, it often serves as a benchmark for quantum algorithms and their resource estimation in solid-state physics~\cite{Yoshioka2022hunting,Babbush2018qubitization,Kivlichan2020improved}.

It is important to note that, while the following analysis focuses on the Hubbard model for simplicity, similar analyses can be directly applied to more general systems, such as extended Hubbard models derived via the ab-initio down-folding method~\cite{Kanno2022,Ivanov2023,Clinton2024,Yoshida2024} and electronic structure problems in quantum chemistry~\cite{McArdle2020}. 
Notably, unlike in Ref.~\cite{Kivlichan2020improved}, our framework does not rely on techniques like the Hamming weight phasing, and therefore, performs well even for inhomogeneous systems such as quantum embedding models~\cite{Bauer2016, Rubin2016, Yamazaki2018, Ma2020, Cao2023ab-initio}.

Let us now move on to detail the resource analysis for the 2D Hubbard model.
First, using the Jordan-Wigner transformation~\cite{Jordan1928}, we can represent the 2D Hubbard model in the form of a linear combination of Pauli string operators as follows:
\begin{equation}
\label{eq:Spin Hamitonian}
\begin{aligned}
      \hat{\mathcal{H}} &= -\frac{t}{2} \sum_{\langle i,j \rangle,\sigma} (\hat{X}_{i,\sigma}\hat{Z}^{\leftrightarrow}_{i,j,\sigma} \hat{X}_{j,\sigma} + \hat{Y}_{i,\sigma} \hat{Z}^{\leftrightarrow}_{i,j,\sigma}\hat{Y}_{j,\sigma})\\
      &\qquad\qquad\qquad\qquad+ \frac{U}{4} \sum_{i} \hat{Z}_{i,\uparrow}\hat{Z}_{i,\downarrow},  
\end{aligned}
\end{equation}
where $\hat{Z}^{\leftrightarrow}_{i,j,\sigma} = \prod_{k = i+1}^{j-1} \hat{Z}_{k,\sigma}$
is the so-called Jordan-Wigner string, which is needed to preserve the appropriate commutation relations between fermionic creation and
annihilation operators.
In the following analyses, we consider the case of the periodic boundary condition, and set the parameters $t$ and $U$ as $t = 1, U = 4$, which are the same as those in previous works~\cite{Yoshioka2022hunting,Kivlichan2020improved}.
Then, the number of terms and the 1-norm of the Hamiltonian are readily calculated as $L=9N_{\text{site}}=\frac92N_{\text{sys}}$ and $\lambda = (4t+U/4)N_{\text{site}}=5N_{\text{site}}=\frac52 N_{\text{sys}}$, respectively. Here, $N_{\text{site}}$ and $N_{\text{sys}}$ are the number of sites and logical qubits required to encode the model, respectively.

Next, let us consider the number of clocks required to execute multi-Pauli rotation gates related to the Pauli strings in Eq.~\eqref{eq:Spin Hamitonian}. For example, the multi-Pauli rotation for  $\hat{Z}_{i,\uparrow}\hat{Z}_{i,\downarrow}$ can be executed in 1 clock with standard lattice surgery techniques~\cite{Litinski2019}. Similarly, the term of $\hat{X}_{i,\sigma}\hat{Z}^{\leftrightarrow}_{i,j,\sigma} \hat{X}_{j,\sigma}$ and $\hat{Y}_{i,\sigma} \hat{Z}^{\leftrightarrow}_{i,j,\sigma}\hat{Y}_{j,\sigma}$ takes 4 and 6 clocks, respectively, supposing that we allocate the up (down)-spin orbitals to the top (bottom) line in Fig.~\ref{fig:footprint}.
Then, we evaluate the average clock number $C_{\text{av}}$ as $C_{\text{av}} = 41/9 \simeq 4.6$.

At last, by determining the minimum code distance to satisfy Eq.~\eqref{eq:code distance condition}, we evaluate the total execution time in Eq.~\eqref{eq:total execution time in physical time unit}. In Table.~\ref{tab:resource estimation}, we show the space-time cost of ground state energy estimation for the 2D Hubbard model with sites from $6\times 6$ to $10\times 10$.
For this estimate, we use the values of Trotter norm provided in Ref.~\cite{Kivlichan2020improved} and set the target accuracy as $\epsilon=0.01$.
The two parameters $\epsilon_{\text{Trotter}}$ and $\epsilon_{\text{QPE}}$ are determined to minimize the total execution time while satisfying Eq.~\eqref{eq:constraint for precision}.
In particular, our results show that the STAR architecture can complete the task for the $(8\times 8)$-sites Hubbard model with less than $4.9\times 10^4$ qubits and an execution time of $2.15\times 10^2$ hours $\simeq 9$ days under $p_{\text{ph}}=10^{-4}$. 
The spatial cost is tens of thousands of physical qubits less than in previous FTQC studies~\cite{Babbush2018qubitization, Yoshioka2022hunting, Kivlichan2020improved}.
In particular, compared to the qubitization approach on full-fledged FTQC~\cite{Babbush2018qubitization,Yoshioka2022hunting}, our framework reduces the number of required physical qubits by roughly one-third.
Such a reduction in the number of qubits is desirable for early-FTQC devices.

Furthermore, the execution time is significantly shorter than the recent estimate on a classical computer with tensor network methods (DMRG and PEPS)~\cite{Yoshioka2022hunting}, which predicts the execution time of $1.5\times 10^9$ seconds $\simeq$ $47.6$ years for the same task.
It is also comparable to the estimates obtained in Ref.~\cite{Kivlichan2020improved}, where the authors evaluated the execution time of the Trotter-based QPE based on a conventional FTQC framework.
However, compared to qubitization-based QPE on FTQC devices~\cite{Yoshioka2022hunting}, our execution time is several orders of magnitude slower.
This is because the qubitization-based QPE is considerably more efficient than the Trotter-based QPE at the algorithmic level.
We expect that, in the future, these gaps could be filled to some extent by formulating an optimal compilation to execute multiple analog rotation gates in parallel.

Here it is noteworthy that our approach using the QCELS algorithm is readily parallelizable across multiple, independent quantum processing units (QPUs).
This is because the QCELS algorithm relies on the repeated and independent execution of a series of shallow Hadamard test circuits.
Therefore, if we can use two QPUs, we can reduce the execution time for our task by half from the values shown in Table.~\ref{tab:resource estimation}.
Given the current state of experimental technologies, it is expected that fabricating multiple independent smaller quantum chips will be less challenging than constructing a single large-scale quantum chip. Therefore, the capability of parallel computation as described above presents a significant advantage in the early-FTQC era.
By leveraging this parallelism, we can close the gap in execution time between our approach and the qubitization method.
In Table.~\ref{tab:resource estimation with parallel QPUs}, we show the minimum execution time achievable when assuming the ideal situation where an arbitrary number of QPUs are available to perform parallel computations for QCELS. In this case, the execution time is evaluated as $4dLC_{\text{av}}N_{\text{max}}$ $\mu$s.





\section{Conclusion}
\label{sec:conclusion}

In this work, we have proposed a novel framework for partially fault-tolerant quantum computing toward a practical quantum advantage in the early-FTQC era.
Our framework is essentially based on the STAR architecture, and we achieved four remarkable reconstructions from its original proposal~\cite{Akahoshi2023}.
First is the proposal of the transversal multi-rotation protocol, which enables the preparation of a resource state for implementing the $R_z(\theta)$ gate with a notably small worst-case error rate of $\order{|\theta| p_\text{ph}}$. This upgrade is essentially important  to enable the STAR architecture to perform practical quantum tasks such as the Trotter simulation. 
Second is the improvement in the success rate of resource state preparation by introducing an optimal post-selection strategy. 
Optimization of post-selection regime and employment of multi-Pauli rotation gate enhanced the success rate to over several orders of magnitude.
Thirdly, we developed error mitigation strategies specialized for our resource state preparation protocol. We proved that we can keep error accumulation as small as possible by appropriately combining the coherent error cancellation method and switching preparation protocols.
Finally, we discussed the adverse effects of control errors on prepared resource states and proposed novel randomized methods to suppress the relative errors in logical rotation angles. These aspects are crucial to the STAR architecture, which were not addressed in the original work~\cite{Akahoshi2023}.

Furthermore, we presented several promising scenarios that demonstrate the potential of our framework.
These include near-term applications such as VQAs and QSCI as well as long-term applications based on the Trotter circuit. 
In particular, as an important example, we illustrated a detailed resource analysis for the ground energy estimation for the Hubbard model. 
As a result, we showed that it is possible to perform QPE for the $(8\times 8)$-sites Hubbard model with less than $4.9\times 10^4$ qubits and the execution time of 9 days under $p_{\text{ph}}=10^{-4}$ with our framework. 
This is significantly faster than the runtime required for tensor network calculations on a classical computer~\cite{Yoshioka2022hunting} and saves more qubits than previous FTQC studies~\cite{Babbush2018qubitization, Yoshioka2022hunting, Kivlichan2020improved}.

Finally, we summarize the important issues that 
have not been addressed in this study. 
These could be an interesting future directions
for our proposal. (i) The logical-level compilation of quantum circuits to leverage the locality and parallelism of our injection protocol to its fullest. 
As already mentioned, in our framework, it is possible to execute multiple rotation gates in parallel without  
allocating extra ancillary patch regions to prepare resource states. This will be very beneficial for achieving further acceleration of quantum computation.
This issue will be addressed in Ref.~\cite{Akahoshi2024}.
(ii) Efficient error tomography of prepared resource states. In the proposed error mitigation strategies, we utilized PEC techniques to mitigate stochastic errors. To realize this, we need to develop a way to estimate the error rate of prepared resource states efficiently and accurately.
Alternatively, it might be another promising approach to explore noise-resilience at the algorithm-level as in Ref.~\cite{Ding2023robust}.
(iii) The overhead of initial state preparation in the QCELS algorithm. In our analyses in Sec.~\ref{sec:resource estimation}, we neglected the computational cost for initial state preparation for simplicity. 
To justify our estimate, we must prove that such a cost is much smaller than, or at least comparable to, that for the QCELS algorithm itself. 
(iv) More concrete discussion on the application of near-term algorithms such as VQA and other modern approaches. Although several promising scenarios have been proposed in this paper, their utility will need to be verified by more detailed resource estimates in the future.

We hope that our proposals will open up a new avenue for achieving practical quantum speedups in the near future, and stimulate
further research aimed at the practical application of early-FTQC devices.

\section{Acknowledgement}

We are grateful to thank Mitsuki Katsuda, Kishi Kaito, Koki Chinzei, Quoc Hoan Tran, Shota Kanasugi, and Norifumi Matsumoto for fruitful discussions. K.F. is supported by MEXT Quantum Leap Flagship Program (MEXT Q-LEAP)
Grant No. JPMXS0120319794, JST COI-NEXT Grant No. JPMJPF2014, and JST
Moonshot R\&D Grant No. JPMJMS2061.

\appendix

\section{Notations in this paper}
\label{Appendix:notations}

In Table.~\ref{tab:notations}, we list the notations frequently used in this paper.

{\renewcommand{\arraystretch}{1.5}
\begin{table*}
    \centering
    \caption{Our notations used in this paper.}
    \begin{tabular}{p{2.5cm}p{14cm}}
    \hline \hline
    Notation\hspace{5mm}  & Meaning\\
    \hline
    $p_{\text{ph}}$ & Physical error rate of the native gate set.\\
    $\ket{m_\theta}_L$ & Logical resource state with rotation angle $\theta$ (Eq.~\eqref{eq:resource_state}). We omit the subscript ``$L$" from Sec.~\ref{sec:Stochastic error mitigation} onwards for simplicity.\\
    $\hat{P}$ & A Pauli string operator, i.e., direct product of Pauli or identity operators on $n$-qubits.\\
    $\hat{R}_P(\theta)$ & A multi-Pauli rotation gate with angle $\theta$ which we define as $\hat{R}_P(\theta)=e^{i\theta \hat{P}}$.\\
    $\hat{R}_{z,i}(\theta)$ & A physical Pauli-Z rotation gate with angle $\theta$ on the $i$-th physical qubit.\\
    $\hat{R}_{z,L}(\theta)$ & A logical Pauli-Z rotation gate with angle $\theta$. We omit the subscript ``$L$" from Sec.~\ref{sec:Stochastic error mitigation} onwards for simplicity. \\
    $d$ & Code distance of a single surface code patch.\\
    $m$ & Weight of multi-Pauli rotation gates that form the transversal multi-Pauli rotation gate.\\
    $k$ & Parameter that counts the number of multi-Pauli rotation gates used in the transversal rotation gate for the transversal multi-rotation protocol.\\
    $\theta$ & Input physical rotation angle in the transversal multi-rotation protocol.\\
    $\theta_*$ & Target logical rotation angle in the transversal multi-rotation protocol (Eq.~\eqref{eq:angle relation}).\\
    $\theta_{\text{error}}$ & Logical rotation angle that the error state has in the transversal multi-rotation protocol (Eq.~\eqref{eq:error angle}). \\
    $\Delta_{\theta_*}$ & Over-rotation angle $\theta_{\text{error}}-\theta_*$ that arises when the error state is realized in the transversal multi-rotation protocol.\\
    $\theta_{\text{total}}$ & Total analog rotation angle required to execute a specific quantum algorithm (Eq.~\eqref{eq: definition of total error and total rotation}).\\
    $p_{\text{suc}}$ & Success rate of the transversal multi-rotation protocol.\\
    $p_{\text{ideal}}$ & Success rate of the transversal multi-rotation protocol in the ideal limit ($p_{\text{ph}}\to 0$) (Eq.~\eqref{eq:ideal success rate}).\\
    $P_{\text{ud}}$ & Total error probability that undetectable errors occur in the transversal multi-rotation protocol.\\
    $P_L$ & Logical error rate of the analog rotation gate prepared using our state preparation protocol when the RUS process succeeds in the first trial.\\
    $\tilde{P}_L$ & Effective logical error rate of the analog rotation gate prepared using our state preparation protocol after averaging over any possible RUS processes (Eq.~\eqref{eq:final error rate}).\\
    $P_{\text{total}}$ & Total error rate that accumulates throughout the overall circuit (Eq.~\eqref{eq: definition of total error and total rotation}).\\
    $\alpha_{\text{RUS}}$ & Factor that represents the error accumulation in the RUS process (Eq.~\eqref{eq:alpha_RUS})\\
    $\mathcal{R}_\theta(\hat{\rho})$ & Quantum channel that corresponds to the ideal logical rotation gate $\hat{R}_{z,L}(\theta)$ (Eq.~\eqref{eq:ideal rotation channel}).\\
    $\mathcal{N}_{\theta_*}(\hat{\rho})$ & Noisy analog rotation channel implemented via the transversal multi-rotation protocol when the RUS process succeeds in the first trial (Eq.~\eqref{eq:noisy rotation channel in the first trial}). This channel can be described with an effective error channel $\mathcal{E}_{\theta_*}(\hat{\rho})$.\\
    $\mathcal{N}_{\theta_*}^K(\hat{\rho})$ & Noisy analog rotation channel implemented via the transversal multi-rotation protocol when the RUS process succeeds in the $K$-th trial (Eq.~\eqref{eq:Kth trial channel}). This channel can be described with an effective error channel $\mathcal{E}_{\theta_*}^K(\hat{\rho})$.\\
    $\tilde{\mathcal{N}}_{\theta_*}(\hat{\rho})$ & Noisy analog rotation channel obtained by averaging the channel $\mathcal{N}_{\theta_*}^K$ over any possible $K$ (Eq.~\eqref{eq:Rotation channel after RUS}). This channel can be described with an effective error channel $\tilde{\mathcal{E}}_{\theta_*}(\hat{\rho})$.\\
    $\varepsilon_{\text{av}}(\mathcal{E})$ & Average error rate of some error channel $\mathcal{E}$. \\
    $\varepsilon_{\diamond}(\mathcal{E})$ & Worst-case error rate of some error channel $\mathcal{E}$.\\
    $\hat{\mathcal{H}}=\sum_{i=1}^L a_i\hat{P}_i$ & Hamiltonian treated in the Trotter simulation or QPE (Eq.~\eqref{eq:Hamiltonian}).\\
     $\psi_0$ & Input state used for the QPE algorithm.\\
     $\eta$ & Overlap between $\psi_0$ and the exact ground state.\\
    \hline \hline
    \end{tabular}
    \label{tab:notations}
\end{table*}
}

\section{State preparation protocol proposed in the original STAR architecture}
\label{Appendix:Akahoshi injection}

\begin{figure*}
    \centering
    \includegraphics[width = \linewidth]{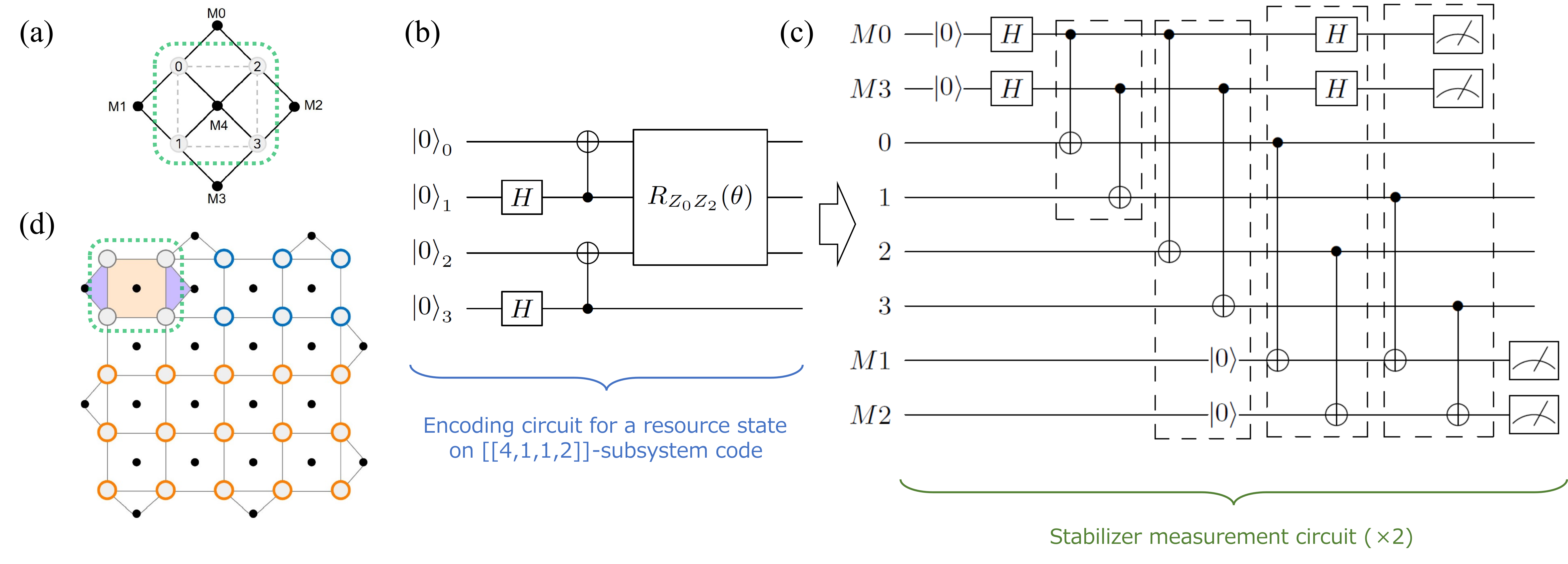}
    \caption{Resource state preparation protocol proposed in Ref.~\cite{Akahoshi2023}. {\bf (a)} Labeling of the physical qubits employed to encode the [[4,1,1,2]]-subsystem stabilizer code. {\bf (b)} In this protocol, we directly encode a resource state on [[4,1,1,2]]-subsystem stabilizer code with a physical $ZZ$-rotation gate $R_{Z_0Z_2}(\theta)$. {\bf (c)} Then, by performing the syndrome measurement twice, we discard noisy states if the measurement outcomes contain unexpected values. {\bf (d)} If the post-selection process completes successfully, we expand the logical patch to the one with any code distance via the standard lattice surgery technique. Here, blue and orange circles denote physical qubits initialized to the $\ket{0}$ and $\ket{+}$ state, respectively.}
    \label{fig:Akahoshi_injection}
\end{figure*}

Here we briefly explain the resource state preparation protocol proposed in Ref.~\cite{Akahoshi2023}. 
In the protocol, the authors employed the $[[4, 1, 1, 2]]$-quantum subsystem code~\cite{Bacon2006}, which is defined with two stabilizer operators
\begin{equation} \label{eq:422stabs}
  \hat{S}_X = \hat{X}_0 \hat{X}_1 \hat{X}_2 \hat{X}_3, \quad \hat{S}_Z = \hat{Z}_0 \hat{Z}_1 \hat{Z}_2 \hat{Z}_3,
\end{equation}
two gauge operators 
\begin{equation} \label{eq:422logops_2}
  \hat{G}_X = \hat{X}_0 \hat{X}_2, \quad \hat{G}_Z = \hat{Z}_0 \hat{Z}_1, 
\end{equation}
and single-qubit logical Pauli operators
\begin{equation} \label{eq:422logops_1}
  \hat{L}_X = \hat{X}_0 \hat{X}_1, \quad \hat{L}_Z = \hat{Z}_0 \hat{Z}_2.
\end{equation}
Following the three steps below, we achieve the non-fault-tolerant preparation of a resource state $\ket{m_\theta}_L$ with reasonably high fidelity:

\begin{enumerate}
    \item {\bf Non-fault-tolerant encoding}: Directly implement a resource state encoded on the $[[4, 1, 1, 2]]$-quantum subsystem code via the non-fault-tolerant encoding circuit in Fig.~\ref{fig:Akahoshi_injection}(b).
    \item {\bf First post-selection in state verification}: Perform a syndrome measurement of the $[[4, 1, 1, 2]]$-quantum subsystem code twice, according to the circuit in Fig.~\ref{fig:Akahoshi_injection}(c). 
    Then, if an error syndrome is detected, reject the output state  and restart the protocol from the step 1.
    \item {\bf Second post-selection in patch expansion}: Extend the [[4, 1, 1, 2]]-subsystem code state, which is equivalent to the surface code with the code distance $d=2$ under a specific gauge fixing, to a desired code distance $d=d_*$ by utilizing the standard lattice surgery technique~\cite{Horsman2012,Litinski2019} (see also Fig.~\ref{fig:Akahoshi_injection}(d)). In this process, we perform stabilizer measurements twice on the code patch with the code distance $d_*$, and reject the output state according to the rule analogous to step 2.
\end{enumerate}
More specifically, in the post-selection process in steps 2 and 3, 
we discard the output state if the measurement outcomes differ from the ones expected in the ideal limit ($p_{\text{ph}}=0$).
This enables us to remove any detectable errors from the output state, thereby increasing the state fidelity.
In step 1, the authors of Ref.~\cite{Akahoshi2023} assumed to execute $\hat{R}_{Z_0Z_2}(\theta)$ gate with (SWAP gates and) an analog two-qubit rotation gate, which could be implemented via a native gate such as the cross-resonance gate~\cite{Rigetti2010,Chow2011}.
As mentioned in the main text, these procedures efficiently reduce the error rate of the prepared resource state to $P_L^{\text{org}}=\frac{2}{15}p_{\text{ph}} + \mathcal{O} (p^2_{\text{ph}})$ under a circuit-level noise model.

\section{Details of numerical simulation for resource state preparation}
\label{appendix:numerical simulation}

In this section, we provide the details of our numerical simulation for the resource state preparation protocol proposed in the main text.

\subsection{Definition of circuit-level noise model}
\label{sec:circuit-level noise model}

First we specify the definitions of our circuit-level noise model used in the Clifford circuit simulation in this paper.
In our model, we assume that we can directly implement qubit initialization and measurement in $Z$-basis, and the gate set $\{H, \text{CNOT}, \text{SWAP}\}$ as native operations.
In addition, we assume that each of these operations suffers from 
the following type of physical errors:
\begin{itemize}
    \item {\bf Qubit initialization and measurement in $Z$-basis}: Bit-flip error with probability $p_{\text{ph}}$,
    \begin{equation}
        \mathcal{E}_{\text{flip}}(\hat{\rho}) = (1-p_{\text{ph}})\hat{\rho} + p_{\text{ph}}\hat{X}\hat{\rho} \hat{X}. 
    \end{equation}
    \item {\bf One-qubit gates}: Single-qubit depolarizing error with probability $p_{\text{ph}}$,
    \begin{equation}
    \label{eq:single depolarizing error}
        \mathcal{E}_{\text{dep,1}}(\hat{\rho}) = (1-p_{\text{ph}})\hat{\rho} + \frac{p_{\text{ph}}}{3}(\hat{X}\hat{\rho} \hat{X} + \hat{Y}\hat{\rho} \hat{Y} + \hat{Z}\hat{\rho} \hat{Z}).
    \end{equation}
    \item {\bf Two-qubit gates}: Two-qubit depolarizing error with probability $p_{\text{ph}}$,
    \begin{equation}
    \label{eq:double depolarizing error}
        \mathcal{E}_{\text{dep,2}}(\rho) = (1-p_{\text{ph}})\rho + \frac{p_{\text{ph}}}{15}\sum_{\hat{E}\in \{\hat{I},\hat{X},\hat{Y},\hat{Z}\}^{\otimes 2}\backslash \{\hat{I}\hat{I}\}} \hat{E}\hat{\rho} \hat{E}.
    \end{equation}
\end{itemize}
Furthermore, as discussed in Sec.~\ref{sec:formulation of injection protocol}, we often consider the case where we can implement a physical $ZZ$-rotation gate $\hat{R}_{zz}(\theta)$ directly with a noise channel $\mathcal{E}_{\text{dep,2}}$, or the case where we can implement a physical $Z$-rotation gate $\hat{R}_{z}(\theta)$ gate with an ignorable noise via the virtual-$Z$ scheme~\cite{Mckay2017}.
The former assumption is valid for typical trapped ion devices and superconducting devices, since the $XX$-rotation gate and
the $ZX$-rotation gate (the cross-resonance gate) can be
directly implemented, respectively.

\subsection{Probabilistic sampling of syndrome subspaces of non-Clifford state}

Next we outline our simulation methodology for resource state
preparation (for details see Ref.~\cite{Choi2023}). 
In the usual simulation of quantum error correction, we can efficiently track the change in quantum states, because the circuit comprises only the Clifford operations and the measurement or qubit initialisation into the computational basis.
Meanwhile, in the simulation of resource state preparation discussed in Sec.~\ref{sec:preparation protocol}, we have to deal with non-Clifford operations, namely analog rotation gates, which generally make the circuit simulations challenging.
In general, such a non-Clifford circuit should be simulated within the full state vector representation. However, in our case, some important features such as infidelity and success rate can be simulated more efficiently by dealing with analog rotating gates as a kind of stochastic process.

To understand this, let us reconsider the action of the transversal rotation gate (Eq.~\eqref{eq:Transversal_rotation}) on the Clifford state $\ket{+}_L$. By introducing the notation 
\begin{equation}
    \hat{Z}^b \equiv \prod_{i:b_i=1} \hat{Z}_i
\end{equation}
with a bit-string $b= b_1b_2\cdots b_d\in \{0,1\}^d$, we can rewrite the transversal rotation gate in Eq.~\eqref{eq:Transversal_rotation} as 
\begin{equation}
    \prod_{i\in Q_z} \hat{R}_{z,i}(\theta) = \sum_{b=0}^{2^d} u_{|b|}  \hat{Z}^b,
\end{equation}
where $u_{n} \equiv i^n\sin^{n}\theta \cos^{d-n}\theta$, and we label each qubit in the qubit set $Q_z$ by $\{1,2,\cdots,d\}$. Here $|b|$ denotes the Hamming weight of the bit-string $b$.
Applying this operator to $\ket{+}_L$, we have
\begin{equation}
\label{eq:bit-wise representation of transversal gate}
\begin{aligned}
    \prod_{i\in Q_z} \hat{R}_{z,i}(\theta)\ket{+}_L &= \sum_{b=0}^{2^d} u_{|b|}  \hat{Z}^b \ket{+}_L\\
    & = \sum_{b=0}^{2^{d-1}}\ket{\psi_b} ,
\end{aligned}
\end{equation}
where we introduce $\ket{\psi_b}\equiv(u_{|b|}  \hat{Z}^b + u_{|\bar{b}|}\hat{Z}^{\bar{b}}) \ket{+}_L$ and $\bar{b}$ denotes bit-wise
negation of the bit-string $b$.
Here note that the quantum state $\ket{\psi_b}$ belongs to a single syndrome subspace that has a unique set of eigenvalues of the stabilizers. This is readily understood from the fact that two quantum states $\hat{Z}^b \ket{+}_L$ and $u_{|\bar{b}|}\hat{Z}^{\bar{b}} \ket{+}_L$ differ from each other by the logical operation $\hat{Z}^b\cdot \hat{Z}^{\bar{b}} = \hat{Z}_L$. 
Meanwhile, for any different bit-strings $b,b'\in \{0,1\}^{2^{d-1}}$, the corresponding states $\ket{\psi_b}$, $\ket{\psi_{\bar{b}}}$  belong to different syndrome subspaces from each other, since the stabilizer code with code distance $d$ can detect the difference by any Pauli string operator with a weight less than $d$.
These observations suggests that, by applying syndrome measurements to the state in Eq.~\eqref{eq:bit-wise representation of transversal gate}, we obtain the quantum state $\ket{\psi_b}$ with probability $|u_{|b|}|^2+|u_{|\bar{b}|}|^2$ for each bit-string $b\in \{0,1\}^{2^{d-1}}$. 

Next let us consider how a Pauli string error $\hat{E}$ modifies the above analyses.
Such an error changes the quantum state in Eq.~\eqref{eq:bit-wise representation of transversal gate} to
\begin{equation}
\label{eq:transversal gate with error}
    \hat{E}\prod_{i\in Q_z} \hat{R}_{z,i}(\theta)\ket{+}_L = \sum_{b=0}^{2^{d-1}}\hat{E}\ket{\psi_b}.
\end{equation}
The new state $\hat{E}\ket{\psi_b}$ always belongs to a different syndrome subspace from that of the original state $\ket{\psi_b}$ as long as the weight of $\hat{E}$ is less than $d$. Importantly, the set of quantum states $\{\hat{E}\ket{\psi_b}\}$ preserves the orthogonality of $\{\ket{\psi_b}\}$, and each state is sampled with probability $|u_{|b|}|^2+|u_{|\bar{b}|}|^2$ when applying  syndrome measurements.
This argument suggests that the occurrence of errors and the sampling of bit-string $b$ via the syndrome measurement are independent as stochastic processes.

Accordingly, instead of performing the full state vector simulation for Eq.~\eqref{eq:transversal gate with error}, we can adopt the following procedures for estimating the statistic of the outcomes of syndrome measurements:
\begin{enumerate}
    \item Sample Pauli string errors $\hat{E}$ related to the preparation circuit for the initial state $\ket{+}_L$: $\ket{+}_L\to \hat{E}\ket{+}_L$.
    \item Sample a single bit-string $b\in \{0,1\}^{2^{d-1}}$ with probability $|u_{|b|}|^2+|u_{|\bar{b}|}|^2$, and virtually construct a quantum state $(u_{|b|}  \hat{Z}^b + u_{|\bar{b}|}\hat{Z}^{\bar{b}})\cdot\hat{E}\ket{+}_L$.
    \item Sample Pauli string errors $\hat{E}'$ related to the transversal rotation gate and the syndrome measurement circuit and determine the measurement outcome.
\end{enumerate}
The first and third procedure can be executed in a straightforward manner via usual Clifford circuit simulation under the circuit-level noise model. 
On the other hand, the second procedure is slightly non-trivial.
On a Clifford circuit, we can never directly construct the non-Clifford state $(u_b  \hat{Z}^b + u_{\bar{b}}\hat{Z}^{\bar{b}})\cdot\hat{E}\ket{+}_L$. Therefore, we instead simulate the state $\hat{Z}^b\hat{E}\ket{+}_L$, which belongs to the same syndrome subspace and is easily simulated on an usual Clifford circuit.
This modification does not change the statistic of the outcomes of syndrome measurements.

These properties remain unchanged as long as we assume a simple Pauli error model like the depolarizing error in Eqs.~\eqref{eq:single depolarizing error} and \eqref{eq:double depolarizing error}.
Furthermore, the above analyses can be extended to the case where we use the transversal multi-rotation gate in Eq.~\eqref{eq:transversal multi rotation} instead of Eq.~\eqref{eq:Transversal_rotation}.

\subsection{Numerical estimation of fidelity}

Eq.~\eqref{eq:final output state} in the main text is easily extended to the form that includes the corrections due to higher-order errors as follows:
\begin{equation}
\label{eq: general form of prepared states}
\begin{aligned}
 \rho_{\text{out}}\simeq \  &\frac{1}{p_{\text{suc}}} \sum_{n=0}^{k-1} q_n\cdot \ket{m_{\theta_n}}\bra{m_{\theta_n}},
\end{aligned}
\end{equation}
where $\ket{m_{\theta_n}}$ is a post-selected non-Clifford state that arises due to errors of order $\order*{p_{\text{ph}}^n}$, and  its angle is determined as 
\begin{equation}
\label{eq: general fom of angle}
    \theta_n(\theta,k)  \  \equiv\   \sin^{-1}\left(
    \frac{u_{k-n}/i}{\sqrt{|u_{k}|^2+|u_{k-n}|^2}} \right),
\end{equation}
where we redefine $u_{n} \equiv i^n\sin^{n}\theta \cos^{d-n}\theta$.
Here $\theta_0$ corresponds to the target value $\theta_*$.
The coefficient $q_n$ ($\sim p_{\text{ph}}^n$) represents the probability where non-Clifford state $\ket{m_{\theta_n}}$ is post-selected via stabilizer measurement, and $p_{\text{suc}}=\sum_n q_n$ is the total success rate of our state preparation protocol.
In more details, $q_n$ can be decomposed into the contributions of all bit-strings that satisfy $|b|=n$  as follows:
\begin{equation}
    q_n=\sum_{b:|b|=n}(|u_{|b|}|^2+|u_{|\bar{b}|}|^2) q_b^{\text{pass}}
    =q_n^{\text{sample}}\cdot q_n^{\text{pass}},
\end{equation}
where we introduced

\begin{equation}
\begin{aligned}
    q_n^{\text{sample}}&\equiv  {}_kC_n(|u_{k}|^2+|u_{k-n}|^2),\\
    q_n^{\text{pass}}&\equiv \frac{1}{{}_kC_n} \sum_{b:|b|=n} q_b^{\text{pass}},
\end{aligned}
\end{equation}
in the last equation.
Here, $q_n^{\text{sample}}$ denotes the total probability of sampling a quantum state with $|b|=n$ from Eq.~\eqref{eq:bit-wise representation of transversal gate}.
In addition, $q_b^{\text{pass}}$ ($\sim p_{\text{ph}}^n$) denotes the probability where the sampled state $\ket{\psi_b}$ passes the post-selection process, and $q_n^{\text{pass}}$ averages it over all bit-strings satisfying $|b|=n$.
For example, $q_0^{\text{sample}}$, $q_1^{\text{sample}}$, $q_0^{\text{pass}}$, and $q_1^{\text{pass}}$  correspond to the parameters $p_{\text{ideal}}$, $kp_{\text{error}}$, $1-Q$, and $P_{\text{ud}}/k$ in the main text, respectively. 

Next let us discuss how to evaluate the infidelity of the prepared state.
As in Eq.~\eqref{eq:state fidelity of multi-rotation protocol}, 
we can calculate the infidelity of the state in Eq.~\eqref{eq: general form of prepared states} as
\begin{equation}
\label{eq: general form of infidelity}
\begin{aligned}
    1-F &\equiv1-\ev{\rho_{\text{out}}}{m_{\theta_*}}\\
    &=  \frac{1}{p_{\text{suc}}}\sum_{n=0}^{k-1} q_n\left(1-|\braket{m_{\theta_*}}{m_{\theta_n}}|^2\right)\\
    & = \frac{1}{p_{\text{suc}}}
    \sum_{n=0}^{k-1} q_n \sin^2(\theta_n-\theta_*).
\end{aligned}
\end{equation}
In our numerical simulation, we sample a bit-string $b\in \{0,1\}^{2^{d-1}}$ and effectively perform the quantum circuit for the transversal multi-rotation protocol. 
Then, if the measurement outcomes satisfy our criteria of post-selection, we record the value of $F_n\equiv \sin^2(\theta_n-\theta_*)$.
We repeat this procedure $N_{\text{shot}}$ times and average the sampled infidelities over all trials that pass the post selection. Here we assume that bit-strings that satisfy $|b|=n$ are sampled  $N_{n,\text{sample}}$ times out of $N_{\text{shot}}$ trials, and then, the non-Clifford state $\ket{m_{\theta_n}}$ is post-selected $N_{n,\text{pass}}$ times after the stabilizer measurements.
In such a situation, because the ratios $N_{n,\text{sample}}/N_{\text{shot}}$ and $N_{n,\text{pass}}/N_{n,\text{sample}}$ coincide with $q_n^{\text{sample}}$ and $q_n^{\text{pass}}$, respectively, in the limit $N_{\text{shot}}\to \infty$,
we can numerically estimate the infidelity of the state prepared via our preparation protocol as follows:
\begin{equation}
\begin{aligned}
    1-F & = \frac{1}{p_{\text{suc}}}
    \sum_{n=0}^{k-1} q_n F_n\\
    &\simeq \frac{1}{p_{\text{suc}}} \sum_{n=0}^{k-1} \frac{N_{n,\text{sample}}}{N_{\text{shot}}}
    \cdot \frac{N_{n,\text{pass}}}{N_{n,\text{sample}}}F_n.\\
\end{aligned}
\end{equation}

In reality, we can numerically confirm that higher-order terms with $n\geq 2$ give almost negligible contributions to the infidelity.
Therefore, the following formula usually gives a good approximation of the infidelity:
\begin{equation}
    1-F\simeq 
    \frac{q_1^{\text{sample}}}{p_{\text{suc}}}  
    \cdot \frac{N_{1,\text{pass}}}{N_{1,\text{sample}}}\sin^2(\theta_{\text{error}}-\theta_*).
\end{equation}
where $p_{\text{suc}}$ is also approximately estimated as 
\begin{eqnarray}
    p_{\text{suc}} \simeq q_0^{\text{sample}} 
    \left(\frac{N_{0,\text{pass}}}{N_{0,\text{sample}}}\right)
    + q_1^{\text{sample}}
    \left(\frac{N_{1,\text{pass}}}{N_{1,\text{sample}}}\right).
\end{eqnarray}
Similarly, the Trace distance in Eq.~\eqref{eq: Trace distance of resource state} can be evaluated up to the leading term in $p_{\text{ph}}$.

\section{Definition of error rate for quantum channels}
\label{Appendix:definition of error rate}

Here we define two important quantities to quantify the error rate of quantum channels.

In the context of quantum information, the error rate of an arbitrary noise channel $\mathcal{E}(\rho)$ is quantified in various manners~\cite{Kliesch2021}.
The most common measure of error rate is the {\it average gate infidelity} (or the {\it average error rate})
\begin{equation}
    \varepsilon_{\text{av}}(\mathcal{E})
    = 1- \int d\psi \bra{\psi} \mathcal{E}(\ket{\psi}\bra{\psi})\ket{\psi}.
\end{equation}
This quantity is experimentally convenient as the value is efficiently estimated via randomized benchmarking~\cite{Emerson2005,Emerson2007,Knill2008,Dankert2009}. 
However, in generic situations where coherent errors could be most dominant than incoherent errors, the above measure is often insufficient to assess the impact of gate errors on arbitrary quantum algorithms.

The {\it diamond distance} from the identity, often referred to as the {\it worst-case error rate}, is known as a more stringent metric of gate errors. It is defined as follows: 
\begin{equation}
    \varepsilon_{\diamond}(\mathcal{E})
    = \frac12 \norm{\mathcal{E}-\mathcal{I}}_\diamond =
    \sup_{\rho} \frac12 \norm{(\mathcal{E}\otimes \mathcal{I}_D - \mathcal{I}_{D^2})(\rho)}_1,
\end{equation}
where $D=2^n$ is the dimension of the target system, $\mathcal{I}_D$ is the identity channel on the $D$-dimensional space, $\norm{A}_1= \Tr\sqrt{ A^\dagger A}$, and the supremum is over all density matrices $\rho$ of dimension $D^2$.
This metric can be intuitively interpreted via a maximum probability of distinguishing two quantum channels~\cite{Kliesch2021}, and
play an important role in the analyses of the rigorous fault-tolerance thresholds~\cite{Aharonov1997threshold,Kueng2016}.
Most importantly, the diamond distance satisfies not only the 
axiom of distance, but also two preferable properties: chaining property and stability~\cite{Kitaev1997_Review,Aharonov1998metric,Gilchrist2005}.
The chaining property states that composing two noisy channel never amplify the gate error by more than the sum of the two individual errors:
\begin{equation}
\label{eq:chaining property}
    \norm{\mathcal{E}_1\mathcal{E}_2-\mathcal{F}_1\mathcal{F}_2}_\diamond
    \ \leq \  \norm{\mathcal{E}_1-\mathcal{F}_1}_\diamond \ +\  \norm{\mathcal{E}_2-\mathcal{F}_2}_\diamond
\end{equation}
This guarantees that we can estimate the worst-case error bound to perform a quantum algorithm as the sum of the diamond distances of individual noisy components from ideal ones, which compose the entire quantum circuit.
The stability means that the diamond distance between two noisy channels is independent on how they are embedded in a larger Hilbert space:
\begin{equation}
    \norm{\mathcal{I}\otimes \mathcal{E}-\mathcal{I}\otimes \mathcal{F}}_\diamond = \norm{\mathcal{E}- \mathcal{F}}_\diamond
\end{equation}

For example, according to the Ref.~\cite{Kueng2016,Huang2019}, the average and the worst-case error rates are calculated as
\begin{equation}
\label{eq: error rate formula}
    \varepsilon_{\text{av}}(\mathcal{E}_1)= 2x/3,\ \ \ \varepsilon_{\diamond}(\mathcal{E}_1)=\sqrt{x^2+y^2}
\end{equation}
for the typical form of single-qubit noise channel:
\begin{equation}
    \mathcal{E}_1(\hat{\rho}) = (1-x)\hat{\rho} + iy(\hat{Z}\hat{\rho} -\hat{\rho} \hat{Z}) + x\hat{Z}\hat{\rho} \hat{Z} 
\end{equation}
This formula suggests that, for depolaring error channel~\eqref{eq:single depolarizing error}, the worst-case error rate equals $ p_{\text{ph}}$~\cite{Kliesch2021}, which has no significant difference from the value of the average infidelity $\frac23 p_{\text{ph}}$.
Meanwhile, considering an over-rotation error channel $\mathcal{E}_{\text{over}}(\hat{\rho}) = e^{i\theta \hat{Z}}\hat{\rho} e^{-i\theta \hat{Z}}$ ($\theta\ll\pi$), which corresponds to $(x,y)=(\sin^2\theta,\sin\theta\cos\theta)$, we notice that the worst-case error rate $\varepsilon_{\diamond}(\mathcal{E}_{\text{over}})=|\sin\theta|\simeq |\theta|$ is much larger than the average error rate $\varepsilon_{\diamond}(\mathcal{E}_{\text{over}})=\frac23 \sin^2\theta\simeq 2\theta^2/3$.
Thus, the average error rate tends to underestimate the effect of the coherent error term, which becomes most dominant in our state preparation protocols in the main text.

\section{Error accumulation in the RUS process}
\label{Appendix:RUS}

This section provide a supplementary analysis of the worst-case error rate of $\tilde{\mathcal{E}}_{\theta_*}$.
The effective error channel $\tilde{\mathcal{E}}_{\theta_*}$ is explicitly described as follows:

\begin{equation}
\begin{aligned}
    \tilde{\mathcal{E}}_{\theta_*}(\rho)\  &\equiv\  \sum_{K=1}^{\infty} \left(\frac12 \right)^K \mathcal{E}_{\theta_*}^K(\rho)\\
    &= (1-\tilde{x}_{\theta_*}) \rho +  i\tilde{y}_{\theta_*}(Z\rho -\rho Z)\\
    & \qquad\qquad\qquad+  \tilde{x}_{\theta_*} Z\rho Z + \order{|\theta_*|^2 p_{\text{ph}}^2},
\end{aligned}
\end{equation}
where 
\begin{equation}
    \tilde{x}_{\theta_*} \equiv \sum_{K=1}^{\infty} \left(\frac12 \right)^K x_{\theta_*}^K,\ \ \ 
    \tilde{y}_{\theta_*} \equiv \sum_{K=1}^{\infty} \left(\frac12 \right)^K y_{\theta_*}^K.
\end{equation}
Using the formula in Eq.~\eqref{eq: error rate formula}, we readily get the worst-case error rate for the channel $\tilde{\mathcal{E}}_{\theta_*}$ as $\varepsilon_\diamond(\tilde{\mathcal{E}}_{\theta_*}) \simeq\sqrt{(\tilde{x}_{\theta_*})^2+(\tilde{y}_{\theta_*})^2} $.
Here, to compare with the results in Fig.~\ref{fig:RUS_factor}, we introduce a prefactor $\alpha_{\text{RUS}}'$ as follows:
\begin{equation}
\label{eq:bare_RUS_factor}
    \varepsilon_\diamond(\tilde{\mathcal{E}}_{\theta_*}) = \alpha_{\text{RUS}}' |\theta_*|p_{\text{ph}}.
\end{equation}

\begin{figure}
    \centering
    \includegraphics[width=8cm]{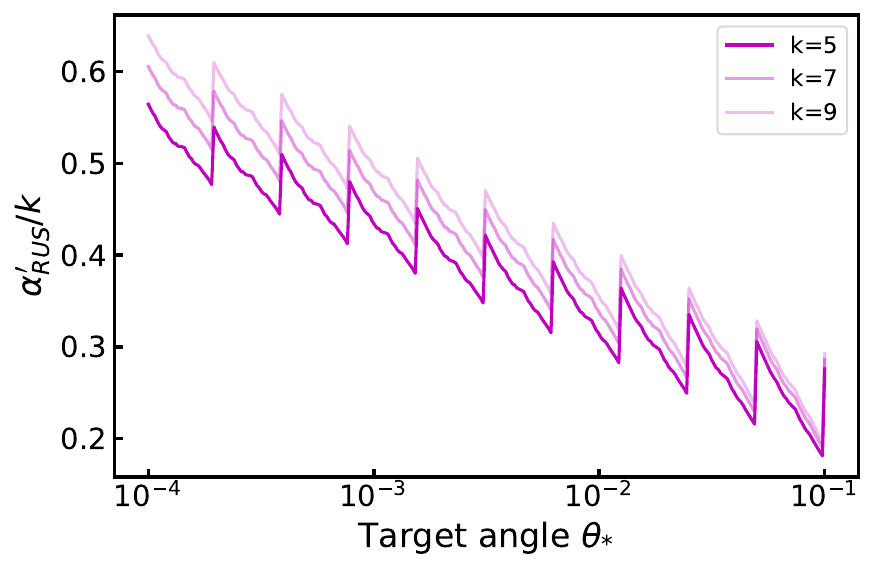}
    \caption{Numerical results for the prefactor $\alpha_{\text{RUS}}'$ in Eq.~\eqref{eq:bare_RUS_factor}. Because it scales almost linearly with $k$, we plot the curves of $\alpha_{\text{RUS}}/k$.}
    \label{fig:bare_RUS_factor}
\end{figure}

Fig.~\ref{fig:bare_RUS_factor} presents numerical results for the prefactor $\alpha_{\text{RUS}}'$, demonstrating its dependence on the parameter $k$ and the target angle $\theta_*$. These results indicate that, in small-angle region, the prefactor $\alpha_{\text{RUS}}'$ becomes slightly larger than those obtained when using probabilistic coherent error cancellation (as shown in Fig.~\ref{fig:RUS_factor}).
Furthermore, it should be noted that $\alpha_{\text{RUS}}'$ has a logarithmic dependence on the target angle $\theta_*$, unlike in $\alpha_{\text{RUS}}$ in the main text.

\section{Efficient implementation of the Hadamard test}
\label{Appendix: Hadamard test}

In this appendix, we show that we can implement the Hadamard test for $U=e^{-it\hat{\mathcal{H}}}$ with the total rotation angle $\theta_{\text{total}}=\frac{t}{2} \sum_{i=1}^L |a_i|$. Here we assume that the Hamiltonian $\hat{\mathcal{H}}$ is defined in Eq.~\eqref{eq:Hamiltonian}.

\begin{figure}[tb]
    \centering
    \vspace{0.3cm}
    \fontsize{11pt}{11pt}\selectfont
    \mbox{
        \Qcircuit @C=1em @R=.7em {
        \lstick{\ket{+}}    & \ctrl{1}  & \gate{\hat{H}}    & \measureD{M_{Z}}  \\
        \lstick{\ket{\psi_0}} & \gate{e^{-it \hat{\mathcal{H}}}}     & \qw & \qw  \\
        }
    }   
    \caption{Usual circuit for the Hadamard test.}
    \label{fig:Hadamard_test}
\end{figure}
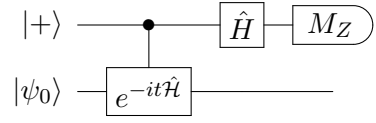

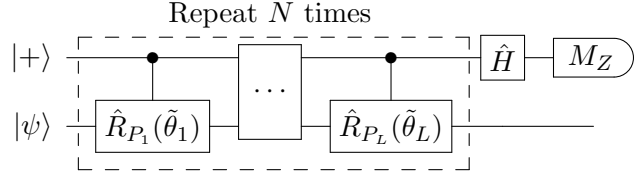
\begin{figure*}[bt]
    \hspace{1cm}
    \fontsize{11pt}{11pt}\selectfont
    \mbox{
        \Qcircuit @C=1.0em @R=0.7em {
        & & \mbox{Repeat $N$ times} & &\\
        \lstick{\ket{+}} & \ctrl{1}  & \multigate{1}{\cdots}  &   \ctrl{1} & \gate{\hat{H}}  &  \measureD{M_{Z}}  \\
        \lstick{\ket{\psi}} & \gate{\hat{R}_{P_1}(\tilde{\theta}_1)} & \ghost{\cdots}  & \gate{\hat{R}_{P_L}(\tilde{\theta}_L)}         & \qw & \qw \gategroup{2}{2}{3}{4}{1.2em}{--}  \\
        }
    }
    \vspace{0.2cm}
    \caption{Quantum circuit equivalent to that in Fig.~\ref{fig:Hadamard_test} under the approximation of the Trotter decomposition. Here, we employ the first-order Trotter decomposition for simplicity, instead of the second-order one discussed in the main text.}
    \label{fig:Hadamard test with First-order Trottter}
\end{figure*}

In the usual implementation of the Hadamard test (Fig.~\ref{fig:Hadamard_test}), we perform a series of unitary transformations
\begin{equation}
\label{eq:Hadamard_test}
\begin{aligned}
    &\ket{+}\ket{\psi} \ \ \xrightarrow{\Lambda(e^{-it\hat{\mathcal{H}}})} \ \ 
    \frac12\left(\ket{0}\ket{\psi} + \ket{1} e^{-it\hat{\mathcal{H}}} \ket{\psi}\right)\\
    &\xrightarrow{\ \hat{H}\ } \ \ \frac12\left(\ket{0}(I+e^{-it\hat{\mathcal{H}}})\ket{\psi}
    +\ket{1}(I-e^{-it\hat{\mathcal{H}}})\ket{\psi}\right),
\end{aligned}
\end{equation}
and then, measure the control qubit in the computational basis.
We then get the measurement outcome of $+1$ with the probability 
\begin{equation}
\label{eq:outcome_of_Hadamard}
    p_+ =   \norm{\frac12(I+e^{-it\hat{\mathcal{H}}})\ket{\psi}}^2 = \frac{1+\Re\ev{e^{-it\hat{\mathcal{H}}}}{\psi}}{2}.
\end{equation}
In an acutual implementation of the controlled gate $\Lambda(e^{-it\hat{\mathcal{H}}})$, we need to decompose it by applying the Trotter decomposition. 
For example, applying the first-order Trotter decomposition with Trotter number $N$, we can approximate the controlled gate as follow:
\begin{equation}
    \Lambda(e^{-it\hat{\mathcal{H}}}) \simeq \left(\prod_{i=1}^L \Lambda(\hat{R}_{P_i}(\tilde{\theta}_i))\right)^N,
\end{equation} 
where we introduce a series of rotation angles $\tilde{\theta}_i \equiv -a_it/N$
This leads to the explicit quantum circuit shown in Fig.~\ref{fig:Hadamard test with First-order Trottter}.
In conclusion, when performing the circuit in Fig.~\ref{fig:Hadamard_test}, we require the
the total rotation angle $\theta_{\text{total}}=N\sum_{i=1}^L|\tilde{\theta}_i|= t \sum_{i=1}^L |a_i|$.

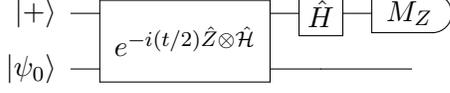
\begin{figure}[tb]
    \centering
    \vspace{0.3cm}
    \fontsize{11pt}{11pt}\selectfont
    \mbox{
        \Qcircuit @C=1em @R=.7em {
        \lstick{\ket{+}}    & \multigate{1}{e^{-i(t/2)\hat{Z}\otimes \hat{\mathcal{H}}}}   & \gate{\hat{H}}   & \measureD{M_{Z}}  \\
        \lstick{\ket{\psi_0}} & \ghost{e^{-i(t/2)\hat{Z}\otimes \hat{\mathcal{H}}}}     & \qw  & \qw \\
        }
    }   
    \caption{Implemetation of the Hadamard test using a rotation with conditional direction.}
    \label{fig:conditional_rotation}
\end{figure}

Another more efficient approach is to utilize rotation gates with conditional directions~\cite{Reiher2017}, as shown in Fig.~\ref{fig:conditional_rotation}. In this approach, we repeatedly use rotation gates in the form of $R_{Z\otimes P}(\theta)=e^{i\theta(\hat{Z}\otimes \hat{P})}$, where $\hat{Z}\otimes \hat{P}$ is a direct product between the Pauli-$Z$ operator of the ancilla qubit and some Pauli string of target systems. First we find that applying the gate $\hat{R}_{Z\otimes P}(\theta)$ to the initial state $\ket{+}\ket{\psi}$ for the Hadamard test leads to 
\begin{equation}
    \hat{R}_{Z\otimes P}(\theta)\ket{+}\ket{\psi} =
    \frac12\left(\ket{0}\hat{R}_{\hat{P}}(\theta)\ket{\psi} + \ket{1} \hat{R}_{\hat{P}}(-\theta)\ket{\psi}\right).
\end{equation}
In the same way, when performing a series of gates $\left(\prod_i \hat{R}_{Z\otimes P_i}(\tilde{\theta}_i/2)\right)^N$, we obtain the following final state:
\begin{equation}
\begin{aligned}
    &\left(\prod_i \hat{R}_{Z\otimes P_i}(\tilde{\theta}_i/2)\right)^N \ket{+}\ket{\psi}\\
    & \ \ \ = \frac12\left[\ket{0}\left(\prod_i \hat{R}_Z(\tilde{\theta}_i/2)\right)^N\ket{\psi}\right.\\
    &\ \ \ \ \ \ \ \ \ \ \ \ \ \ \ \ \ \ \ \ 
    \left. + \ket{1}\left(\prod_i \hat{R}_Z(-\tilde{\theta}_i/2)\right)^N\ket{\psi}
    \right].\\
    & \ \ \ \simeq \frac12\left(
    \ket{0} e^{-it\hat{\mathcal{H}}/2}\ket{\psi} 
    + \ket{1} e^{it\hat{\mathcal{H}}/2}\ket{\psi} 
    \right), \\
\end{aligned}
\end{equation}
Finally, we perform the Hadamard gate to the ancilla qubit, 
\begin{equation}
\begin{aligned}
    \xrightarrow{\ \hat{H}\ } \ \ \ &\frac12\left(
    \ket{0} (e^{it\hat{\mathcal{H}}/2}+e^{-it\hat{\mathcal{H}}/2})\ket{\psi} \right.\\
    &\qquad\qquad\qquad \left. + \ket{1} (e^{it\hat{\mathcal{H}}/2}-e^{-it\hat{\mathcal{H}}/2})\ket{\psi} 
    \right),
\end{aligned}
\end{equation}
and then, measure the control qubit in the computational basis, yielding the outcome of $+1$ with the probability 
\begin{equation}
    p_+ =   \norm{\frac12(e^{it\hat{\mathcal{H}}/2}+e^{-it\hat{\mathcal{H}}/2})\ket{\psi}}^2 = \frac{1+\Re\ev{e^{-it\hat{\mathcal{H}}}}{\psi}}{2}.
\end{equation}
Comparing with Eq.~\eqref{eq:outcome_of_Hadamard}, we notice that this distribution of the outcome equals the one obtained with the usual Hadamard test for $\hat{U}=e^{-it\hat{\mathcal{H}}}$. However, the present case requires the total rotation $\theta_{\text{total}}=\frac{t}{2} \sum_i |a_i|$, which is the half of that discussed in the previous approach.

Furthermore, if the target Hamiltonian satisfies some reasonable conditions, we can implement the Hadamard test without directly controlling the time evolution operator~\cite{Lin2022,Dong2022preparation}.
For example, such an implementation is achieved by controlling specific simple gates that obey the anti-commutation relation with the target Hamiltonian~\cite{Dong2022preparation}.
These techniques offer the advantage of reducing the total execution time by enhancing the parallelism of rotation gates within the STAR architecture.

\section{Details of the Hubbard model}

Here we briefly explain how to derive the Hamiltonian in Eq.~\eqref{eq:Spin Hamitonian}.
The Hamiltonian of the 2D Hubbard model is originally defined as 
\begin{equation}
\label{eq:Usual Hubbard Hamiltonian}
    \hat{\mathcal{H}} = -t \sum_{\langle i,j \rangle, \sigma} (\hat{c}^{\dag}_{i, \sigma}\hat{c}_{j, \sigma} + {\rm h.c.}) + U \sum_{i} \hat{n}_{i, \uparrow} \hat{n}_{i, \downarrow}, 
\end{equation}
where $\hat{c}^{(\dag)}_{i, \sigma} \ (\sigma = \uparrow, \downarrow)$ are fermionic creation (annihilation) operators, and $\hat{n}_{i, \sigma} = \hat{c}^{\dag}_{i, \sigma} \hat{c}_{i, \sigma}$ is the number operator for the corresponding fermionic mode. 
The notation $\langle i,j \rangle$ indicates any pairs of adjacent sites on the 2D square lattice. 
The parameters $t$ and $U$ indicate the strength of the hopping and on-site Coulomb interaction, respectively. 

Here let us consider the decomposition of the on-site Coulomb interaction as follows~\cite{Campbell2021early}:
\begin{equation}
\begin{aligned}
    \sum_{i} \hat{n}_{i, \uparrow} \hat{n}_{i, \downarrow} = \sum_{i} (\hat{n}_{i, \uparrow}-\hat{I}/2) (\hat{n}_{i, \downarrow}-\hat{I}/2) + \frac12 \hat{N}  + \frac14 \hat{I},
\end{aligned}
\end{equation}
where $\hat{N}\equiv \sum_{i}(\hat{n}_{i, \uparrow}+\hat{n}_{i, \downarrow})$ is the total electron number operator. Since the last two terms are constant in the Hilbert subspace with a constant particle number, we can remove them from the Hamiltonian for QPE. 
Therefore, we will deal with the following Hamiltonian instead of Eq.~\eqref{eq:Usual Hubbard Hamiltonian}: 
\begin{equation}
\label{eq:Shifted Hubbard Hamiltonian}
\begin{aligned}
    \hat{\mathcal{H}} &= -t \sum_{\langle i,j \rangle, \sigma} (\hat{c}^{\dag}_{i, \sigma}\hat{c}_{j, \sigma} + {\rm h.c.})\\
    &\qquad \qquad + U \sum_{i} (\hat{n}_{i, \uparrow}-\hat{I}/2) (\hat{n}_{i, \downarrow}-\hat{I}/2).
\end{aligned}
\end{equation}
Finally, by performing the Jordan-Wigner transformation,
\begin{eqnarray} \label{eq:jwt}
    \hat{c}_{i, \sigma} = \frac{1}{2}\left( \hat{X}_i + i \hat{Y}_i \right) \hat{Z}_{i-1} \cdots \hat{Z}_{1}, \\
    \hat{c}^{\dag}_{i, \sigma} = \frac{1}{2}\left( \hat{X}_i - i \hat{Y}_i \right) \hat{Z}_{i-1} \cdots \hat{Z}_{1},
\end{eqnarray}
we can rewrite the Hamiltonian in Eq.~\eqref{eq:Shifted Hubbard Hamiltonian} 
into its spin representation, yielding Eq.~\eqref{eq:Spin Hamitonian} in the main text.

\bibliographystyle{apsrev4-2}
\bibliography{citation}

\begin{thebibliography}{208}
\bibitem [{com()}]{comment}%
  \BibitemOpen
  \href@noop {} {}\bibinfo {note} {Please note that, throughout this paper, we use a notation of rotation gates that is slightly different from the one used commonly, i.e., $\hat{R}_{z,i}'(\theta)=e^{-i\theta\hat{Z}_i/2}$ to make simpler our formulae obtained below.}\BibitemShut {Stop}%
\end{thebibliography}

\end{document}